\documentclass[superscriptaddress,twocolumn,amsmath,amssymb,prd,floatfix,nofootinbib]{revtex4-2}
\usepackage[dvipdfmx]{graphicx}% Include figure files
\usepackage{dcolumn}% Align table columns on decimal point
\usepackage[varg]{txfonts}
\usepackage{bm}% bold math
\usepackage{empheq}%
\usepackage{multirow}

\usepackage{color}% draft version

\newcommand{\sY}[2]{{}_{#1}\hspace*{-0.1ex}Y_{#2}}

\begin{document}
%\begin{CJK}{UTF8}{} % Use default fonts from CJK (see below)

\title{Integrated perturbation theory for cosmological tensor
  fields. III. Projection effects
%  \\ \ecom{preliminary draft (do not distribute)}
}

\author{Takahiko Matsubara} \email{tmats@post.kek.jp}
\affiliation{%
  Institute of Particle and Nuclear Studies, High Energy
  Accelerator Research Organization (KEK), Oho 1-1, Tsukuba 305-0801,
  Japan}%
\affiliation{%
  The Graduate Institute for Advanced Studies, SOKENDAI,
  Tsukuba 305-0801, Japan}%

\date{\today}% It is always \today, today,

\begin{abstract}
  The integrated perturbation theory (iPT) is a set of methods in
  nonlinear perturbation theory for the structure formation in the
  Universe. In Papers~I and II \cite{PaperI,PaperII}, the basic
  formalism and technical methods of the iPT for cosmological tensor
  fields are developed, generalizing the corresponding theory for
  scalar fields. In previous papers, methods to predict statistical
  quantities, such as power spectra, correlation functions, etc., of
  three-dimensional tensor fields are developed based on the iPT.
  However, observations of tensors, such as angular momenta and shapes
  of galaxies, etc., are only possible after the three-dimensional
  tensors are projected onto the two-dimensional sky. In this paper,
  power spectra and correlation functions of projected two-dimensional
  tensors are related to those of original three-dimensional tensors,
  so that one can make predictions for the observable statistics of
  projected tensor fields from the iPT. The relations are consistently
  represented on the basis of irreducible decomposition of both two-
  and three-dimensional tensors.
\end{abstract}

%\pacs{
% 98.80.-k,
% 98.65.-r,
% 98.80.Cq,
% 98.80.Es
% }% PACS, the Physics and Astronomy

\maketitle

%\end{CJK}

\section{\label{sec:Introduction}%
  Introduction
}

The statistics of the large-scale structure of the Universe play an
important role in observationally constraining possible cosmological
models. Significant progress has been made in recent years in redshift
surveys of galaxies, and the large amount of survey data expected in
the near future will allow us to test exquisite details of the
conceivable cosmological models. The galaxies are biased tracers of
mass, and observations of the spatial distribution of galaxies are used as
powerful tools to constrain the models.

The spatial positions of galaxies are not the only information that
can be extracted from galaxy surveys. In weak-lensing surveys, galaxy
shapes are precisely measured, and their spatial correlations are used
as a probe of the clustering of dark matter along the lines of sight
between the galaxies and an observer
\cite{Gunn1967,Miralda1991,Kaiser:1991qi,Bartelmann:1999yn}. The
spatial correlations of galaxy shapes are determined not only by the
clustering of dark matter along the lines of sight, but also by the
intrinsic alignment of the galaxies, which is not a lensing-induced
effect. While the intrinsic alignment can potentially be a source of
systematic errors in signals of weak lensing
\cite{Croft:2000gz,Heavens:2000ad,Catelan:2000vm,Jing:2002mm,Hirata:2004gc},
the statistical properties of the intrinsic alignment itself can
provide another observational way to constrain cosmological models
\cite{Okumura:2008du,Blazek:2011xq,Chisari:2013dda,Schmidt:2015xka,Joachimi:2015mma,Chisari:2016xki,Kogai:2018nse,Okumura:2019ned,Okumura:2019ozd,Okumura:2020hhr,Taruya:2020tdi,Akitsu:2020jvx,Kogai:2020vzz,Okumura:2021xgc,Kurita:2022agh,Kurita:2023qku}.
Statistics of angular momenta or spins of galaxies also contain
cosmological information
\cite{Peebles:1969jm,White:1984uf,Catelan:1996hv,Catelan:1996hw,Lee:1999ii,Lee:2000br,Crittenden:2000wi,Yu:2019bsd,Motloch:2020qvx,Motloch:2021mfz}.
As the galaxy surveys advance in both quality and quantity, the
importance of statistics beyond the position clustering of galaxies
increases, such as the galaxy spins and shapes, etc.

On one hand, intrinsic alignments of galaxies are quantitatively
characterized by the moments of galaxy shapes or images, and thus
tensor values are attributed to every galaxy. On the other hand, the
angular momenta of galaxies are (pseudo)vectors, and thus vector
values are attributed to every galaxy. Directional correlations of
these vectors and tensors can be useful in statistically constraining
cosmological models. One can also consider the scalar values are
attributed to every galaxy, such as the mass, size, color, etc.
Generalizing these, there are statistics in which a tensor value is
attributed to each galaxy. In the above example, the rank-$l$ of the
tensor is given by $l=0$ for the mass, size, and color, $l=1$ for the
angular momentum, and $l=2$ for the intrinsic alignment. In principle,
the higher-order moments of galaxy shape are also used for
characterizing the intrinsic alignments \cite{Kogai:2020vzz}, and the
rank-$l$ of tensor values can take an arbitrary value of non-negative
integer, aside from the observability of higher-rank shape tensors in
practice. Theoretical characterizations and predictions for
statistical properties of galaxies are important for cosmological
applications. Recently, the correlation statistics of intrinsic
alignment are analytically modeled by applying the nonlinear
perturbation theory in different ways
\cite{Blazek:2015lfa,Blazek:2017wbz,Schmitz:2018rfw,Vlah:2019byq,Vlah:2020ovg,Taruya:2021jhg}.
These developments generalize the nonlinear perturbation theory to
describe the correlation statistics of galaxy shapes. In the above
context, they are statistics of rank-2 tensors.

In Papers~I and II \cite{PaperI,PaperII}, we generalize the integrated
perturbation theory (iPT)
\cite{Matsubara:2011ck,Matsubara:2012nc,Matsubara:2013ofa,Matsubara:2016wth}
to include the case in the presence of tensor-valued bias, using the
irreducible decomposition of the tensor degrees of freedom attributed
to astronomical objects like galaxies and halos. In this theory,
arbitrary rank-$l$ of the tensor attached to the objects is described
in a single formalism of nonlinear perturbation theory. Therefore, the
correlation statistics of the number density of galaxies, angular
momentum and shapes of galaxies, and so forth, are described in a
unified formalism of the iPT. In Paper~I, the basic formalism of the
theory is explained in detail, and in Paper~II, useful methods of
calculating nonlinear effects, i.e., loop corrections, are developed
and explained with working examples. These previous papers provide the
method to calculate the correlation statistics of three-dimensional
tensor fields by the perturbation theory.

However, all the components of a three-dimensional tensor are not
observable quantities, because the observable components of tensors
attributed to objects are two-dimensionally projected components onto
the sky. Therefore, one needs to transform the correlation statistics
of three-dimensional tensor fields to those of projected tensor fields
on the sky. While the spatial positions of objects are still
three dimensional in redshift surveys, the tensors attributed to
observed objects are projected onto the two-dimensional surface
perpendicular to lines of sight. Therefore, we need to consider the
projection effects of the tensor fields in order to derive the
observational quantities from the theoretical predictions of the iPT
of three-dimensional tensor fields. This paper addresses this problem,
and we derive analytically useful equations for that purpose.

The projected two-dimensional tensor fields are distributed in
three-dimensional space. As the three-dimensional tensors are
decomposed into irreducible tensors in previous Papers~I and II, the
projected two-dimensional tensors are correspondingly decomposed into
irreducible tensors as well. The rank of the projected tensors are
frequently called the {\em spin} $s$, and we use the same wording in
this paper, although we should not confuse the term spin here with the
angular momentum or the angular velocity of galaxies. In observational
cosmology, the spin-2 fields frequently appear, as the polarization
field of the cosmic microwave background (CMB), and the galaxy shape
fields in the contexts of weak lensing and intrinsic alignment.
Irreducible decomposition of the spin-2 field is well investigated in
the previous study. In this paper, we first formally generalize the
previous work of spin-2 fields in literature to describe tensor fields
of arbitrary ranks and spins, and then derive numerous formulas for
the correlation statistics in the distant-observer limit. The results
are related to the invariant power spectra introduced in Papers~I and
II, with which the results of the iPT for tensor fields can be applied
in order to predict the correlation statistics of projected fields.

This paper is organized as follows. In Sec.~\ref{sec:FullSky}, we
generally derive the relation between the three-dimensional
irreducible tensor and their projected two-dimensional irreducible
tensor in full-sky, spherical coordinates. Symmetric properties, such
as complex conjugate, rotation, and parity are comprehensively given
for related statistics of tensor fields in the full sky. In
Sec.~\ref{sec:FlatSky}, the distant-observer and flat-sky limits of
the projected tensor fields are considered. Explicit relations are
derived between irreducible tensors in spherical coordinates of the
full-sky and those in Cartesian coordinates of the distant-observer,
flat-sky limit, by which approximation the iPT for tensor fields
developed in Papers~I and II can be applied. In
Sec.~\ref{sec:PowerSpec}, taking the distant-observer limit, relations
between the full-sky power spectra in spherical coordinates and the
distant-observer power spectra in Cartesian coordinates are derived,
and the latter spectra are represented by invariant spectra of
originally three-dimensional tensor fields. In
Sec.~\ref{sec:CorrFunc}, the correlation function of projected tensor
fields is similarly represented by invariant spectra. In
Sec.~\ref{sec:IPT}, the results of substituting the invariant spectra
from the iPT are explicitly presented in the lowest-order
approximations. The conclusions are summarized in
Sec.~\ref{sec:Conclusions}. In the Appendix, a proof of a formula of
the spin-weighted spherical harmonics in the flat-sky limit is given.

\section{\label{sec:FullSky}%
   Projected tensor fields in the full sky
}

\subsection{%
  Decomposition of projected tensors into the spherical basis
}

\subsubsection{%
  Irreducible decomposition of projected tensors
}

In Papers~I and II, a tensor field in general is denoted by
$F_{Xi_1\cdots i_l}(\bm{r})$, where $X$ specifies the kind of objects
such as galaxies we observe and $i_1,\ldots,i_l$ are Cartesian indices
of the rank-$l$ tensor which is attributed to the objects. For a given
tensor field, the corresponding projected field of rank-$s$ is defined
by
\begin{equation}
  f_{Xi_1\cdots i_s}(\bm{r}) \equiv
  \mathcal{P}_{i_1j_1}(\hat{\bm{r}}) \cdots
  \mathcal{P}_{i_sj_s}(\hat{\bm{r}})
  F_{Xj_1\cdots j_s}(\bm{r}),
  \label{eq:1}
\end{equation}
where 
\begin{equation}
    \mathcal{P}_{ij}(\hat{\bm{r}}) \equiv
   \delta_{ij} - \hat{r}_i \hat{r}_j
    \label{eq:2}
\end{equation}
is the projection tensor, and $\hat{\bm{r}} \equiv \bm{r}/r$ is a unit
vector parallel to the radial coordinates $\bm{r}$. Observationally,
the projected field corresponds to the two-dimensional tensor on the
sky, where $\hat{\bm{r}}$ corresponds to the direction to the line of
sight. The projected tensor does not have a radial component and can
have nonzero value only for tangential components on the sky. That
is, projected tensors are essentially two-dimensional tensors on the
sky.

In general, the two-dimensional symmetric tensor $t_{i_1\cdots i_s}$ of
rank-$s \geq 2$ can be uniquely decomposed into traceless tensors as
\begin{align}
  t_{ij} &= t^{(2)}_{ij} + \frac{1}{2} \delta_{ij} t^{(0)},
  \label{eq:3}\\
  t_{ijk} &= t^{(3)}_{ijk} + \frac{3}{4}\delta_{(ij} t^{(1)}_{k)},
  \label{eq:4}\\
  t_{ijkl} &= t^{(4)}_{ijkl} + \delta_{(ij} t^{(2)}_{kl)}
             + \frac{3}{8} \delta_{(ij} \delta_{kl)} t^{(0)},
  \label{eq:5}
\end{align}
and so forth, where the parentheses in the subscripts indicate the
symmetrization of the indices inside them, and
$t^{(l)}_{i_1\cdots i_s}$ are traceless parts of an original symmetric
tensor $t_{i_1\cdots i_s}$. Any traces of the traceless parts vanish
as $t^{(l)}_{i_1\cdots j\cdots j\cdots i_s} = 0$, where we adopt
Einstein summation convention and repeated indices are implicitly
summed over without summation symbols here and throughout this paper.
Traceless parts are also symmetric tensors. A general way of
decomposing a three-dimensional symmetric tensor into traceless parts
is described in Paper~I \cite{PaperI}, and the same method also
applies to two-dimensional tensors. Any two-dimensional symmetric
tensor of rank-$s$ can be decomposed into
\begin{equation}
  t_{i_1i_2\cdots i_s} =
  t^{(s)}_{i_1i_2\cdots i_s}
  + \frac{s}{4} \delta_{(i_1i_2} t^{(s-2)}_{i_3\cdots i_s)} + \cdots.
  \label{eq:6}
\end{equation}
The traceless parts $t^{(s)}_{i_1\cdots i_s}$ are generally given
by
\begin{multline}
  t^{(s)}_{i_1\cdots i_s} =
  s \sum_{k=0}^{[s/2]} \frac{(-1)^k (s-k-1)!}{4^k k! (s-2k)!}
  \delta_{(i_1i_2} \cdots \delta_{i_{2k-1}i_{2k}}
  t^{(s:k)}_{i_{2k+1}\cdots i_{s})},
  \label{eq:7}
\end{multline}
where $[s/2]$ is the Gauss symbol, i.e., $[s/2]=s/2$ if $s$ is an even
integer and $[s/2] = (s-1)/2$ if $s$ is an odd integer, and
\begin{align}
  t^{(s:k)}_{i_{2k+1}\cdots i_{s}}
  &= \delta_{i_1i_2} \delta_{i_3i_4}\cdots \delta_{i_3\cdots i_{2k-1}i_{2k}}
    t_{i_1\cdots i_{2k-1}i_{2k}\cdots i_s}
  \nonumber\\
  &=  t_{j_1j_1j_2j_2\cdots j_k j_k i_{2k+1}\cdots i_s}
  \label{eq:8}
\end{align}
are traces of the original tensor taken $k$ times.
Equation~(\ref{eq:7}) can be proven by similar considerations given in,
e.g., an Appendix of Ref.~\cite{Kogai:2020vzz}. Tensors of rank-0 and
rank-1 are already traceless as any trace cannot be taken from them.
Recursive applications of Eq.~(\ref{eq:7}) derive explicit
decomposition of Eqs.~(\ref{eq:3})--(\ref{eq:6}) for symmetric
tensors of any higher-order rank. The rank-$s$ of a two-dimensional
projected tensor is usually called ``spin,'' and we follow this usage
of the word in the following. One should not confuse the spin of a
projected tensor with the spin of a galaxy, which usually refers to
the angular momentum or angular velocity of the galaxy.

\subsubsection{%
  Two-dimensional tensor decomposition on the spherical basis
}

We decompose the projected field $f_{Xi_1\cdots i_s}(\bm{r})$ of
Eq.~(\ref{eq:1}) according to the above procedure, and denote the
traceless parts as $f^{(s)}_{Xi_1\cdots i_s}(\bm{r})$. Any traceless,
symmetric tensor can be decomposed into a two-dimensional spherical
basis. In the case of intrinsic alignment with spin-2 tensors, the
decomposition is explicitly given in Ref.~\cite{Vlah:2020ovg}. For
projected tensors of general spin-$s$, the decomposition is given by
\begin{equation}
  f^{(s)}_{Xi_1\cdots i_s} =
%  i^s \sqrt{\alpha_s}
%  \left(
    \tilde{f}^{(+s)}_X \tilde{\mathrm{m}}^+_{i_1} \cdots \tilde{\mathrm{m}}^+_{i_s} +
    \tilde{f}^{(-s)}_X \tilde{\mathrm{m}}^-_{i_1} \cdots
    \tilde{\mathrm{m}}^-_{i_s},
%  \right),
  \label{eq:9}
\end{equation}
where $\tilde{\mathrm{m}}^\pm_i$ are components of vectors
\begin{equation}
    \tilde{\mathbf{m}}^{\pm} \equiv
    \mp \frac{\mathbf{e}_\theta \mp i \mathbf{e}_\phi}{\sqrt{2}},
    \label{eq:10}
\end{equation}
and $\bm{\mathrm{e}}_\theta$ and $\bm{\mathrm{e}}_\phi$ are unit bases
of spherical coordinates $(\theta,\phi)$ located in the direction of
$\hat{\bm{r}}$. In this paper, the tildes on variables in most cases
indicate that the corresponding variables are defined in the full-sky
spherical coordinates. In later sections, we consider the flat-sky
limit of those variables, in which corresponding variables are more or
less changed from the full-sky case. The spherical bases satisfy
$\tilde{\mathbf{m}}^\pm \cdot \tilde{\mathbf{m}}^\pm = 0$ and
$\tilde{\mathbf{m}}^\pm \cdot \tilde{\mathbf{m}}^\mp = -1$. Thus the
decomposition of Eq.~(\ref{eq:9}) is inverted as
\begin{align}
  \tilde{f}^{(\pm s)}_X
  &=
    (-1)^s f^{(s)}_{Xi_1\cdots i_s}
    \tilde{\mathrm{m}}^\mp_{i_1} \cdots \tilde{\mathrm{m}}^\mp_{i_s}
    \label{eq:11}\\
  &=
    (-1)^s f_{Xi_1\cdots i_s}
    \tilde{\mathrm{m}}^\mp_{i_1} \cdots \tilde{\mathrm{m}}^\mp_{i_s}.
    \label{eq:12}
\end{align}
The second equality above holds because all the trace parts of
$f_{Xi_1\cdots i_s}$ contain Kronecker's delta symbols [as
Eqs.~(\ref{eq:3})--(\ref{eq:6})] and we have
$\tilde{\mathbf{m}}^\pm \cdot \tilde{\mathbf{m}}^\pm = 0$, and thus
the trace parts $f^{(s)}_{Xi_1\cdots i_s}$ can be replaced by the
original tensor field $f_{Xi_1\cdots i_s}$ in the expression. One
should note that the replacement is only valid if the rank $s$ of the
traceless part $f^{(s)}_{Xi_1\cdots i_s}$ is the same as the rank of
the original tensor $f_{Xi_1\cdots i_s}$, and does not hold for trace
parts, such as the second and subsequent terms on the right-hand sides
of Eqs.~(\ref{eq:3})--(\ref{eq:6}).

\subsubsection{%
  Relations to the three-dimensional tensor decomposition
}

As explicitly described in Paper~I, three-dimensional tensors are also
uniquely decomposed into traceless parts in general. Corresponding to
Eqs.~(\ref{eq:3})--(\ref{eq:5}) in the two-dimensional case, the
decompositions of three-dimensional tensors are given by
\begin{align}
  T_{ij} &= T^{(2)}_{ij} + \frac{1}{3} \delta_{ij} T^{(0)},
  \label{eq:13}\\
  T_{ijk} &= T^{(3)}_{ijk} + \frac{3}{5}\delta_{(ij} T^{(1)}_{k)},
  \label{eq:14}\\
  T_{ijkl} &= T^{(4)}_{ijkl} + \frac{6}{7} \delta_{(ij} T^{(2)}_{kl)}
             + \frac{1}{5} \delta_{(ij} \delta_{kl)} T^{(0)},
  \label{eq:15}
\end{align}
and so forth. The traceless part $F^{(l)}_{Xi_1\cdots i_l}(\bm{r})$ of
a three-dimensional tensor field $F_{Xi_1\cdots i_l}(\bm{r})$ is
decomposed by a three-dimensional spherical basis as
\begin{equation}
  F^{(l)}_{Xi_1\cdots i_l}(\bm{r}) =
  A_l  
  \tilde{F}_{Xlm}(\bm{r})
  \tilde{\mathsf{Y}}^{(m)}_{i_1\cdots i_l}(\hat{\bm{r}}),
  \label{eq:16}
\end{equation}
where the prefactor $A_l$ is the normalization convention of
spherical tensors defined in Paper~I, with
\begin{equation}
  A_l \equiv \sqrt{\frac{l!}{(2l-1)!!}}.
  \label{eq:17}
\end{equation}
The Einstein summation convention is applied and the index $m$ on the
right-hand side (rhs) of Eq.~(\ref{eq:16}) is summed over
$m=0,\pm 1, \cdots \pm l$. The spherical bases
$\tilde{\mathsf{Y}}^{(m)}_{i_1\cdots i_l}$ are defined and constructed
by a set of basis
\begin{equation}
  \tilde{\mathbf{e}}^0 = \mathbf{e}_r, \quad
  \tilde{\mathbf{e}}^\pm =
  \mp \frac{\mathbf{e}_\theta \mp i\mathbf{e}_\phi}{\sqrt{2}},
  \label{eq:18}
\end{equation}
where $\mathbf{e}_r = \hat{\bm{r}}$ is a unit vector of radial
direction.
For example,
\begin{equation}
  \tilde{\mathsf{Y}}^{(0)}(\hat{\bm{r}}) = 1
  \label{eq:19}
\end{equation}
for tensors of rank-0,
\begin{equation}
  \tilde{\mathsf{Y}}^{(0)}_i(\hat{\bm{r}}) =
  {\tilde{\mathrm{e}}^0\phantom{}}_i, \quad
  \tilde{\mathsf{Y}}^{(\pm 1)}_i(\hat{\bm{r}}) =
  {\tilde{\mathrm{e}}^\pm\phantom{}}_i
  \label{eq:20}
\end{equation}
for tensors of rank-1, and
\begin{align}
  &
    \tilde{\mathsf{Y}}^{(0)}_{ij}(\hat{\bm{r}})
    = \sqrt{\frac{3}{2}}
    \left(
    {\tilde{\mathrm{e}}^0\phantom{}}_i\,
    {\tilde{\mathrm{e}}^0\phantom{}}_j - \frac{1}{3} \delta_{ij}
    \right),
  \label{eq:21}\\
  &
    \tilde{\mathsf{Y}}^{(\pm 1)}_{ij}(\hat{\bm{r}})
    = \sqrt{2}\,
    {\tilde{\mathrm{e}}^0\phantom{}}_{(i}\,
    {\tilde{\mathrm{e}}^\pm\phantom{}}_{j)},
  \label{eq:22}\\
  &
    \tilde{\mathsf{Y}}^{(\pm 2)}_{ij}(\hat{\bm{r}})
    =
    {\tilde{\mathrm{e}}^\pm\phantom{}}_i\,
    {\tilde{\mathrm{e}}^\pm\phantom{}}_j
  \label{eq:23}
\end{align}
for tensors of rank-2, where
${\tilde{\mathrm{e}}^m\phantom{}}_i = [\tilde{\mathbf{e}}^m]_i$ with
$m=0,\pm$ are the Cartesian components of the spherical basis. The
spherical bases in Cartesian coordinates,
$\mathsf{Y}^{(m)}_{i_1\cdots i_l}$, are given in Paper~I in terms of
the spherical basis $\mathbf{e}^m$ in the Cartesian coordinates
system, which is shortly defined below. The relations between
spherical bases
$\tilde{\mathsf{Y}}^{(m)}_{i_1\cdots i_l}(\hat{\bm{r}})$ and the basis
$\tilde{\mathbf{e}}^m$ in spherical coordinates here are just the same
as the relations between those in Cartesian coordinates in Paper~I.
The spherical bases $\tilde{\mathsf{Y}}^{(m)}_{i_1\cdots i_l}$ in
spherical coordinates depend on the radial direction $\hat{\bm{r}}$,
while the bases $\mathsf{Y}^{(m)}_{i_1\cdots i_l}$ in Cartesian
coordinates do not.

In Paper~I, the spherical basis $\mathsf{Y}^{(m)}_{i_1\cdots i_l}$ is
constructed from the spherical basis in Cartesian coordinates,
\begin{equation}
  \mathbf{e}^0 = \hat{\mathbf{e}}_3, \quad
  \mathbf{e}^\pm =
  \mp \frac{\hat{\mathbf{e}}_1 \mp i \hat{\mathbf{e}}_2}{\sqrt{2}},
\label{eq:24}
\end{equation}
where $\hat{\mathbf{e}}_i$ ($i=1,2,3$) are unit vectors of Cartesian
basis. The spherical basis $\tilde{\mathbf{e}}^m$ of Eq.~(\ref{eq:18})
is obtained by rotating the spherical basis $\mathbf{e}^m$ of Paper~I.
The relations between unit vectors in spherical and Cartesian
coordinates systems are given by
\begin{equation}
  \begin{pmatrix}
    \mathbf{e}_\theta \\ \mathbf{e}_\phi \\ \mathbf{e}_r
  \end{pmatrix}
  = S(\hat{\bm{r}})
  \begin{pmatrix}
    \hat{\mathbf{e}}_1 \\ \hat{\mathbf{e}}_2 \\ \hat{\mathbf{e}}_3
  \end{pmatrix},
  \label{eq:25}
\end{equation}
where $S(\hat{\bm{r}})$ is a $3\times 3$ matrix,
\begin{equation}
  S(\hat{\bm{r}}) =
  \begin{pmatrix}
    \cos\theta\,\cos\phi & \cos\theta\,\sin\phi & -\sin\theta \\
    -\sin\phi & \cos\phi & 0 \\
    \sin\theta\,\cos\phi & \sin\theta\,\sin\phi & \cos\theta
  \end{pmatrix},
  \label{eq:26}
\end{equation}
and $(\theta,\phi)$ is the spherical coordinates of the unit vector
$\hat{\bm{r}}$. As one can explicitly see, the relation between the
components of spherical bases, Eqs.~(\ref{eq:18}) and (\ref{eq:24})
are given by
\begin{equation}
  {\tilde{\mathrm{e}}^m\phantom{}}_i =
  {\mathrm{e}^m}_j\, S_{ji}(\hat{\bm{r}}),
  \label{eq:27}
\end{equation}
where $S_{ji}$ are the components of the matrix $S$, and
${\mathrm{e}^m}_i = [\mathbf{e}^m]_i$ with $m=0,\pm$ are the Cartesian
components of the spherical basis of Eq.~(\ref{eq:24}). Therefore,
the spherical basis in spherical coordinates and that in Cartesian
coordinates are related by
\begin{equation}
  \tilde{\mathsf{Y}}^{(m)}_{i_1\cdots i_l}(\hat{\bm{r}})
  = 
  {\mathsf{Y}}^{(m)}_{j_1\cdots j_l} S_{j_1i_1}(\hat{\bm{r}})
  \cdots S_{j_1i_1}(\hat{\bm{r}}).
  \label{eq:28}
\end{equation}
That is, the spherical tensors
$\tilde{\mathsf{Y}}^{(m)}_{i_1\cdots i_l}(\hat{\bm{r}})$ are obtained
by passively rotating the tensors $\mathsf{Y}^{(m)}_{i_1\cdots i_l}$
with a rotation of axes according to Eq.~(\ref{eq:25}). Because of the
orthogonality relations described in Paper~I of the spherical tensors
in Cartesian coordinates, the same relations hold in spherical
coordinates as well:
\begin{equation}
  \tilde{\mathsf{Y}}^{(m)*}_{i_1i_2\cdots i_l}
  \tilde{\mathsf{Y}}^{(m')}_{i_1i_2\cdots i_l} = \delta_m^{m'}.
  \label{eq:29}
\end{equation}
The decomposition of Eq.~(\ref{eq:16}) is inverted as
\begin{align}
  \tilde{F}_{Xlm}(\bm{r})
  &=
    \frac{1}{A_l}   
    F^{(l)}_{Xi_1\cdots i_l}(\bm{r})
  \tilde{\mathsf{Y}}^{(m)*}_{i_1\cdots i_l}(\hat{\bm{r}})
  \label{eq:30}\\
  &=
    \frac{1}{A_l}   
    F_{Xi_1\cdots i_l}(\bm{r})
  \tilde{\mathsf{Y}}^{(m)*}_{i_1\cdots i_l}(\hat{\bm{r}}).
  \label{eq:31}
\end{align}
The second equality above holds because all the trace parts of
$F_{Xi_1\cdots i_l}$ contain Kronecker's delta symbols, and
spherical tensors $\tilde{\mathsf{Y}}^{(m)}_{i_1\cdots i_l}$ are traceless.

From the construction of the spherical tensor basis
$\mathsf{Y}^{(m)}_{i_1\cdots i_l}$ described in Appendix~A of Paper~I
\cite{PaperI}, and the property of the spherical harmonics
$Y_{lm}(\theta,\phi)$ with components $l=s, m=\pm s$
\cite{Khersonskii:1988krb},
\begin{equation}
  Y_{s,\pm s}(\theta,\phi)
  = (\mp 1)^s\sqrt{\frac{(2s+1)!!}{4\pi\,(2s)!!}}\,
  (\sin\theta)^s e^{\pm is\phi},
\label{eq:32}
\end{equation}
it is straightforward to show that
\begin{equation}
  \mathsf{Y}^{(\pm s)}_{i_1\cdots i_s}
  = \mathrm{m}^{\pm}_{i_1} \cdots \mathrm{m}^{\pm}_{i_s},
  \label{eq:33}
\end{equation}
where $\mathbf{m}^\pm = \mathbf{e}^\pm$ is the two-dimensional
spherical basis in Cartesian coordinates. Transforming the above
identity in the Cartesian coordinates into that in spherical
coordinates, we have
\begin{equation}
  \tilde{\mathrm{m}}^{\pm}_{i_1} \cdots \tilde{\mathrm{m}}^{\pm}_{i_s}
  = \tilde{\mathsf{Y}}^{(\pm s)}_{i_1\cdots i_s}(\theta,\phi).
  \label{eq:34}
\end{equation}
Substituting Eq.~(\ref{eq:1}) into Eq.~(\ref{eq:12}), using an
identity
$\tilde{\mathrm{m}}^\pm_i\mathcal{P}_{ij} = \tilde{\mathrm{m}}^\pm_j$,
$\tilde{\mathbf{m}}^{\pm *}= -\tilde{\mathbf{m}}^{\mp}$ and
Eqs.~(\ref{eq:31}) and (\ref{eq:34}), we derive a simple relation
\begin{equation}
  \tilde{f}^{(\pm s)}_X(\bm{r}) =
    A_s 
  \tilde{F}_{Xs,\pm s}(\bm{r}),
    \label{eq:35}
\end{equation}
where $\tilde{F}_{Xs,\pm s}$ on the rhs corresponds to
$\tilde{F}_{Xlm}$ with indices substituted by $l=s$ and $m=\pm s$. One
should note that the above formula is only valid if the rank $s$ of
the traceless part $f^{(s)}_{Xi_1\cdots i_s}$ is the same as the rank
of the original tensor $f_{Xi_1\cdots i_s}$, and does not hold for
trace parts, for the reason given below Eq.~(\ref{eq:12}). The above
equation plays a central role in relating the irreducible tensors in
three dimensions to those in two dimensions, and in calculating the
statistics of projected two-dimensional tensor fields from those of
three-dimensional tensor fields.

\subsubsection{%
  Multipole expansion of angular coordinates
}

The angular dependence of the irreducible representation for the
projected field $\tilde{f}^{(\pm s)}_X(\bm{r})$ of the spin-$s$ tensor
is naturally expanded into multipoles by the spin-weighted spherical
harmonics \cite{Newman:1966ub,Goldberg:1966uu},
$\sY{\pm s}{lm}(\theta,\phi)$, as
\begin{equation}
  \tilde{f}^{(\pm s)}_X(\bm{r}) =
  \sum_l
  \sqrt{\frac{4\pi}{2l+1}}\,
  \tilde{f}^{(\pm s)m}_{Xl}(r) \,\sY{\pm s}{lm}(\hat{\bm{r}}),
  \label{eq:36}
\end{equation}
where $\tilde{f}^{(\pm s)m}_{Xl}(r)$ are the coefficients of the
expansion which depend on the radial coordinate $r$, and the arguments
of the spin-weighted spherical harmonics, $(\theta,\phi)$, are the
direction of $\bm{r}$, which is denoted by a unit vector
$\hat{\bm{r}}=\bm{r}/r$ for simplicity in the following. The index $m$
is summed over without a summation symbol as we employ the Einstein
summation convention for azimuthal indices $m,m',\ldots$, following
Paper~I. The spin-weighted spherical harmonics $\sY{\pm s}{lm}$
vanishes when $s>l$, thus the summation over the integer $l$ is given
by $l=s,s+1,\ldots,\infty$. Putting the normalization factor
$\sqrt{4\pi/(2l+1)}$ in Eq.~(\ref{eq:36}) is convenient for our
purpose.

For our convenience, we employ a notation,
\begin{equation}
  \tilde{f}^{(\pm s)}_{Xlm}(r) \equiv
  g^{(l)}_{mm'} \tilde{f}^{(\pm s)m'}_{Xl}(\bm{r}),
  \label{eq:37}
\end{equation}
where 
\begin{equation}
  g^{(l)}_{mm'} = (-1)^m \delta_{m,-m'},
  \label{eq:38}
\end{equation}
with $0\leq m,m' \leq l$, is a metric tensor of spherical tensors
defined in Paper~I. The inverse metric $g_{(l)}^{mm'}$ which satisfies
$g_{(l)}^{mm'} g^{(l)}_{m'm''} = \delta^m_{m''}$ has the same elements
as $g_{(l)}^{mm'} =g^{(l)}_{mm'}$. Using the metric tensor introduced
above, the complex conjugate of spin-weighted spherical harmonics is
given by
\begin{align}
  \sY{\pm s}{lm}^*(\theta,\phi)
  &= (-1)^{m+s}\,\sY{\mp s}{l,-m}(\theta,\phi)
    \nonumber\\
  &= (-1)^s g_{(l)}^{mm'} \,\sY{\mp s}{lm'}(\theta,\phi),
  \label{eq:39}
\end{align}
and thus the expansion of Eq.~(\ref{eq:36}) is equivalently
represented by
\begin{equation}
  \tilde{f}^{(\pm s)}_X(\bm{r}) =
  (-1)^s \sum_l
    \sqrt{\frac{4\pi}{2l+1}}\,
  \tilde{f}^{(\pm s)}_{Xlm}(r) \,\sY{\mp s}{lm}^*(\hat{\bm{r}}).
  \label{eq:40}
\end{equation}
Because of the orthonormality relation of the spin-weighted spherical
harmonics,
\begin{equation}
  \int \sin\theta\,d\theta\,d\phi
  \,\sY{\pm s}{lm}^{*}(\theta,\phi)
  \,\sY{\pm s}{l'm'}(\theta,\phi) = \delta_{ll'} \delta_{m'}^{m},
  \label{eq:41}
\end{equation}
the expansion coefficients in Eqs.~(\ref{eq:36}) and (\ref{eq:40})
are inverted as
\begin{align}
  \tilde{f}^{(\pm s)m}_{Xl}(r)
  &=
    \sqrt{\frac{2l+1}{4\pi}}
  \int d^2\hat{r}
  \,\sY{\pm s}{lm}^{*}(\hat{\bm{r}})
  \tilde{f}^{(\pm s)}_X(\bm{r}),
  \label{eq:42}\\
  \tilde{f}^{(\pm s)}_{Xlm}(r)
  &= (-1)^s
    \sqrt{\frac{2l+1}{4\pi}}
  \int d^2\hat{r}
  \,\sY{\mp s}{lm}(\hat{\bm{r}})
  \tilde{f}^{(\pm s)}_X(\bm{r}),
  \label{eq:43}
\end{align}
where $d^2\hat{r} = \sin\theta\,d\theta\,d\phi$ is an element of
angular integrations over the radial direction $\hat{\bm{r}}$.

In cosmology, the coefficients of expansion are conveniently
represented by the so-called E and B modes \cite{Newman:1966ub}, first
in a context of polarization field of CMB \cite{Zaldarriaga:1996xe}.
In our context of general tensors, they are defined by\footnote{In
  Ref.~\cite{Zaldarriaga:1996xe}, the E/B modes of the polarization in
  CMB (spin-2 field) are defined with minus signs with respect to the
  definition here with
  $a^\mathrm{E/B}_{lm} = -[4\pi/(2l+1)]^{1/2}
  \tilde{f}^{\mathrm{E/B}{(2)}}_{Xlm}$. }
\begin{equation}
  \tilde{f}^{(\pm s)}_{Xlm}(r) 
  = \tilde{f}^{\mathrm{E}(s)}_{Xlm}(r) \pm i \tilde{f}^{\mathrm{B}(s)}_{Xlm}(r), 
  \label{eq:44}
\end{equation}
or equivalently 
\begin{align}
  \tilde{f}^{\mathrm{E}(s)}_{Xlm}(r)
  &= \frac{1}{2}
    \left[
    \tilde{f}^{(+s)}_{Xlm}(r) + \tilde{f}^{(-s)}_{Xlm}(r)
    \right],
  \label{eq:45}\\
  \tilde{f}^{\mathrm{B}(s)}_{Xlm}(r)
  &= \frac{1}{2i}
    \left[
    \tilde{f}^{(+s)}_{Xlm}(r) - \tilde{f}^{(-s)}_{Xlm}(r)
    \right].
  \label{eq:46}
\end{align}
The index $m$ in the E/B modes above can be also raised by the metric
tensor as
$\tilde{f}^{\mathrm{E/B}(s)m}_{Xl}(r) =
g_{(l)}^{mm'}\tilde{f}^{\mathrm{E/B}(s)}_{Xlm'}(r)$.

\subsection{%
  Symmetries
  \label{subsec:Symmetry}
}

In this subsection, we consider the symmetric properties of projected
tensor fields and their decomposition.

\subsubsection{Complex conjugate
}

As in Paper~I, we assume the original tensor field
$F_{Xi_1\cdots i_l}(\bm{r})$ is a real-valued tensor. The projected
field $f_{Xi_1\cdots i_s}(\bm{r})$ defined by Eq.~(\ref{eq:1}) is also
real. Because of Eq.~(\ref{eq:12}) and
$\tilde{\mathbf{m}}^{\mp *} = -\tilde{\mathbf{m}}^{\pm}$, the
corresponding irreducible tensor satisfy
\begin{equation}
  \tilde{f}^{(\pm s) *}_X(\bm{r})
  =  (-1)^s \tilde{f}^{(\mp s)}_X(\bm{r}).
  \label{eq:47}
\end{equation}
Using Eqs.~(\ref{eq:39}) and (\ref{eq:47}), the decomposition
coefficients of Eq.~(\ref{eq:40}) satisfy
\begin{equation}
  \tilde{f}^{(\pm s)*}_{Xlm}(r) =
  \tilde{f}^{(\mp s)m}_{Xl}(r).
  \label{eq:48}
\end{equation}
Accordingly, the E/B modes both satisfies
\begin{equation}
  \tilde{f}^{\mathrm{E/B}(s)*}_{Xlm}(r) = 
  \tilde{f}^{\mathrm{E/B}(s)m}_{Xl}(r).
  \label{eq:49}
\end{equation}

\subsubsection{Rotation}

Under the rotation of the spherical coordinates system,
$R:(\theta,\phi) \rightarrow (\theta',\phi')$, the
two-dimensional irreducible tensor of Eq.~(\ref{eq:11}) transforms
as
\begin{equation}
  \tilde{f}^{(\pm s)}_X(\bm{r}) \xrightarrow{\mathbb{R}}
  \tilde{f}^{\prime(\pm s)}_X(\bm{r}') =
  e^{\mp is\gamma} \tilde{f}^{(\pm s)}_X(\bm{r}),
\label{eq:50}
\end{equation}
where $\gamma$ is some angle determined by the coordinates rotation.
In a special case when the coordinates system is rotated around the
radial axis $\hat{\bm{r}}$ by an angle $\gamma$, the basis
$\tilde{\mathbf{m}}^\pm$ transforms as
$\tilde{\mathbf{m}}^\pm \rightarrow \tilde{\mathbf{m}}^{\pm\prime} =
e^{\pm i\gamma} \tilde{\mathbf{m}}^\pm$, and therefore the
two-dimensional irreducible tensor of Eq.~(\ref{eq:12}) transforms
as in the above. In general cases, the angle $\gamma$ is determined by
the coordinates rotation $R$ in a somehow complicated way
\cite{Newman:1966ub,Boyle:2013nka}, as derived from the relations of
Euler angles in successive rotations \cite{Khersonskii:1988krb}.
Accordingly, the spin-weighted spherical harmonics transform as
\cite{Boyle:2013nka}
\begin{equation}
  \sY{\pm s}{lm}(\theta,\phi) \xrightarrow{\mathbb{R}}
  \sY{\pm s}{lm}(\theta',\phi') =
  e^{\mp is\gamma}
  \,\sY{\pm s}{lm'}(\theta,\phi)D_{(l)m}^{m'}(R),
  \label{eq:51}
\end{equation}
where $D_{(l)m}^{m'}(R)$ is the Wigner's rotation matrix, and the
rotation $R$ of the coordinates system is considered a passive
rotation.

Because of the properties of rotation, Eqs.~(\ref{eq:50}) and
(\ref{eq:51}), the dependence on the phase $e^{\mp is\gamma}$ appears
exactly in the same way in the transformation, and thus the
transformations of coefficients of Eqs.~(\ref{eq:36}) and
(\ref{eq:37}) do not receive this phase factor, as we have
\begin{align}
  \tilde{f}^{(\pm s)m}_{Xl}(r)
  &\xrightarrow{\mathbb{R}}
  \tilde{f}^{\prime(\pm s)m}_{Xl}(r) =
   D^{m}_{(l)m'}(R^{-1})\tilde{f}^{(\pm s)m'}_{Xl}(r),
  \label{eq:52}\\
  \tilde{f}^{(\pm s)}_{Xlm}(r)
  &\xrightarrow{\mathbb{R}}
  \tilde{f}^{\prime(\pm s)}_{Xlm}(r) =
  \tilde{f}^{(\pm s)}_{Xlm'}(r)  D^{m'}_{(l)m}(R),
  \label{eq:53}
\end{align}
where $D^m_{(l)m'}(R^{-1}) = D^{m'*}_{(l)m}(R)$ is the rotation matrix
of the inverse rotation, and Wigner's rotation matrix is unitary.
Obviously from Eqs.~(\ref{eq:45}), (\ref{eq:46}),
(\ref{eq:52}) and (\ref{eq:53}), the E/B modes transform under
the rotation of the coordinates system as
\begin{align}
  \tilde{f}^{\mathrm{E/B}(s)m}_{Xl}(r)
  &\xrightarrow{\mathbb{R}}
  \tilde{f}^{\prime\mathrm{E/B}(s)m}_{Xl}(r) =
  D^{m}_{(l)m'}(R^{-1})\tilde{f}^{\mathrm{E/B}(s)m'}_{Xl}(r),
  \label{eq:54}\\
  \tilde{f}^{\mathrm{E/B}(s)}_{Xlm}(r)
  &\xrightarrow{\mathbb{R}}
  \tilde{f}^{\prime\mathrm{E/B}(s)}_{Xlm}(r) =
  \tilde{f}^{\mathrm{E/B}(s)}_{Xlm'}(r)  D^{m'}_{(l)m}(R).
  \label{eq:55}
\end{align}

\subsubsection{Parity
}

We consider a parity transformation of the tensor field,
\begin{equation}
  F_{Xi_1\cdots i_l}(\bm{r}) \xrightarrow{\mathbb{P}}
  F'_{Xi_1\cdots i_l}(\bm{r}) =
  (-1)^{p_X+l}F_{Xi_1\cdots i_l}(-\bm{r}),
  \label{eq:56}
\end{equation}
where $p_X=0$ for ordinary tensors and $p_X=1$ for pseudotensors. The
angular momentum is a typical example of a pseudotensor of rank-1. We
consider the parity transformation as an active one, and the
coordinate axes are fixed and not flipped, keeping the left-handedness
of the coordinates system. In the following, while the parity
transformation does not change the basis, the basis at $-\bm{r}$ is
needed to represent the irreducible tensor $f^{(\pm s)}_X$ at the same
position. Provided that the spherical coordinates and their basis are,
respectively, given by $(r,\theta,\phi)$ and
$(\mathbf{e}_r,\mathbf{e}_\theta,\mathbf{e}_\phi)$ at a position
$\bm{r}$, the same quantities at an opposite position
$\bm{r}'=-\bm{r}$ are given by
$(r',\theta',\phi') = (r,\pi-\theta,\phi+\pi)$ and
$(\mathbf{e}_r',\mathbf{e}_\theta',\mathbf{e}_\phi') =
(-\mathbf{e}_r,\mathbf{e}_\theta,-\mathbf{e}_\phi)$, respectively. The
corresponding spherical bases are given by
$\tilde{\mathbf{e}}^{\prime m} = -\tilde{\mathbf{e}}^{m}$ and
$\tilde{\mathbf{m}}^{\prime\pm} = -\tilde{\mathbf{m}}^{\mp}$.

Accordingly, the projected tensor of Eq.~(\ref{eq:1})
transforms as
\begin{equation}
  f_{Xi_1\cdots i_s}(\bm{r}) \xrightarrow{\mathbb{P}}
  f'_{Xi_1\cdots i_s}(\bm{r}) =
  (-1)^{p_X+s}f_{Xi_1\cdots i_s}(-\bm{r}),
  \label{eq:57}
\end{equation}
because $\mathcal{P}_{ij}(-\hat{\bm{r}}) =
\mathcal{P}_{ij}(\hat{\bm{r}})$, and from Eq.~(\ref{eq:12}), the
projected spherical tensor transforms as
\begin{equation}
  \tilde{f}^{(\pm s)}_X(\bm{r}) \xrightarrow{\mathbb{P}}
  \tilde{f}'^{(\pm s)}_X(\bm{r}) =
  (-1)^{p_X} \tilde{f}^{(\mp s)}_X(-\bm{r}).
  \label{eq:58}
\end{equation}
Because of the parity property of spin-weighted spherical harmonics,
$\sY{\pm s}{lm}(-\hat{\bm{r}}) = (-1)^l \,\sY{\mp
  s}{lm}(\hat{\bm{r}})$, the expansion coefficients of
Eq.~(\ref{eq:36}) transform as
\begin{equation}
  \tilde{f}^{(\pm s)}_{Xlm}(r) \xrightarrow{\mathbb{P}}
  \tilde{f}'^{(\pm s)}_{Xlm}(r) =
  (-1)^{p_X+l} \tilde{f}^{(\mp s)}_{Xlm}(r).
  \label{eq:59}
\end{equation}
On one hand, the parity transformation of the spherical tensors is
mixed between the two modes $\pm s$. On the other hand, the E/B modes
transform as
\begin{align}
  \tilde{f}^{\mathrm{E}(s)}_{Xlm}(r)
  &\xrightarrow{\mathbb{P}} \tilde{f}'^{\mathrm{E}(s)}_{Xlm}(r)
  = (-1)^{p_X+l}\tilde{f}^{\mathrm{E}(s)}_{Xlm}(r),
  \label{eq:60}\\
  \tilde{f}^{\mathrm{B}(s)}_{Xlm}(r)
  &\xrightarrow{\mathbb{P}} \tilde{f}'^{\mathrm{B}(s)}_{Xlm}(r)
  = (-1)^{p_X+l+1}\tilde{f}^{\mathrm{B}(s)}_{Xlm}(r),
  \label{eq:61}
\end{align}
i.e., the E/B modes transform independently under the parity, while
the signs are oppositely appearing between the two modes. The last
property is one of the main reasons why E/B modes are popular in
cosmology. 

\subsection{%
  Power spectrum
  \label{subsec:PowerSpecFullSky}
}

We consider the two-point statistic of the multipole coefficients,
\begin{equation}
  \left\langle
    \tilde{f}^{(\sigma_1)}_{X_1lm}(r_1)
    \tilde{f}^{(\sigma_2)}_{X_2l'm'}(r_2)
  \right\rangle,
  \label{eq:62}
\end{equation}
where $\sigma_1=\pm s_1$ and $\sigma_2=\pm s_2$ can take both positive
and negative integers as well as zero. The above quantity should be
independent of the choice of the coordinates system in a
statistically isotropic Universe, and thus should be invariant under
the coordinates rotation of Eq.~(\ref{eq:53}). An integral over a
product of rotation matrices is given by \cite{Edmonds:1955fi}
\begin{equation}
  \frac{1}{8\pi^2} \int [dR]
  D_{(l_1)m_1}^{m_1'}(R) D_{(l_2)m_2}^{m_2'}(R)
  = \frac{\delta_{l_1l_2}}{2l_1+1}
  g^{(l_1)}_{m_1m_2} g_{(l_2)}^{m_1'm_2'},
  \label{eq:63}
\end{equation}
where $\int [dR]\cdots$ represents the integrals over Euler angles of
the rotation $R$ and satisfies $\int [dR] = 8\pi^2$. Using the above
integral and the rotational invariance of Eq.~(\ref{eq:62}), we
find the two-point function of Eq.~(\ref{eq:62}) in a statistically
isotropic Universe should have a form,
\begin{equation}
  \left\langle
    \tilde{f}^{(\sigma_1)}_{X_1lm}(r_1)
    \tilde{f}^{(\sigma_2)}_{X_2l'm'}(r_2)
  \right\rangle
  = \delta_{ll'} g^{(l)}_{mm'} C^{(\sigma_1\sigma_2)}_{X_1X_2l}(r_1,r_2),
  \label{eq:64}
\end{equation}
where
\begin{equation}
  C^{(\sigma_1\sigma_2)}_{X_1X_2l}(r_1,r_2) =
  \frac{1}{2l+1}
  \left\langle
    \tilde{f}^{(\sigma_1)}_{X_1lm}(r_1)
    \tilde{f}^{(\sigma_2)m}_{X_2l}(r_2)
  \right\rangle.
  \label{eq:65}
\end{equation}
The power spectrum $C^{(\sigma_1\sigma_2)}_{X_1X_2l}(r_1,r_2)$ is
invariant under the coordinates rotation. The property of
Eq.~(\ref{eq:48}) indicates that the complex conjugate of the angular
power spectrum is given by
\begin{equation}
  C^{(\sigma_1\sigma_2)*}_{X_1X_2l}(r_1,r_2)
  = C^{(-\sigma_1,-\sigma_2)}_{X_1X_2l}(r_1,r_2).
  \label{eq:66}
\end{equation}
The property of Eq.~(\ref{eq:59}) indicates that the parity
transformation of the angular spectrum is given by
\begin{equation}
  C^{(\sigma_1\sigma_2)}_{X_1X_2l}(r_1,r_2)
  \xrightarrow{\mathbb{P}}
  (-1)^{p_{X_1} + p_{X_2}}
  C^{(-\sigma_1,-\sigma_2)}_{X_1X_2l}(r_1,r_2).
  \label{eq:67}
\end{equation}

For the E/B modes of Eqs.~(\ref{eq:45}) and (\ref{eq:46}), the
two-point statistics similarly have forms,
\begin{align}
  \left\langle
    \tilde{f}^{\mathrm{E}(s_1)}_{X_1lm}(r_1)
    \tilde{f}^{\mathrm{E}(s_2)}_{X_2l'm'}(r_2)
  \right\rangle
  &= \delta_{ll'} g^{(l)}_{mm'} C^{\mathrm{EE}(s_1s_2)}_{X_1X_2l}(r_1,r_2),
  \label{eq:68}\\
  \left\langle
    \tilde{f}^{\mathrm{B}(s_1)}_{X_1lm}(r_1)
    \tilde{f}^{\mathrm{B}(s_2)}_{X_2l'm'}(r_2)
  \right\rangle
  &= \delta_{ll'} g^{(l)}_{mm'} C^{\mathrm{BB}(s_1s_2)}_{X_1X_2l}(r_1,r_2),
  \label{eq:69}\\
  \left\langle
    \tilde{f}^{\mathrm{E}(s_1)}_{X_1lm}(r_1)
    \tilde{f}^{\mathrm{B}(s_2)}_{X_2l'm'}(r_2)
  \right\rangle
  &= \delta_{ll'} g^{(l)}_{mm'} C^{\mathrm{EB}(s_1s_2)}_{X_1X_2l}(r_1,r_2),
  \label{eq:70}\\
  \left\langle
    \tilde{f}^{\mathrm{B}(s_1)}_{X_1lm}(r_1)
    \tilde{f}^{\mathrm{E}(s_2)}_{X_2l'm'}(r_2)
  \right\rangle
  &= \delta_{ll'} g^{(l)}_{mm'} C^{\mathrm{BE}(s_1s_2)}_{X_1X_2l}(r_1,r_2),
  \label{eq:71}
\end{align}
where $s_1\geq 0$ and $s_2\geq 0$. The invariant power spectra,
$C^{\mathrm{EE}(s_1s_2)}_{X_1X_2l}$,
$C^{\mathrm{BB}(s_1s_2)}_{X_1X_2l}$,
$C^{\mathrm{EB}(s_1s_2)}_{X_1X_2l}$,
$C^{\mathrm{BE}(s_1s_2)}_{X_1X_2l}$ are given by similar relations of
Eq.~(\ref{eq:65}), which are obviously derived by contracting indices
of Eqs.~(\ref{eq:68}) and (\ref{eq:69}). The property of
Eq.~(\ref{eq:49}) indicates that the complex conjugates of the E/B
power spectra are given by
\begin{align}
  C^{\mathrm{EE}(s_1s_2)*}_{X_1X_2l}(r_1,r_2)
  &= C^{\mathrm{EE}(s_1s_2)}_{X_1X_2l}(r_1,r_2),
  \label{eq:72}\\
  C^{\mathrm{BB}(s_1s_2)*}_{X_1X_2l}(r_1,r_2)
  &= C^{\mathrm{BB}(s_1s_2)}_{X_1X_2l}(r_1,r_2),
  \label{eq:73}\\
  C^{\mathrm{EB}(s_1s_2)*}_{X_1X_2l}(r_1,r_2)
  &= C^{\mathrm{EB}(s_1s_2)}_{X_1X_2l}(r_1,r_2),
  \label{eq:74}\\
  C^{\mathrm{BE}(s_1s_2)*}_{X_1X_2l}(r_1,r_2)
  &= C^{\mathrm{BE}(s_1s_2)}_{X_1X_2l}(r_1,r_2),
  \label{eq:75}
\end{align}
i.e., they are real functions. The properties of
Eqs.~(\ref{eq:60}) and (\ref{eq:61}) indicate that the parity
transformations of the E/B spectra are given by
\begin{align}
  C^{\mathrm{EE}(s_1s_2)}_{X_1X_2l}(r_1,r_2)
  &\xrightarrow{\mathbb{P}}
  (-1)^{p_{X_1} + p_{X_2}}
  C^{\mathrm{EE}(s_1s_2)}_{X_1X_2l}(r_1,r_2),
  \label{eq:76}\\
  C^{\mathrm{BB}(s_1s_2)}_{X_1X_2l}(r_1,r_2)
  &\xrightarrow{\mathbb{P}}
  (-1)^{p_{X_1} + p_{X_2}}
  C^{\mathrm{BB}(s_1s_2)}_{X_1X_2l}(r_1,r_2).
  \label{eq:77}\\
  C^{\mathrm{EB}(s_1s_2)}_{X_1X_2l}(r_1,r_2)
  &\xrightarrow{\mathbb{P}}
  (-1)^{p_{X_1} + p_{X_2}+1}
  C^{\mathrm{EB}(s_1s_2)}_{X_1X_2l}(r_1,r_2),
  \label{eq:78}\\
  C^{\mathrm{BE}(s_1s_2)}_{X_1X_2l}(r_1,r_2)
  &\xrightarrow{\mathbb{P}}
  (-1)^{p_{X_1} + p_{X_2}+1}
  C^{\mathrm{BE}(s_1s_2)}_{X_1X_2l}(r_1,r_2),
  \label{eq:79}
\end{align}
When $p_{X_1}+p_{X_2}=\mathrm{even}$ as in mostly interested cases,
the EB and BE power spectra are parity odd and vanish in the Universe
with parity symmetry. Other EE and BB spectra are parity even. When
$p_{X_1}+p_{X_2}=\mathrm{odd}$, however, the EB and BE spectra are
parity even and others are parity odd.

The power spectra of E/B modes are related to the power spectra of
$\pm s$ modes by, in an abbreviated notation, 
\begin{align}
  C^{\mathrm{EE}}_{l}
  &= \frac{1}{4}
    \left[
    \left(C^{++}_{l} + C^{--}_{l}\right) + 
    \left(C^{+-}_{l} + C^{-+}_{l}\right)
    \right],
  \label{eq:80}\\
  C^{\mathrm{BB}}_{l}
  &= -\frac{1}{4}
    \left[
    \left(C^{++}_{l} + C^{--}_{l}\right) - 
    \left(C^{+-}_{l} + C^{-+}_{l}\right)
    \right],
  \label{eq:81}\\
  C^{\mathrm{EB}}_{l}
  &= -\frac{i}{4}
    \left[
    \left(C^{++}_{l} - C^{--}_{l}\right) - 
    \left(C^{+-}_{l} - C^{-+}_{l}\right)
    \right],
  \label{eq:82}\\
  C^{\mathrm{BE}}_{l}
  &= -\frac{i}{4}
    \left[
    \left(C^{++}_{l} - C^{--}_{l}\right) + 
    \left(C^{+-}_{l} - C^{-+}_{l}\right)
    \right],
  \label{eq:83}
\end{align}
where the abbreviations,
\begin{align}
  &
    C^{\pm \pm}_{l} = C^{(\pm s_1,\pm s_2)}_{X_1X_2l}(r_1,r_2), \quad
    C^{\pm \mp}_{l} = C^{(\pm s_1,\mp s_2)}_{X_1X_2l}(r_1,r_2),
  \label{eq:84}\\
  &
    C^{\mathrm{EE}}_{l} = C^{\mathrm{EE}(s_1s_2)}_{X_1X_2l}(r_1,r_2), \quad
    C^{\mathrm{BB}}_{l} = C^{\mathrm{BB}(s_1s_2)}_{X_1X_2l}(r_1,r_2),
  \label{eq:85}\\
  &
    C^{\mathrm{EB}}_{l} = C^{\mathrm{EB}(s_1s_2)}_{X_1X_2l}(r_1,r_2), \quad
    C^{\mathrm{BE}}_{l} = C^{\mathrm{BE}(s_1s_2)}_{X_1X_2l}(r_1,r_2),
  \label{eq:86}
\end{align}
are adopted just for simplicity of presentation.

\section{\label{sec:FlatSky}%
  Distant-observer limit of projected tensor fields }

\subsection{%
  Projected spherical tensor in Cartesian coordinates }

The flat-sky limit is valid in the cases where the angular scales of
clustering we are interested in are small. In the case of the power
spectrum in the previous section, the flat-sky limit is a good
approximation for large multiples $l \gg 1$. In this limit, the
direction of the line of sight is fixed in the observed space. It is
convenient to fix the direction of the line of sight to be the third
axis, $\hat{\bm{z}} = \mathbf{e}_3$ in Cartesian coordinates. In
addition, we consider the radial differences between $r_1$ and $r_2$
in the power spectrum are much smaller than absolute radii $r_1$ and
$r_2$ themselves, i.e., $|r_1-r_2| \ll r_1, r_2$. We refer to these
approximations or limits by the distant-observer approximation, or the
distant-observer limit, which implicitly include the flat-sky limit.
In this approximation, we can put the Cartesian coordinates in the
volume of observation, and the origin of Cartesian coordinates can
conveniently be taken in the corresponding volume of observation. In
the following, we consider the origin to be located at
\begin{equation}
  \bm{r}_0 = r_0 \hat{\mathbf{e}}_3 =(0,0,r_0)
  \label{eq:87}
\end{equation}
in the observer's coordinates. The distant-observer, Cartesian
coordinates are denoted by $\bm{x}$, and the relation to the observer's
spherical coordinates is given by
\begin{equation}
  \bm{x} = \bm{r} - r_0 \hat{\mathbf{e}}_3 =(r_1,r_2,r_3 - r_0).
  \label{eq:88}
\end{equation}

Instead of the spherical basis $\tilde{\mathbf{m}}^\pm$ of
Eq.~(\ref{eq:10}) in spherical coordinates, the basis in the
Cartesian coordinates in the distant-observer limit is defined by
\begin{equation}
    \mathbf{m}^{\pm} \equiv
    \mp \frac{\hat{\mathbf{e}}_1 \mp i \hat{\mathbf{e}}_2}{\sqrt{2}},
    \label{eq:89}
\end{equation}
where $\hat{\mathbf{e}}_1$ and $\hat{\mathbf{e}}_2$ are unit vectors
along the first and second axes, respectively. The flat-sky limit is
obtained by taking a limit that the polar angle is small,
$\theta \ll 1$, in the spherical coordinates. Taking this limit in
Eqs.~(\ref{eq:25}) and (\ref{eq:26}), the unit vectors of axes are
related by
\begin{equation}
  \begin{pmatrix}
    \mathbf{e}_\theta \\ \mathbf{e}_\phi
  \end{pmatrix}
  =
  \begin{pmatrix}
    \cos\phi & \sin\phi \\
    -\sin\phi & \cos\phi
  \end{pmatrix}
  \begin{pmatrix}
    \hat{\mathbf{e}}_1 \\ \hat{\mathbf{e}}_2
  \end{pmatrix}, \quad
  \mathbf{e}_r = \hat{\mathbf{e}}_3,
  \label{eq:90}
\end{equation}
where $\phi$ is the azimuthal angle of $\bm{r}$. Thus the spherical
bases in spherical coordinates, Eqs.~(\ref{eq:10}) and
(\ref{eq:18}), are related to the spherical bases in Cartesian
coordinates, Eqs.~(\ref{eq:89}) and (\ref{eq:24}) by
\begin{equation}
  \tilde{\mathbf{m}}^\pm =
  e^{\pm i\phi} \mathbf{m}^\pm, \quad
  \tilde{\mathbf{e}}^0 =
  \mathbf{e}^0, \quad
  \tilde{\mathbf{e}}^\pm =
  e^{\pm i\phi} \mathbf{e}^\pm
  \label{eq:91}
\end{equation}
in the flat-sky limit.

The decomposition of the projected tensors into spherical tensors are
defined by similar equations to Eq.~(\ref{eq:9}) but with spherical
basis of Cartesian coordinates,
\begin{equation}
  f^{(s)}_{Xi_1\cdots i_s} =
    f^{(+s)}_X \mathrm{m}^+_{i_1} \cdots \mathrm{m}^+_{i_s} +
    f^{(-s)}_X \mathrm{m}^-_{i_1} \cdots
    \mathrm{m}^-_{i_s}.
  \label{eq:92}
\end{equation}
In this case, the spin basis $\mathbf{m}^\pm$ in Cartesian coordinates
does not depend on the position, unlike in the case of spherical
coordinates. Because of the relation between the spin bases in
Eq.~(\ref{eq:91}), the projected tensor fields in spherical
coordinates and those in Cartesian coordinates in the distant-observer
approximation are related by
\begin{equation}
  f^{(\pm s)}_X(\bm{x})
  = e^{\pm is\phi} \tilde{f}^{(\pm s)}_X(\bm{x}+r_0\mathbf{e}_3), \quad
  (|\bm{x}| \ll r_0),
  \label{eq:93}
\end{equation}
where the origin of the Cartesian coordinates $\bm{x}$ is given by
Eq.~(\ref{eq:87}). The phase factor in front of the rhs stems from
the difference that the basis in spherical coordinates depends on the
position while that in Cartesian coordinates does not. The azimuthal
angle $\phi$ of $\bm{r}=\bm{x} + r_0\mathbf{e}_3$ is the same as that
of $\bm{x}$, because the first and second components of $\bm{x}$ and
$\bm{r}$ are the same. In the distant-observer limit, the
Eq.~(\ref{eq:35}) straightforwardly reduces to
\begin{equation}
    f^{(\pm s)}_X(\bm{x}) =
    A_s   
    F_{Xs,\pm s}(\bm{x}),
    \label{eq:94}
\end{equation}
where the last factor in the rhs is $F_{Xlm}(\bm{x})$ with
substitutions $l=s$ and $m=\pm s$, and is the irreducible tensor field
on the fixed Cartesian basis $(\mathbf{e}^0,\mathbf{e}^\pm)$, i.e.,
\begin{align}
  F_{Xlm}(\bm{x})
  &=
    \frac{1}{A_l}   
    F^{(l)}_{Xi_1\cdots i_l}(\bm{x})
  \mathsf{Y}^{(m)*}_{i_1\cdots i_l}
  \label{eq:95}\\
  &=
    \frac{1}{A_l}   
    F_{Xi_1\cdots i_l}(\bm{x})
    \mathsf{Y}^{(m)*}_{i_1\cdots i_l},
  \label{eq:96}
\end{align}
where $\mathsf{Y}^{(m)}_{i_1\cdots i_l}$ is the Cartesian spherical
basis defined in Paper~I.

\subsection{%
  Symmetries}

\subsubsection{%
  Complex conjugate}

Taking the complex conjugate of Eq.~(\ref{eq:93}) and using
Eq.~(\ref{eq:47}), we have
\begin{equation}
  f^{(\pm s)*}_X(\bm{x}) = (-1)^s f^{(\mp s)}_X(\bm{x}).
  \label{eq:97}
\end{equation}
This property is derived as well from Eq.~(\ref{eq:94}) and the
property of the three-dimensional tensor field described in Paper~I,
$F^*_{Xlm}(\bm{x}) = (-1)^{l+m}F_{Xl,-m}(\bm{x})$.

\subsubsection{%
  Rotation}

We consider the rotational symmetry by the passive transformation of
rotation, i.e., rotation of the coordinates axes, instead of rotation
of the physical system. In this case, the rotational transformation in
the distant-observer limit is more naturally taken around the origin
of the observed volume, $\bm{r}=r_0\mathbf{e}_3$ ($\bm{x} = \bm{0}$),
rather than around the observer's position, $\bm{r}=\bm{0}$.
Accordingly, the position of the observer is displaced and rotated in
the transformation, keeping the direction of the line of sight toward the
third axis in the new basis\footnote{When we consider active rotations
  of the physical system, however, the rotation around the position of
  the observer is naturally considered, fixing the third axis of the
  observer.}.

The rotational symmetry of the three-dimensional tensor $F_{Xlm}$ of
Eq.~(\ref{eq:96}) is fully described in Paper~I. Under the passive
rotation of the Cartesian basis,
$\hat{\mathbf{e}}_i \rightarrow \hat{\mathbf{e}}'_i =
R_{ij}\hat{\mathbf{e}}_j$, the components of the coordinate vector
$\bm{x}$ transform as $x_i \rightarrow x'_i = x_jR_{ji}$, or
$\bm{x} \rightarrow \bm{x}' = R^{-1}\bm{x}$ in short, where
$R_{ij} =(R^{-1})_{ji}$ is the elements of the rotation matrix. The
three-dimensional tensor field transforms as
\begin{equation}
  F_{Xlm}(\bm{x}) \xrightarrow{\mathbb{R}}
  F'_{Xlm}(\bm{x}') = F_{Xlm'}(\bm{x}) D^{m'}_{(l)m}(R).
  \label{eq:98}
\end{equation}
Because the rotation is considered passive, the position of the
observer is also rotated to remain along the third axis at
$\bm{x}' = (0,0,-r_0)$ in terms of the rotated basis. Therefore, the
projected tensor of Eq.~(\ref{eq:94}) transforms as
\begin{equation}
  f^{(\pm s)}_X(\bm{x}) \xrightarrow{\mathbb{R}}
  f'^{(\pm s)}_X(\bm{x}') =
  A_s
  F_{Xsm}(\bm{x}) D^{m}_{(s)\pm s}(R).
  \label{eq:99}
\end{equation}

As the direction of projection (i.e., the line of sight) rotates, the
projected tensor acquires other components of the three-dimensional
tensor than the projected components on the original basis. There are
two cases that the projected tensor transforms within the projected
components. The first case is that the rotation axis is parallel to
the line of sight. The Euler angles of the rotation $R$ in this case
are given by $(\alpha,\beta,\gamma)=(0,0,\gamma)$, whose rotation is
denoted by $R_\gamma$ below. The Wigner's matrix for this rotation is
given by $D^{m'}_{(l)m}(R_\gamma) = e^{-im\gamma}\delta_{m'm}$, and
thus we have
\begin{equation}
  f^{(\pm s)}_X(\bm{x}) \xrightarrow{\mathbb{R}_\gamma}
  f'^{(\pm s)}_X(\bm{x}') = e^{\mp is\gamma}f^{(\pm s)}_X(\bm{x}),
  \label{eq:100}
\end{equation}
which is consistent with Eq.~(\ref{eq:50}) in the spherical
coordinates.

The second case of the rotation in which the tensor is transformed
within the projected components is that the position vector $\bm{x}$
is flipped by 180 degrees after three-dimensional rotation. In this case,
the Euler angles are given by
$(\alpha,\beta,\gamma)=(\phi,\pi,-\phi)$, where $\phi$ is the
azimuthal angle of the position vector
$\bm{x}=(x\sin\theta\cos\phi, x\sin\theta\sin\phi,x\cos\theta)$. The
rotation matrix of the rotation is given by
\begin{equation}
  D^{m'}_{(l)m}(\phi,\pi,-\phi) = (-1)^{l+m} e^{2im\phi}\delta_{m,-m'},
  \label{eq:101}
\end{equation}
and Eq.~(\ref{eq:99}) in this case reduces to
\begin{equation}
  f^{(\pm s)}_X(\bm{x}) \xrightarrow{\mathbb{F}}
  f^{\prime(\pm s)}_X(\bm{x}) = e^{\pm 2is\phi} f^{(\mp s)}_X(-\bm{x}),
  \label{eq:102}
\end{equation}
where the transformation is represented by the change of functional
form. We call this specific rotation a flipping rotation, which
depends on the azimuthal angle of position vector $\bm{x}$.

\subsubsection{%
  Parity}

In order to discuss the parity symmetry below in the distant-observer
approximation, it is useful to consider the distant-observer
approximation near the opposite pole of the spherical coordinates,
$\pi - \theta \ll 1$, where $\theta$ is the longitudinal angle of the
spherical coordinates $\bm{r}$. In this case, the unit vectors of
spherical axes are replaced by
$\mathbf{e}_\theta \rightarrow - \mathbf{e}_\theta$,
$\mathbf{e}_\phi \rightarrow \mathbf{e}_\phi$,
$\mathbf{e}_r \rightarrow -\mathbf{e}_r$ in Eq.~(\ref{eq:90}), and
these replacements are equivalent to
$\tilde{\mathbf{m}}^\pm \rightarrow \tilde{\mathbf{m}}^\mp$,
$\tilde{\mathbf{e}}^0 \rightarrow -\tilde{\mathbf{e}}^0$, and
$\tilde{\mathbf{e}}^\pm \rightarrow \tilde{\mathbf{e}}^\mp$.
Therefore, the signs of spin modes in the spherical coordinates are
flipped as $\tilde{f}^{(\pm s)}_X \rightarrow \tilde{f}^{(\mp s)}_X$,
and the same calculations above are applied to this replacement in
the case of $\pi-\theta \ll 1$. Thus, Eq.~(\ref{eq:93}) now becomes
\begin{equation}
  f^{(\pm s)}_X(\bm{x})
  = e^{\pm is\phi}
  \tilde{f}^{(\mp s)}_X(\bm{x}-r_0\mathbf{e}_3), \quad
  (|\bm{x}|\ll r_0),
  \label{eq:103}
\end{equation}
where $\phi$ is the azimuthal angle of $\bm{x}$ and also of $\bm{r}
=\bm{x}-r_0\mathbf{e}_3$.

We consider the active parity transformation, i.e., the physical
system is flipped with respect to the origin of the spherical
coordinates, i.e., the position of the observer. The parity
transformation in the spherical coordinates is given by
Eq.~(\ref{eq:58}). In this transformation, the longitudinal angle
$\theta$ of $\bm{r}$ changes from that which satisfies
$\pi-\theta \ll 1$ to that which satisfies $\theta \ll 1$. Applying
Eq.~(\ref{eq:58}) to Eq.~(\ref{eq:103}), using Eq.~(\ref{eq:93}), and
noting that the azimuthal angle of $-\bm{x}$ is given by $\phi + \pi$,
we have
\begin{equation}
  f^{(\pm s)}_X(\bm{x}) \xrightarrow{\mathbb{P}}
  f^{\prime (\pm s)}_X(\bm{x})
  = (-1)^{p_X+s}f^{(\pm s)}_X(-\bm{x}).
  \label{eq:104}
\end{equation}
This property of parity is also derived from Eq.~(\ref{eq:94}) and
the property of the three-dimensional tensor field described in
Paper~I,
$F_{Xlm}(\bm{x}) \rightarrow F_{Xlm}'(\bm{x}) = (-1)^{p_X+l}
F_{Xlm}(-\bm{x})$, and therefore they are consistent with each other.

Unlike in the case of spherical coordinates, Eq.~(\ref{eq:58}), the
signs of the spin are not flipped by the parity transformation in this
case of Cartesian coordinates, because the spherical basis depends on
the position in the former, while it does not in the latter. A more
similar transformation to the parity in spherical coordinates is
provided by a combination of parity and flipping transformations,
Eqs.~(\ref{eq:104}) and (\ref{eq:102}):
\begin{equation}
  f^{(\pm s)}_X(\bm{x}) \xrightarrow{\mathbb{PF}}
  f^{\prime (\pm s)}_X(\bm{x})
  = (-1)^{p_X+s} e^{\pm 2is\phi} f^{(\mp s)}_X(\bm{x}),
  \label{eq:105}
\end{equation}
where $\phi$ is the azimuthal angle of $\bm{x}$.

\subsection{%
  Multipole expansion and Fourier modes
}

\subsubsection{%
  Distant-observer approximation of the multipole expansion
}

As shown in the Appendix, the spin-weighted spherical harmonics in the
flat-sky limit are asymptotically represented by derivatives of usual
spherical harmonics
\begin{equation}
  \sY{\pm s}{lm}(\theta,\phi) \approx
    \frac{(\mp 1)^s}{(l+1/2)^s}
  e^{\mp is\phi}
  \left(\partial_1 \pm i\partial_2\right)^s
  Y_{lm}(\theta,\phi),
  \label{eq:106}
\end{equation}
where $\partial_i = \partial/\partial \theta_i$ with
$(\theta_1,\theta_2) = (\theta\cos\phi,\theta\sin\phi)$ are Cartesian
derivatives along the first and second axes on the unit sphere. The
spherical harmonics in the flat-sky limit of $\theta \ll 1$ and
$l\gg 1$ is given by
\begin{equation}
  Y_{lm}(\theta,\phi) \approx
  (-1)^m
     e^{im\phi} \sqrt{\frac{2l+1}{4\pi}}
     J_m\left[\left(l+\frac{1}{2}\right)\theta\right].
  \label{eq:107}
\end{equation}
It is convenient below to introduce a shifted variable,
\begin{equation}
  \bar{l} \equiv l + \frac{1}{2}.
  \label{eq:108}
\end{equation}

Following Refs.~\cite{White:1997wq,Hu:2000ee}, we define a
weighted sum of the multipole expansion coefficients in
Eq.~(\ref{eq:36}) as
\begin{equation}
  \tilde{g}^{(\pm s)}_{X}(\bar{\bm{l}},r) \equiv
  \frac{2\pi}{\bar{l}}
  \sum_m i^m \tilde{f}^{(\pm s)m}_{Xl}(r)
  e^{im\phi_l},
  \label{eq:109}
\end{equation}
where the two-dimensional vector $\bar{\bm{l}}$ perpendicular to the
line of sight is given by
$\bar{\bm{l}} \equiv (\bar{l}\cos\phi_l,\bar{l}\sin\phi_l,0)$ in the
flat-sky limit and $\phi_l$ is the azimuthal angle of the vector
$\bar{\bm{l}}$. The inverse relation of the above is given by
\begin{equation}
  \tilde{f}^{(\pm s)m}_{Xl}(r)
  =
  \frac{\bar{l}}{2\pi}  (-i)^m
  \int \frac{d\phi_l}{2\pi} e^{-im\phi_l}
  \tilde{g}^{(\pm s)}_{X}(\bar{\bm{l}},r).
  \label{eq:110}
\end{equation}
With the vector $\bar{\bm{l}}$ introduced above, the Rayleigh expansion of
two-dimensional plane wave and its flat-sky approximation is given by
\begin{align}
  e^{i\bar{\bm{l}}\cdot\bm{\theta}}
  &= \sum_m i^m
    J_m\left(\bar{l}\,\theta\right)
    e^{im(\phi-\phi_l)}
    \nonumber\\
  &\approx
    \sqrt{\frac{2\pi}{\bar{l}}}
    \sum_m (-i)^m Y_{lm}(\theta,\phi) e^{-im\phi_l}.
    \label{eq:111}
\end{align}
where $\bm{\theta} = (\theta\cos\phi,\theta\sin\phi,0)$, and
Eq.~(\ref{eq:107}) is substituted in the second approximation.

Substituting Eqs.~(\ref{eq:106}), (\ref{eq:107}) and (\ref{eq:110})
into the multipole expansion of Eq.~(\ref{eq:36}), the following
result is derived:
\begin{equation}
  \tilde{f}^{(\pm s)}_X(\bm{r}) \approx
  (\mp i)^s
  \int
  \frac{d^2\bar{l}}{(2\pi)^2}
  e^{i\bar{\bm{l}}\cdot\bm{\theta}} e^{\pm is(\phi_l-\phi)}
  \tilde{g}^{(\pm s)}_X(\bar{\bm{l}},r),
  \label{eq:112}
\end{equation}
where a summation of $l$ is replaced by an integral,
$\sum_l \approx \int d\bar{l}$, which is justified in the flat-sky limit,
$l\gg 1$. The above equation with $s=0,1,2$ is found in the literature
\cite{Zaldarriaga:1996xe,Hu:2000ee}. The relation between spherical
tensors $\tilde{f}^{(\pm s)}_X$ in spherical coordinates and
$f^{(\pm s)}_X$ in Cartesian coordinates is given by
Eq.~(\ref{eq:93}). Correspondingly, we naturally define
\begin{equation}
  g^{(\pm s)}_X(\bar{\bm{l}},r) \equiv
  e^{\pm is\phi_l} \tilde{g}^{(\pm s)}_X(\bar{\bm{l}},r),
  \label{eq:113}
\end{equation}
and Eq.~(\ref{eq:112}) is equivalent to
\begin{equation}
  f^{(\pm s)}_X(\bm{r}-r_0\hat{\mathbf{e}}_3) \approx
  (\mp i)^s
  \int \frac{d^2\bar{l}}{(2\pi)^2}
  e^{i\bar{\bm{l}}\cdot\bm{\theta}}
  g^{(\pm s)}_X(\bar{\bm{l}},r).
  \label{eq:114}
\end{equation}
In the flat-sky limit, the correspondence between Cartesian
coordinates and spherical coordinates is approximately given by
$\bm{r} \approx (r\theta_1,r\theta_2,r)$, where $|\bm{\theta}| \ll 1$.
In the same limit, the wave vector is given by
$\bm{k} \approx (\bar{l}_1/r, \bar{l}_2/r, k_3)$. We apply a Fourier
transform of Eq.~(\ref{eq:93}) in this approximately flat space,
resulting in\footnote{We use the same symbol $f^{(\pm s)}_X$ for the
  function both in configuration space and in Fourier space for
  simplicity, while one should note that they are not the same
  function in fact. They should be distinguished by the type of
  arguments, $\bm{r}$ or $\bm{x}$ for configuration space and $\bm{k}$
  for Fourier space. The same simplified notations apply to other
  functions in the following, as long as they are not confusing.}
\begin{equation}
  f^{(\pm s)}_X(\bm{k}) \approx
  (\mp i)^s \int r^2 dr\, e^{-ik_3 (r-r_0)}
  g^{(\pm s)}_X(\bar{\bm{l}},r),
    \label{eq:115}
\end{equation}
where $\bar{\bm{l}} = (k_1r,k_2r,0)$. 

The properties of complex conjugate, Eq.~(\ref{eq:97}), rotations,
Eqs.~(\ref{eq:99}) and (\ref{eq:100}), and parity,
Eq.~(\ref{eq:104}), are translated into those in Fourier space. They
are given by
\begin{equation}
  f^{(\pm s)*}_X(\bm{k}) = (-1)^s f^{(\mp s)}_X(-\bm{k}),
  \label{eq:116}
\end{equation}
\begin{equation}
  f^{(\pm s)}_X(\bm{k}) \xrightarrow{\mathbb{R}}
  f'^{(\pm s)}_X(\bm{k}') =
  A_s
  F_{Xsm}(\bm{k}) D^{m}_{(s)\pm s}(R),
  \label{eq:117}
\end{equation}
\begin{equation}
  f^{(\pm s)}_X(\bm{k}) \xrightarrow{\mathbb{R}_\gamma}
  f'^{(\pm s)}_X(\bm{k}') = e^{\mp is\gamma} f^{(\pm s)}_X(\bm{k}),
  \label{eq:118}
\end{equation}
and
\begin{equation}
  f^{(\pm s)}_X(\bm{k}) \xrightarrow{\mathbb{P}}
  f^{\prime (\pm s)}_X(\bm{k})
  = (-1)^{p_X+s}f^{(\pm s)}_X(-\bm{k}).
  \label{eq:119}
\end{equation}
One can confirm that the properties of complex conjugate and parity,
Eqs.~(\ref{eq:116}) and (\ref{eq:119}), are explicitly satisfied
in the expression of Eq.~(\ref{eq:115}).

In Fourier space, the flipping rotation is defined as well by the
three-dimensional rotation in which the vector $\bm{k}$ is flipped by 180
degrees. The Euler angles of the rotation are given by
$(\alpha,\beta,\gamma)=(\phi_k,\pi,-\phi_k)$, where $\phi_k$ is the
azimuthal angle of the position vector
$\bm{k}=(k\sin\theta_k\cos\phi_k,
k\sin\theta_k\sin\phi_k,k\cos\theta_k)$. As in the configuration
space, the rotation matrix, Eq.~(\ref{eq:101}), is applied to
Eq.~(\ref{eq:117}), resulting in
\begin{equation}
  f^{(\pm s)}_X(\bm{k}) \xrightarrow{\mathbb{F}}
  f^{\prime(\pm s)}_X(\bm{k}) = e^{\pm 2is\phi_k} f^{(\mp s)}_X(-\bm{k}).
  \label{eq:120}
\end{equation}
The combined transformation of flip and parity is given by
\begin{equation}
  f^{(\pm s)}_X(\bm{k}) \xrightarrow{\mathbb{PF}}
  f^{\prime(\pm s)}_X(\bm{k}) = (-1)^{p_X + s}
  e^{\pm 2is\phi_k} f^{(\mp s)}_X(\bm{k}).
  \label{eq:121}
\end{equation}
As the last transformation does not change
the sign of the argument $\bm{k}$, the combined transformation is more
convenient than the pure parity transformation in Cartesian
coordinates in the following considerations.

\subsubsection{%
  E/B decomposition in Cartesian coordinates
}

The decomposition of the projected tensor fields into E/B modes in
spherical coordinates is given by Eqs.~(\ref{eq:44})--(\ref{eq:46}).
In the transition from the field $\tilde{f}^{(\pm s)}_X$ in spherical
coordinates system to $f^{(\pm s)}_X$ in Cartesian coordinates system
in the flat-sky limit, the phase factor $(\mp i)^s e^{\pm is\phi_k}$
is acquired through Eqs.~(\ref{eq:113})--(\ref{eq:115}) for each
$\pm s$ mode, respectively. In Fourier space of the distant-observer
approximation, the corresponding E/B modes are thus naturally given by
\begin{equation}
  f^{(\pm s)}_X(\bm{k}) =
  (\mp i)^s e^{\pm is\phi_k}
  \left[
    f^{\mathrm{E}(s)}_X(\bm{k})
    \pm i f^{\mathrm{B}(s)}_X(\bm{k})
  \right],
  \label{eq:122}
\end{equation}
or equivalently, 
\begin{align}
  f^{\mathrm{E}(s)}_X(\bm{k})
  &=
    \frac{i^s}{2}
    \left[
    e^{-is\phi_k} f^{(+s)}_X(\bm{k}) +
    (-1)^s e^{is\phi_k} f^{(-s)}_X(\bm{k})
    \right],
  \label{eq:123}\\
  f^{\mathrm{B}(s)}_X(\bm{k})
  &=
    \frac{i^{s-1}}{2}
    \left[
    e^{-is\phi_k} f^{(+s)}_X(\bm{k}) -
    (-1)^s e^{is\phi_k} f^{(-s)}_X(\bm{k})
    \right],
  \label{eq:124}
\end{align}
for non-negative integers, $s\geq 0$. The decomposed fields
$f^{\mathrm{E/B}(s)}_X(\bm{k})$ are invariant under the
two-dimensional passive rotation around the line of sight,
Eq.~(\ref{eq:118}), because the azimuthal angle transforms as
$\phi_k \rightarrow \phi_k - \gamma$. From Eqs.~(\ref{eq:116}),
(\ref{eq:123}), and (\ref{eq:124}), the complex conjugates of the E/B
modes are given by
\begin{equation}
  f^{\mathrm{E/B}(s)*}_X(\bm{k}) =
  (-1)^s f^{\mathrm{E/B}(s)}_X(-\bm{k}).
  \label{eq:125}
\end{equation}
The parity transformation, Eq.~(\ref{eq:119}) of the Fourier modes is
translated to those of E/B modes as
\begin{equation}
  f^{\mathrm{E/B}(s)}_X(\bm{k})
  \xrightarrow{\mathbb{P}}
  (-1)^{p_X+s}f^{\mathrm{E/B}(s)}_X(-\bm{k}).
  \label{eq:126}
\end{equation}
Similarly, the flipping rotation, Eq.~(\ref{eq:120}), is translated
to
\begin{equation}
  f^{\mathrm{E}(s)}_X(\bm{k})
  \xrightarrow{\mathbb{F}}
  f^{\mathrm{E}(s)}_X(-\bm{k}), \quad
  f^{\mathrm{B}(s)}_X(\bm{k})
  \xrightarrow{\mathbb{F}}
  -f^{\mathrm{B}(s)}_X(-\bm{k}),
  \label{eq:127}
\end{equation}
and the combination of parity and flipping, Eq.~(\ref{eq:121}), is
translated to
\begin{align}
  f^{\mathrm{E}(s)}_X(\bm{k})
  &\xrightarrow{\mathbb{PF}}
  (-1)^{p_X+s}  f^{\mathrm{E}(s)}_X(\bm{k}),
  \label{eq:128}\\
  f^{\mathrm{B}(s)}_X(\bm{k})
  &\xrightarrow{\mathbb{PF}}
  (-1)^{p_X+s+1} f^{\mathrm{B}(s)}_X(\bm{k}).
  \label{eq:129}
\end{align}

\section{\label{sec:PowerSpec}%
  The Power spectrum of the projected tensor field
}

\subsection{The power spectrum in the distant-observer limit
} 

The two-point correlation for the variables of Eq.~(\ref{eq:109}) in
the flat-sky limit is given by
\begin{equation}
  \left\langle
    \tilde{g}^{(\sigma_1)}_{X_1}(\bar{\bm{l}},r)
    \tilde{g}^{(\sigma_2)}_{X_2}(\bar{\bm{l}}',r')
  \right\rangle \approx
  \frac{4\pi}{2l+1}
  (2\pi)^2 \delta^2_\mathrm{D}(\bar{\bm{l}} + \bar{\bm{l}}')
  C^{(\sigma_1\sigma_2)}_{X_1X_2l}(r,r'),
  \label{eq:130}
\end{equation}
where $\sigma_1$ and $\sigma_2$ take either positive and negative
integers, and zero as well. The above equation is derived by noting an
identity,
\begin{equation}
  \frac{4\pi}{2l+1}
  \delta_{ll'} \sum_m (-1)^m e^{im(\phi_l-\phi_{l'})}
  \approx (2\pi)^2 \delta^2_\mathrm{D}(\bar{\bm{l}} + \bar{\bm{l}}'),
  \label{eq:131}
\end{equation}
in the flat-sky limit, which is shown by several methods, e.g., using
the Fourier decomposition of the two-dimensional delta function and
Eq.~(\ref{eq:111}).

In the distant-observer limit, the dependencies on two radial distances
$r$ and $r'$ in the angular power spectrum
$C^{(\sigma_1\sigma_2)}_{X_1X_2l}(r,r')$ can be decomposed into their
relative distance $x=r-r'$ and the fiducial distance $r_0$ introduced
in Eqs.~(\ref{eq:87}) and (\ref{eq:88}), and the angular
spectrum is denoted as
$C^{(\sigma_1\sigma_2)}_{X_1X_2l}(r,r') =
\bar{C}^{(\sigma_1\sigma_2)}_{X_1X_2l}(x;r_0)$, where $|x| \ll r_0$.
Combining Eqs.~(\ref{eq:113}), (\ref{eq:115}) and (\ref{eq:130}) in
the distant-observer approximation, we derive
\begin{equation}
  \left\langle
    f^{(\sigma_1)}_{X_1}(\bm{k}) f^{(\sigma_2)}_{X_2}(\bm{k}')
  \right\rangle
  = (2\pi)^3 \delta_\mathrm{D}^3(\bm{k}+\bm{k}')
  P^{(\sigma_1\sigma_2)}_{X_1X_2}(\bm{k}),
  \label{eq:132}
\end{equation}
where
\begin{multline}
  P^{(\sigma_1\sigma_2)}_{X_1X_2}(\bm{k}) =
  (-i)^{\sigma_1-\sigma_2}
  e^{i(\sigma_1+\sigma_2)\phi_k}
  \\
  \times
  \frac{4\pi{r_0}^2}{2l+1}
  \int dx\,e^{-ik_3x} \bar{C}^{(\sigma_1\sigma_2)}_{X_1X_2l}(x;r_0),
  \label{eq:133}
\end{multline}
and
\begin{equation}
  \phi_k =
  \arccos\left(\frac{k_1}{\sqrt{{k_1}^2  + {k_2}^2}}\right), \quad
  l = r_0\sqrt{{k_1}^2  + {k_2}^2}.
  \label{eq:134}
\end{equation}
The appearance of the three-dimensional delta function of the
wave vector in Eq.~(\ref{eq:132}) is a manifestation of the
translational invariance in the distant-observer limit, which is not
explicitly seen in the full-sky formulation with the spherical
coordinates. The distant-observer power spectrum
$P^{(\sigma_1\sigma_2)}_{X_1X_2}(\bm{k})$ depends explicitly on the
fiducial distance to the sample, $r_0$, which is omitted in the
argument above for simplicity. This dependence on $r_0$ is interpreted
as a redshift dependence of the power spectrum in the distant-observer
approximation.

Because of the relation of Eq.~(\ref{eq:94}) between the projected
tensor and three-dimensional tensor, the power spectrum
$P^{(\sigma_1\sigma_2)}_{X_1X_2}(\bm{k})$ is related to the power
spectrum $P^{(l_1l_2)}_{X_1X_2m_1m_2}(\bm{k})$ of the
three-dimensional tensor, defined in Paper~I, by
\begin{equation}
  \left\langle
    F_{X_1l_1m_1}(\bm{k}) F_{X_2l_2m_2}(\bm{k}')
  \right\rangle =
  (2\pi)^3 \delta_\mathrm{D}^3(\bm{k}+\bm{k}')
  P^{(l_1l_2)}_{X_1X_2m_1m_2}(\bm{k}).
  \label{eq:135}
\end{equation}
Simply substituting $l_1=|\sigma_1|$, $l_2=|\sigma_2|$, $m_1=\sigma_1$
and $m_2=\sigma_2$ in the above, and comparing with
Eq.~(\ref{eq:94}), we find the two power spectra are related by
\begin{equation}
  P^{(\sigma_1\sigma_2)}_{X_1X_2}(\bm{k}) =
  A_{|\sigma_1|} A_{|\sigma_2|}
  P^{(|\sigma_1||\sigma_2|)}_{X_1X_2\sigma_1\sigma_2}(\bm{k}).
  \label{eq:136}
\end{equation}

The transformation properties of complex conjugate, rotations, parity,
and others in Fourier modes, Eqs.~(\ref{eq:116})--(\ref{eq:121})
are trivially translated into the corresponding transformations of the
power spectrum. Among others, we explicitly write down just some of
them which are useful in the following considerations. The complex
conjugate is given by
\begin{equation}
  P^{(\sigma_1\sigma_2)*}_{X_1X_2}(\bm{k}) =
  (-1)^{\sigma_1+\sigma_2} P^{(-\sigma_1,-\sigma_2)}_{X_1X_2}(-\bm{k}).
  \label{eq:137}
\end{equation}
The rotation along the line of sight is given by
\begin{equation}
  P^{(\sigma_1\sigma_2)}_{X_1X_2}(\bm{k})
  \xrightarrow{\mathbb{R}_\gamma}
  P^{\prime (\sigma_1\sigma_2)}_{X_1X_2}(\bm{k}') =
  e^{-i(\sigma_1+\sigma_2)\gamma}
  P^{(\sigma_1\sigma_2)}_{X_1X_2}(\bm{k}),
  \label{eq:138}
\end{equation}
the flipping rotation is given by
\begin{equation}
  P^{(\sigma_1\sigma_2)}_{X_1X_2}(\bm{k})
  \xrightarrow{\mathbb{F}}
  P^{\prime(\sigma_1\sigma_2)}_{X_1X_2}(\bm{k}) =
  e^{2i(\sigma_1+\sigma_2)\phi_k}
  P^{(-\sigma_1,-\sigma_2)}_{X_1X_2}(-\bm{k}).
  \label{eq:139}
\end{equation}
and the combined transformation of parity and flipping is given by
\begin{multline}
  P^{(\sigma_1\sigma_2)}_{X_1X_2}(\bm{k})
  \xrightarrow{\mathbb{PF}}
  P^{\prime(\sigma_1\sigma_2)}_{X_1X_2}(\bm{k}) =
  (-1)^{p_{X_1} + p_{X_2} + \sigma_1 + \sigma_2}
  \\ \times
  e^{2i(\sigma_1+\sigma_2)\phi_k}
  P^{(-\sigma_1,-\sigma_2)}_{X_1X_2}(\bm{k}).
  \label{eq:140}
\end{multline}

In the Universe with statistical isotropy, which we assume throughout
this paper, the functional form of the power spectrum should not
change under the flipping rotation. Therefore, we have
$P^{\prime(\sigma_1\sigma_2)}_{X_1X_2}(\bm{k}) =
P^{(\sigma_1\sigma_2)}_{X_1X_2}(\bm{k})$ in Eq.~(\ref{eq:139}), or
equivalently,
\begin{equation}
  P^{(\sigma_1\sigma_2)}_{X_1X_2}(-\bm{k}) =
  e^{2i(\sigma_1+\sigma_2)\phi_k}
  P^{(-\sigma_1,-\sigma_2)}_{X_1X_2}(\bm{k}).
  \label{eq:141}
\end{equation}
When the above equality is satisfied, the property of the complex
conjugate of Eq.~(\ref{eq:137}) reduces to
\begin{equation}
  P^{(\sigma_1\sigma_2)*}_{X_1X_2}(\bm{k}) =
  (-1)^{\sigma_1+\sigma_2} e^{-2i(\sigma_1+\sigma_2)\phi_k}
  P^{(\sigma_1\sigma_2)}_{X_1X_2}(\bm{k}).
  \label{eq:142}
\end{equation}

\subsection{%
  The E/B power spectra
}

The power spectra of E/B modes,
$P^{\mathrm{EE}(s_1s_2)}_{X_1X_2}(\bm{k})$,
$P^{\mathrm{EB}(s_1s_2)}_{X_1X_2}(\bm{k})$,
$P^{\mathrm{BE}(s_1s_2)}_{X_1X_2}(\bm{k})$, and
$P^{\mathrm{BB}(s_1s_2)}_{X_1X_2}(\bm{k})$ are defined by
\begin{align}
  \left\langle
    f^{\mathrm{E}(s_1)}_{X_1}(\bm{k}) f^{\mathrm{E}(s_2)}_{X_2}(\bm{k}')
  \right\rangle
  &=
    (2\pi)^3 \delta_\mathrm{D}^3(\bm{k}+\bm{k}')
    P^{\mathrm{EE}(s_1s_2)}_{X_1X_2}(\bm{k}),
  \label{eq:143}\\
  \left\langle
    f^{\mathrm{E}(s_1)}_{X_1}(\bm{k}) f^{\mathrm{B}(s_2)}_{X_2}(\bm{k}')
  \right\rangle
  &=
    (2\pi)^3 \delta_\mathrm{D}^3(\bm{k}+\bm{k}')
    P^{\mathrm{EB}(s_1s_2)}_{X_1X_2}(\bm{k}),
  \label{eq:144}\\
  \left\langle
    f^{\mathrm{B}(s_1)}_{X_1}(\bm{k}) f^{\mathrm{E}(s_2)}_{X_2}(\bm{k}')
  \right\rangle
  &=
    (2\pi)^3 \delta_\mathrm{D}^3(\bm{k}+\bm{k}')
    P^{\mathrm{BE}(s_1s_2)}_{X_1X_2}(\bm{k}),
  \label{eq:145}\\
  \left\langle
    f^{\mathrm{B}(s_1)}_{X_1}(\bm{k}) f^{\mathrm{B}(s_2)}_{X_2}(\bm{k}')
  \right\rangle
  &=
    (2\pi)^3 \delta_\mathrm{D}^3(\bm{k}+\bm{k}')
    P^{\mathrm{BB}(s_1s_2)}_{X_1X_2}(\bm{k}),
  \label{eq:146}
\end{align}
where $s_1\geq 0, s_2 \geq 0$.
Substituting Eqs.~(\ref{eq:123}) and (\ref{eq:124}) into the above
equations, the relations between the power spectra of E/B modes and
those of Eq.~(\ref{eq:132}) are given by
\begin{align}
  P^{\mathrm{EE}}
  &= \frac{1}{4}
    \left[
    \left(Q^{++} + Q^{--}\right) + 
    \left(Q^{+-} + Q^{-+}\right)
    \right],
  \label{eq:147}\\
  P^{\mathrm{BB}}
  &= -\frac{1}{4}
    \left[
    \left(Q^{++} + Q^{--}\right) - 
    \left(Q^{+-} + Q^{-+}\right)
    \right],
  \label{eq:148}\\
  P^{\mathrm{EB}}
  &= -\frac{i}{4}
    \left[
    \left(Q^{++} - Q^{--}\right) - 
    \left(Q^{+-} - Q^{-+}\right)
    \right],
  \label{eq:149}\\
  P^{\mathrm{BE}}
  &= -\frac{i}{4}
    \left[
    \left(Q^{++} - Q^{--}\right) + 
    \left(Q^{+-} - Q^{-+}\right)
    \right],
  \label{eq:150}
\end{align}
where the abbreviations
\begin{align}
  &
    Q^{\pm\pm} =
    Q^{\pm\pm(s_1s_2)}_{X_1X_2}(\bm{k})
    \equiv (\pm i)^{s_1-s_2} e^{\mp i(s_1+s_2)\phi_k} 
    P^{(\pm s_1,\pm s_2)}_{X_1X_2}(\bm{k}),
  \label{eq:151}\\
  &
    Q^{\pm \mp} =
    Q^{\pm\mp(s_1s_2)}_{X_1X_2}(\bm{k})
    \equiv (\pm i)^{s_1+s_2} e^{\mp i(s_1-s_2)\phi_k} 
    P^{(\pm s_1,\mp s_2)}_{X_1X_2}(\bm{k}),
  \label{eq:152}\\
  &
    P^{\mathrm{EE}} = P^{\mathrm{EE}(s_1s_2)}_{X_1X_2}(\bm{k}), \quad
    P^{\mathrm{EB}} = P^{\mathrm{EB}(s_1s_2)}_{X_1X_2}(\bm{k}),
  \label{eq:153}\\
  &
    P^{\mathrm{BE}} = P^{\mathrm{BE}(s_1s_2)}_{X_1X_2}(\bm{k}), \quad
    P^{\mathrm{BB}} = P^{\mathrm{BB}(s_1s_2)}_{X_1X_2}(\bm{k})
  \label{eq:154}
\end{align}
are adopted just for simplicity of representation.

Assuming the invariance under the flipping rotation,
Eq.~(\ref{eq:141}), the property of the complex conjugate in
Eq.~(\ref{eq:142}) is translated to that of the functions $Q^{\pm\pm}$
and $Q^{\pm\mp}$ simply as
\begin{align}
  Q^{\pm\pm(s_1s_2)*}_{X_1X_2}(\bm{k})
  &=  Q^{\pm\pm(s_1s_2)}_{X_1X_2}(\bm{k}),
  \label{eq:155}\\
  Q^{\pm\mp(s_1s_2)*}_{X_1X_2}(\bm{k})
  &=  Q^{\pm\mp(s_1s_2)}_{X_1X_2}(\bm{k}).
  \label{eq:156}
\end{align}
That is, they are real functions. Therefore, the EE and BB power
spectra, Eqs.~(\ref{eq:147}) and (\ref{eq:148}), are both real, while
the EB and BE power spectra, Eqs.~(\ref{eq:149}) and (\ref{eq:150}),
are both pure imaginary.

Under the rotation along the line of sight of Eq.~(\ref{eq:138}), the
azimuthal angle of $\bm{k}$ transforms as
$\phi_k \rightarrow \phi_k -\gamma$, and the above functions are shown
to transform as
\begin{align}
  Q^{\pm\pm(s_1s_2)}_{X_1X_2}(\bm{k})
  &\xrightarrow{\mathbb{R}_\gamma}
    Q^{\pm\pm(s_1s_2)}_{X_1X_2}(\bm{k}),
  \label{eq:157}\\
  Q^{\pm\mp(s_1s_2)}_{X_1X_2}(\bm{k})
  &\xrightarrow{\mathbb{R}_\gamma}
   Q^{\pm\mp(s_1s_2)}_{X_1X_2}(\bm{k}),
  \label{eq:158}
\end{align}
that is, they are invariant under the rotation along the line of
sight. Therefore, the functions $Q^{\pm\pm}$ and $Q^{\pm\mp}$ do not
depend on the azimuthal angle $\phi_k$ of the wave vector $\bm{k}$,
and depend only on the magnitude $k$ and the polar angle $\theta_k$,
or the direction cosine $\mu = \cos\theta_k$. Therefore, the E/B power
spectra of Eqs.~(\ref{eq:147})--(\ref{eq:150}) do not depend on the
azimuthal angle $\phi_k$ either, and are functions of only $(k,\mu)$.

As for the combined transformation of parity and flipping rotation,
Eq.~(\ref{eq:140}), the corresponding transformations are given by
\begin{align}
  Q^{\pm\pm(s_1s_2)}_{X_1X_2}(\bm{k})
  &\xrightarrow{\mathbb{PF}}
  (-1)^{p_{X_1}+p_{X_2}}Q^{\mp\mp(s_1s_2)}_{X_1X_2}(\bm{k}),
  \label{eq:159}\\
  Q^{\pm\mp(s_1s_2)}_{X_1X_2}(\bm{k})
  &\xrightarrow{\mathbb{PF}}
  (-1)^{p_{X_1}+p_{X_2}}Q^{\mp\pm(s_1s_2)}_{X_1X_2}(\bm{k}).
  \label{eq:160}
\end{align}
Assuming the invariance of the power spectrum under the flipping
rotation, the combined transformations above are equivalent to the
purely parity transformation as noted before. Using the above
transformations in Eqs.~(\ref{eq:147})--(\ref{eq:150}), the parity
transformations of the E/B power spectra are found to be given by
\begin{align}
  P^{\mathrm{EE}(s_1s_2)}_{X_1X_2}(\bm{k})
  &\xrightarrow{\mathbb{PF}}
  (-1)^{p_{X_1}+p_{X_2}}P^{\mathrm{EE}(s_1s_2)}_{X_1X_2}(\bm{k}),
  \label{eq:161}\\
  P^{\mathrm{BB}(s_1s_2)}_{X_1X_2}(\bm{k})
  &\xrightarrow{\mathbb{PF}}
  (-1)^{p_{X_1}+p_{X_2}}P^{\mathrm{BB}(s_1s_2)}_{X_1X_2}(\bm{k}),
  \label{eq:162}\\
  P^{\mathrm{EB}(s_1s_2)}_{X_1X_2}(\bm{k})
  &\xrightarrow{\mathbb{PF}}
  (-1)^{p_{X_1}+p_{X_2}+1}P^{\mathrm{EB}(s_1s_2)}_{X_1X_2}(\bm{k}),
  \label{eq:163}\\
  P^{\mathrm{BE}(s_1s_2)}_{X_1X_2}(\bm{k})
  &\xrightarrow{\mathbb{PF}}
  (-1)^{p_{X_1}+p_{X_2}+1}P^{\mathrm{BE}(s_1s_2)}_{X_1X_2}(\bm{k}).
  \label{eq:164}
\end{align}
In the Universe with parity symmetry, the EB and BE power spectra
vanish if $p_{X_1}+p_{X_2}=\mathrm{even}$, while EE and BB power
spectra vanish if $p_{X_1}+p_{X_2}=\mathrm{odd}$. As is well known for
the spin-2 cases such as statistics of the CMB polarization and weak
lensing with $p_X=0$, the EB cross power spectra vanish in the
Universe with parity symmetry. However, in cases of cross spectra
between normal and pseudotensors, the EB spectrum does not vanish even
when the Universe has the parity symmetry and the EE and BB spectra
vanish instead.

\subsection{%
  Relation to the rotationally invariant power spectrum }

In Paper~I, assuming the statistical isotropy of the Universe, the
power spectra of three-dimensional tensor fields are represented by
rotationally invariant spectra in the distant-observer approximation,
and theoretical predictions of invariant spectra are consistently
calculated from the iPT of tensor fields. Below we explicitly
represent the power spectra of the projected field in terms of the
invariant spectra both in real space and in redshift space.

\subsubsection{Real space}

In real space, the power spectrum of the three-dimensional tensor
fields is represented by the invariant spectrum as derived in Paper~I:
\begin{equation}
  P^{(l_1l_2)}_{X_1X_2m_1m_2}(\bm{k}) =
  i^{l_1+l_2}
  \sum_l \left(l_1\,l_2\,l\right)_{m_1m_2}^{\phantom{m_1m_2}m}
  C_{lm}(\hat{\bm{k}}) P^{l_1l_2;l}_{X_1X_2}(k),
  \label{eq:165}
\end{equation}
where $P^{l_1l_2;l}_{X_1X_2}(k)$ is the invariant power spectrum of
tensor fields, $(l_1\,l_2\,l_3)_{m_1m_2m_3}$ is the $3j$-symbol
following the notation of Paper~I, and azimuthal indices are raised or
lowered by the spherical metric, as
$(l_1\,l_2\,l)_{m_1m_2}^{\phantom{m_1m_2}m} = (l_1\,l_2\,l)_{m_1m_2m'}
g_{(l)}^{m'm}$ and so forth. The functions $C_{lm}(\hat{\bm{k}})$ are
the spherical harmonics with Racah's normalization, which are defined
by
\begin{equation}
  C_{lm}(\theta,\phi)
  = \sqrt{\frac{4\pi}{2l+1}}\, Y_{lm}(\theta,\phi).
  \label{eq:166}
\end{equation}
The $3j$-symbol in the sum on the rhs of Eq.~(\ref{eq:165}) is
nonzero only when $m=m_1+m_2$ is satisfied. The invariant power
spectrum is a real function,
\begin{equation}
  P^{l_1l_2;l*}_{X_1X_2}(k) = P^{l_1l_2;l}_{X_1X_2}(k),
  \label{eq:167}
\end{equation}
and invariant under any three-dimensional rotation of coordinates axes.
The parity transformation of the function is given by
\begin{equation}
  P^{l_1l_2;l}_{X_1X_2}(k)
  \xrightarrow{\mathbb{P}}
  (-1)^{p_{X_1}+p_{X_2}+l_1+l_2+l}
  P^{l_1l_2;l}_{X_1X_2}(k).
  \label{eq:168}
\end{equation}

Substituting the above expression into the relation of
Eq.~(\ref{eq:136}), we have
\begin{multline}
  P^{(\sigma_1\sigma_2)}_{X_1X_2}(\bm{k}) =
  (-i)^{|\sigma_1|+|\sigma_2|}
  \sum_l
  \begin{pmatrix}
    |\sigma_1| & |\sigma_2| & l \\
    \sigma_1 & \sigma_2 & -\sigma_{12}
  \end{pmatrix}
  \\ \times
  Y_{l\sigma_{12}}(\hat{\bm{k}}) \hat{P}^{|\sigma_1||\sigma_2|;l}_{X_1X_2}(k),
  \label{eq:169}
\end{multline}
where we denote
\begin{equation}
  \sigma_{12} \equiv \sigma_1 + \sigma_2
  \label{eq:170}
\end{equation}
for simplicity, the functions $Y_{l\sigma_{12}}$ are the spherical
harmonics $Y_{lm}$ with $m=\sigma_{12}$, and we define the normalized
invariant spectrum,
\begin{equation}
  \hat{P}^{l_1l_2;l}_{X_1X_2}(k)
  \equiv
  A_{l_1}A_{l_2}\sqrt{\frac{4\pi}{2l+1}}\, P^{l_1l_2;l}_{X_1X_2}(k).
  \label{eq:171}
\end{equation}
The $3j$-symbol in the above Eq.~(\ref{eq:169}) is nonzero only
when $l\leq |\sigma_1|+|\sigma_2|$ for the triangle inequality and
$l \geq |\sigma_{12}|$ for the third row of the $3j$-symbol. In
particular, when the signs of $\sigma_1$ and $\sigma_2$ coincide, or
either of them is zero (i.e., when $\sigma_1\sigma_2 \geq 0$), we have
$|\sigma_{12}| = |\sigma_1|+|\sigma_2|$, and there is only one
possibility of $l = |\sigma_1|+|\sigma_2|$ that satisfies the
conditions. On the contrary, when $\sigma_1\sigma_2 < 0$, there are
multiple possibilities of
$l = |\sigma_{12}|,\ldots,|\sigma_1|+|\sigma_2|$.

The symmetries of Eq.~(\ref{eq:137})--(\ref{eq:142}) are explicitly
satisfied by the above expression, as shown by using the symmetries of
the $3j$-symbol and the spherical harmonics, which are conveniently
listed in Paper~I. The methods of theoretically evaluating the
invariant spectrum by the iPT are detailed in Papers~I and II. The
angular dependence of the power spectrum on the direction
$\hat{\bm{k}}$ is uniquely determined by the spherical harmonics
$Y_{l\sigma_{12}}(\hat{\bm{k}})$ in Eq.~(\ref{eq:169}) due to the
symmetry. In this sense, the power spectrum
$P^{(\sigma_1\sigma_2)}_{X_1X_2}(\bm{k})$ has redundant information on
the directional dependence, due to the statistical isotropy of the
Universe.

\subsubsection{Redshift space}

In redshift space, the corresponding expression of the power spectrum
of tensor fields in terms of the invariant power spectrum is given by
(Paper~I)
\begin{multline}
  P^{(l_1l_2)}_{X_1X_2m_1m_2}(\bm{k};\hat{\bm{z}}) =
  i^{l_1+l_2}
  \sum_{l,l_z,L}
  (-1)^L \sqrt{2L+1}
  \left(l_1\,l_2\,L\right)_{m_1m_2}^{\phantom{m_1m_2}M}
  \\ \times
  X^{l_zl}_{LM}(\hat{\bm{z}},\hat{\bm{k}})
  P^{l_1l_2;l\,l_z;L}_{X_1X_2}(k,\mu),
  \label{eq:172}
\end{multline}
where $P^{l_1l_2;l\,l_z;L}_{X_1X_2}(k,\mu)$ is the invariant power
spectrum in redshift space, the direction to the line of sight
$\hat{\bm{z}}$ is arbitrarily oriented, and
$\mu=\hat{\bm{z}}\cdot\hat{\bm{k}}$ is the direction cosine of the
wave vector with respect to the line of sight, and
\begin{equation}
  X^{l_zl}_{LM}(\hat{\bm{z}},\hat{\bm{k}})
  = \left(L\,l_z\,l\right)_{M}^{\phantom{M}m_zm}
    C_{l_zm_z}(\hat{\bm{z}}) C_{lm}(\hat{\bm{k}})
  \label{eq:173}
\end{equation}
are the bipolar spherical harmonics \cite{Khersonskii:1988krb,PaperI}.
In Eq.~(\ref{eq:172}) derived in Paper~I, the direction of the line of
sight $\hat{\bm{z}}$ is not fixed and arbitrarily oriented in the
coordinates system. However, the direction of the projection in the
distant-observer limit of our context is fixed to the third axis,
$\hat{\bm{z}} = \mathbf{e}_3$, in this paper, and we have
\begin{equation}
  C_{l_zm_z}(\hat{\bm{z}}) =
  \sqrt{\frac{4\pi}{2l_z+1}}\, Y_{l_zm_z}(0,\phi) = 
  \delta_{m_z0}.
  \label{eq:174}
\end{equation}

As explained in Papers~I and II, if the dependence on the direction
cosine $\mu$ is completely expanded and the invariant spectrum does
not depend on $\mu$, the expansion is uniquely determined. In this
case, the number of terms in the series can be infinite if the
dependence of $\mu$ in the spectrum is nonpolynomial, such as the
nonlinear effect of the fingers of God. In practice, it makes analytical
calculations easier to include the dependence on $\mu$ in the
invariant spectrum as much as possible (see Paper~II for more details
of practical calculations). Below, we consider the general case that
the invariant spectrum can explicitly depend on the direction cosine
$\mu$ of the wave vector.

In Paper~I, it is shown that an equality
\begin{equation}
  P^{l_1l_2;l\,l_z;L}_{X_1X_2}(k,\mu) =
  (-1)^{l_z} P^{l_1l_2;l\,l_z;L}_{X_1X_2}(k,-\mu)
  \label{eq:175}
\end{equation}
holds as the system is invariant under the flipping of the direction
of line of sight, and that the complex conjugate of the invariant
spectrum is given by
\begin{equation}
  P^{l_1l_2;l\,l_z;L\,*}_{X_1X_2}(k,\mu)
  = P^{l_1l_2;l\,l_z;L}_{X_1X_2}(k,\mu).
  \label{eq:176}
\end{equation}
The parity transformation is given by
\begin{equation}
  P^{l_1l_2;l\,l_z;L}_{X_1X_2}(k,\mu)
  \xrightarrow{\mathbb{P}}
  (-1)^{p_{X_1}+p_{X_2}+l_1+l_2+l+l_z}
  P^{l_1l_2;l\,l_z;L}_{X_1X_2}(k,\mu). 
  \label{eq:177}
\end{equation}

Substituting Eqs.~(\ref{eq:172})--(\ref{eq:174}) into the relation
of Eq.~(\ref{eq:136}), we have
\begin{multline}
  P^{(\sigma_1\sigma_2)}_{X_1X_2}(\bm{k}) =
  i^{|\sigma_1|+|\sigma_2|}
  \sum_{L,l,l_z} (-1)^L\sqrt{2L+1}
  \begin{pmatrix}
    |\sigma_1| & |\sigma_2| & L \\
    \sigma_1 & \sigma_2 & -\sigma_{12}
  \end{pmatrix}
  \\ \times
  \begin{pmatrix}
    L & l_z & l \\
    \sigma_{12} & 0 & -\sigma_{12}
  \end{pmatrix}
  Y_{l\sigma_{12}}(\hat{\bm{k}})
  \hat{P}^{|\sigma_1||\sigma_2|;l\,l_z;L}_{X_1X_2}(k,\mu),
  \label{eq:178}
\end{multline}
where we define a normalized invariant spectrum
\begin{equation}
  \hat{P}^{l_1l_2;l\,l_z;L}_{X_1X_2}(k,\mu)
  \equiv
  A_{l_1}A_{l_2}\sqrt{\frac{4\pi}{2l+1}}\, 
  P^{l_1l_2;l\,l_z;L}_{X_1X_2}(k,\mu).
  \label{eq:179}
\end{equation}
The first $3j$-symbol on the rhs of Eq.~(\ref{eq:178}) is nonzero
only when $L\leq |\sigma_1|+|\sigma_2|$ for the triangle inequality
and $L \geq |\sigma_{12}|$ for the third row of the $3j$-symbol. In
particular, when the signs of $\sigma_1$ and $\sigma_2$ coincide, or
either of them is zero (i.e., when $\sigma_1\sigma_2 \geq 0$), we have
$|\sigma_{12}| = |\sigma_1|+|\sigma_2|$, and there is only one
possibility of $L = |\sigma_1|+|\sigma_2|$ that satisfies the
conditions. On the contrary, when $\sigma_1\sigma_2 < 0$, there are
multiple possibilities of
$L = |\sigma_{12}|,\ldots,|\sigma_1|+|\sigma_2|$. The summations over
$l_z$ and $l$ generally remain, unless either of them is zero.

The symmetric properties of Eqs.~(\ref{eq:137})--(\ref{eq:142}) are
explicitly shown to be automatically satisfied by the above
expression, using the symmetries of the $3j$-symbol and the spherical
harmonics, as well as in the case of redshift space. The methods of
theoretically evaluating the invariant spectrum in redshift space by
the iPT are detailed in Paper~I and II. The angular dependence of the
power spectrum on the direction $\hat{\bm{k}}$ is uniquely determined
by the spherical harmonics $Y_{l\sigma_{12}}(\hat{\bm{k}})$ in
Eq.~(\ref{eq:178}) due to the symmetry. In this sense, the power
spectrum $P^{(\sigma_1\sigma_2)}_{X_1X_2}(\bm{k})$ has redundant
information on the directional dependence in redshift space as well,
due to the rotational symmetry.

\subsection{\label{subsec:EBdecomp}%
  The E/B power spectra in terms of the invariant power spectrum}

The E/B power spectra are also represented by the invariant spectrum
of the three-dimensional tensor field. The expressions are
straightforwardly derived by substituting Eq.~(\ref{eq:169}) in real
space, or Eq.~(\ref{eq:178}) in redshift space, into
Eqs.~(\ref{eq:147})--(\ref{eq:154}). As we have shown in
Eqs.~(\ref{eq:157}) and (\ref{eq:158}), the E/B power spectra do not
depend on the azimuthal angle $\phi_k$ of the wave vector $\bm{k}$.
Therefore the phase factor $e^{i\sigma_{12}\phi_k}$ of the spherical
harmonics $Y_{l,\sigma_{12}}(\hat{\bm{k}})$ should be canceled in the
corresponding expressions of E/B spectra. The spherical harmonics
without the phase factor is equivalent to the associated Legendre
polynomials, $P_l^m(\cos\theta)$. In the case of power spectra
involving the density and intrinsic alignment with spin-0 and spin-2,
it is already shown that the associated Legendre polynomials with
$m=0,2,4$ naturally appear in the projected power spectra
\cite{Kurita:2022agh}, which is fully consistent with the above
consideration because the index $m$ of the associated Legendre
polynomials corresponds to $\sigma_{12}$, and the absolute values can
take only $|\sigma_{12}|=0,2,4$ in this case.

It is convenient to introduce normalized associated Legendre
polynomials,
\begin{equation}
  \Theta_l^m(\cos\theta) \equiv
  \sqrt{2\pi}\, Y_{lm}(\theta,0) =
  \sqrt{\frac{2l+1}{2}\cdot\frac{(l-m)!}{(l+m)!}}\,
  P_l^m(\cos\theta).
  \label{eq:180}
\end{equation}
The Condon-Shortley phase is included in our
convention of the associated Legendre polynomials,
\begin{equation}
  P_l^m(x) =
  \frac{(-1)^m}{2^l\,l!}
  \left( 1-x^2 \right)^{m/2}
  \frac{d^{l+m}}{dx^{l+m}}
  \left( 1-x^2 \right)^l.
  \label{eq:181}
\end{equation}
The normalized functions above satisfy
\begin{equation}
  \Theta_l^{-m}(x) =
  (-1)^m \Theta_l^m(x), \quad
  \Theta_l^{m}(-x) =
  (-1)^{l+m} \Theta_l^m(x),
  \label{eq:182}
\end{equation}
as derived from a standard property of associated Legendre polynomials
or spherical harmonics. An orthogonality relation of the associated
Legendre polynomials in terms of the normalized ones is given by
\begin{equation}
  \int_{-1}^1 dx\,\Theta_l^m(x)\,\Theta_{l'}^m(x)
  =\delta_{ll'},
  \label{eq:183}
\end{equation}
where the common azimuthal index $m$ on the left-hand side (LHS) is
arbitrary and is not summed over in the above equation.

\subsubsection{Real space}

In real space, Eq.~(\ref{eq:169}) is substituted into
Eqs.~(\ref{eq:151}) and (\ref{eq:152}). Using identities of
$3j$-symbols, the results are given by
\begin{align}
  Q^{\pm\pm(s_1s_2)}_{X_1X_2}(\bm{k})
  &= \frac{(-1)^{s_2}}{\sqrt{2\pi\,(2s^+_{12}+1)}}
  \Theta_{s^+_{12}}^{s^+_{12}}(\mu)
  \hat{P}^{s_1s_2;s^+_{12}}_{X_1X_2}(k),
  \label{eq:184}\\
  Q^{\pm\mp(s_1s_2)}_{X_1X_2}(\bm{k})
  &= \frac{(\pm 1)^{s^+_{12}}}{\sqrt{2\pi}}
    \sum_{l=|s^-_{12}|}^{s^+_{12}} (\pm 1)^l
    \Theta_l^{s^-_{12}}(\mu)
    \nonumber \\
  & \hspace{3.6pc} \times
  \begin{pmatrix}
    s_1 & s_2 & l \\
    s_1 & -s_2 & -s^-_{12}
  \end{pmatrix}
  \hat{P}^{s_1s_2;l}_{X_1X_2}(k),
  \label{eq:185}
\end{align}
where simplified notations
\begin{equation}
  s^\pm_{12} \equiv s_1 \pm s_2
  \label{eq:186}
\end{equation}
are employed, and
$\mu \equiv \cos\theta_k = \mathbf{e}_3\cdot\hat{\bm{k}}$ is the
cosine of the polar angle of wave vector $\bm{k}$. In
Eq.~(\ref{eq:184}), the summation over $l$ does not appear because of
the same signs of the modes as described below Eq.~(\ref{eq:171}).
Using Eqs.~(\ref{eq:167}), (\ref{eq:168}) and (\ref{eq:182}), one can
see that the symmetric properties Eqs.~(\ref{eq:155})--(\ref{eq:160}),
are all explicitly satisfied. Because of Eq.~(\ref{eq:184}), we have
identically $Q^{++} = Q^{--}$ in real space. This property does not
generally hold in redshift space, as we shortly see below. The E/B
power spectra in real space are given by substituting
Eqs.~(\ref{eq:184}) and (\ref{eq:185}) into
Eqs.~(\ref{eq:147})--(\ref{eq:150}).

To extract separately the contributions of Eqs.~(\ref{eq:184}) and
(\ref{eq:185}), it is natural to define combinations of the power
spectra,
\begin{equation}
  P^\pm \equiv P^\mathrm{EE} \pm  P^\mathrm{BB}, \qquad
  Q^\pm \equiv -i
  \left(P^\mathrm{EB} \pm  P^\mathrm{BE}\right),
  \label{eq:187}
\end{equation}
in the abbreviated notation. Substituting
Eqs.~(\ref{eq:147})--(\ref{eq:150}) into the above, we have
$P^+ = (Q^{+-} + Q^{-+})/2$, $P^- = (Q^{++} + Q^{--})/2$,
$Q^+ = -(Q^{++} - Q^{--})/2$, and $Q^- = (Q^{+-} - Q^{-+})/2$, or
explicitly,
\begin{align}
  \begin{Bmatrix}
    P^+ \\ Q^-
  \end{Bmatrix}
  &= \frac{1}{\sqrt{2\pi}}
  \sum_l
  \frac{1 \pm (-1)^{s^+_{12}+ l}}{2}\,
  \Theta_l^{s^-_{12}}(\mu)
    \nonumber \\
  & \hspace{6pc} \times
  \begin{pmatrix}
    s_1 & s_2 & l \\
    s_1 & -s_2 & -s^-_{12}
  \end{pmatrix}
  \hat{P}^{s_1s_2;l}_{X_1X_2}(k),
  \label{eq:188}\\
  P^-
  &= 
  \frac{(-1)^{s_2}}{\sqrt{2\pi\,(2s^+_{12}+1)}}
  \Theta_{s^+_{12}}^{s^+_{12}}(\mu)
  \hat{P}^{s_1s_2;s^+_{12}}_{X_1X_2}(k),
  \label{eq:189}\\
  Q^+
  &= 0,
  \label{eq:190}
\end{align}
where the double sign in Eq.~(\ref{eq:188}) corresponds to the
upper and lower elements, respectively, in the curly brackets on the
LHS. From the above, the E/B power spectra are given by
$P^\mathrm{EE} = (P^+ + P^-)/2$, $P^\mathrm{BB} = (P^+ - P^-)/2$,
$P^\mathrm{EB} = i(Q^+ + Q^-)/2$, $P^\mathrm{BE} = i(Q^+ - Q^-)/2$.
Equation~(\ref{eq:190}) indicates that
$P^\mathrm{EB}=-P^\mathrm{BE} = iQ^-/2$ in real space. However, this
property does not generally hold in redshift space, as we shortly see
below.

In a Universe with parity symmetry, i.e., when the invariant spectrum
is invariant under Eq.~(\ref{eq:168}), we have
$(-1)^{s^+_{12}+l} =(-1)^{p_{X_1}+p_{X_2}}$ in Eq.~(\ref{eq:188}).
Also Eq.~(\ref{eq:189}) is nonzero only when
$p_{X_1}+p_{X_2}=\mathrm{even}$. Therefore, we have
$P^\mathrm{EB} = P^\mathrm{BE} = 0$ when
$p_{X_1}+p_{X_2}=\mathrm{even}$, and
$P^\mathrm{EE} = P^\mathrm{BB} =0$ when $p_{X_1}+p_{X_2}=\mathrm{odd}$
in a parity-conserved Universe.

Because of the orthogonality of the associated Legendre polynomials,
Eq.~(\ref{eq:183}), one can extract the invariant spectrum by angular
integrations. When $s^+_{12}+l=\mathrm{even}$, from
Eq.~(\ref{eq:188}), we have
\begin{multline}
  \hat{P}^{s_1s_2;l}_{X_1X_2}(k) =
  \sqrt{2\pi}\,
  \begin{pmatrix}
    s_1 & s_2 & l \\
    s_1 & -s_2 & -s^-_{12}
  \end{pmatrix}^{-1}
  \int_{-1}^1 d\mu\, 
  \Theta_l^{s^-_{12}}(\mu)
  \\ \times
  \left[
    P^{\mathrm{EE}(s_1s_2)}_{X_1X_2}(k,\mu) +
    P^{\mathrm{BB}(s_1s_2)}_{X_1X_2}(k,\mu)
  \right],
  \label{eq:191}
\end{multline}
where we present the relation in original notations rather than the
abbreviated notations, and we note that the power spectra
$P^\mathrm{EE/BB}$ do not depend on the azimuthal angle $\phi_k$ of
the wave vector $\bm{k}$. Similar expressions are obtained from
Eq.~(\ref{eq:188}) when $s^+_{12}+l=\mathrm{odd}$,
\begin{multline}
  \hat{P}^{s_1s_2;l}_{X_1X_2}(k) =
  \sqrt{2\pi}\,
  \begin{pmatrix}
    s_1 & s_2 & l \\
    s_1 & -s_2 & -s^-_{12}
  \end{pmatrix}^{-1}
  \int_{-1}^1 d\mu\, 
  \Theta_l^{s^-_{12}}(\mu)
  \\ \times
  \frac{1}{i}
  \left[
    P^{\mathrm{EB}(s_1s_2)}_{X_1X_2}(k,\mu) -
    P^{\mathrm{BE}(s_1s_2)}_{X_1X_2}(k,\mu)
  \right],
  \label{eq:192}
\end{multline}
where we note EB/BE spectra are pure imaginary. One notices that
Eq.~(\ref{eq:191}) in the case of $l=s^+_{12}$ is proportional to
Eq.~(\ref{eq:189}), and thus a consistency relation between angular
integrals of $P^\pm$ can be derived. If we write the relation in
original notations rather than the abbreviated notations, we have
\begin{multline}
  \int_{-1}^1 d\mu\,P_{s^+_{12}}^{s^-_{12}}(\mu)
  \left[
    P^{\mathrm{EE}(s_1s_2)}_{X_1X_2}(k,\mu) +
    P^{\mathrm{BB}(s_1s_2)}_{X_1X_2}(k,\mu)
  \right]
  \\
  = (-1)^{s_2} \frac{(2s_1)!}{(2s^+_{12})!}
  \int_{-1}^1 d\mu\,P_{s^+_{12}}^{s^+_{12}}(\mu)
  \\ \times
  \left[
    P^{\mathrm{EE}(s_1s_2)}_{X_1X_2}(k,\mu) -
    P^{\mathrm{BB}(s_1s_2)}_{X_1X_2}(k,\mu)
  \right],
  \label{eq:193}
\end{multline}
where $P_l^m(\mu)$ is the original associated Legendre polynomials.
However, the similar consistency relation in redshift space is not
found in general.

\subsubsection{Redshift space}

In redshift space, we substitute Eq.~(\ref{eq:178}) into
Eqs.~(\ref{eq:151}) and (\ref{eq:152}). Following the similar
calculations to those in real space above, we have
\begin{multline}
  Q^{\pm\pm}
  = \frac{(-1)^{s_2} (\pm 1)^{s^+_{12}}}{\sqrt{2\pi}}
  \sum_{l} (\pm 1)^{l}\,
  \Theta_l^{s^+_{12}}(\mu)
  \sum_{l_z} (\pm 1)^{l_z}
  \\ \times
  \begin{pmatrix}
    s^+_{12} & l_z & l \\
    s^+_{12} & 0 & -s^+_{12}
  \end{pmatrix}
  \hat{P}^{s_1s_2;l\,l_z;s^+_{12}}_{X_1X_2}(k,\mu),
  \label{eq:194}
\end{multline}
and
\begin{multline}
  Q^{\pm\mp} =
  \frac{(\mp 1)^{s^+_{12}}}{\sqrt{2\pi}}
  \sum_{l} (\pm 1)^{l}\,
  \Theta_l^{s^-_{12}}(\mu)
  \sum_{l_z,L} (\pm 1)^{l_z} (-1)^L \sqrt{2L+1}
  \\ \times
  \begin{pmatrix}
    s_1 & s_2 & L \\
    s_1 & -s_2 & -s^-_{12}
  \end{pmatrix}
  \begin{pmatrix}
    L & l_z & l \\
    s^-_{12} & 0 & -s^-_{12}
  \end{pmatrix}
  \hat{P}^{s_1s_2;l\,l_z;L}_{X_1X_2}(k,\mu).
  \label{eq:195}
\end{multline}
As a consistency check, putting $l_z=0$ reproduces the above results
of Eqs.~(\ref{eq:184}) and (\ref{eq:185}) in real space with a
correspondence
$\hat{P}^{(s_1s_2);l\,0;l}_{X_1X_2}(k,0) \leftrightarrow
\hat{P}^{(s_1s_2);l}_{X_1X_2}(k)$. Using
Eqs.~(\ref{eq:175})--(\ref{eq:177}) and (\ref{eq:182}), one can
see that the symmetric properties
Eqs.~(\ref{eq:155})--(\ref{eq:160}), are all explicitly satisfied.

The E/B spectra in redshift space are derived by substituting
Eqs.~(\ref{eq:194}) and (\ref{eq:195}) into
Eqs.~(\ref{eq:147})--(\ref{eq:150}). The combinations of
Eq.~(\ref{eq:187}) are useful also in redshift space. The results
are straightforward, and explicitly given by
\begin{align}
  \begin{Bmatrix}
    P^+ \\ Q^-
  \end{Bmatrix}
  &=
    \frac{(-1)^{s^+_{12}}}{\sqrt{2\pi}}
    \sum_{l,l_z}
    \frac{1 \pm (-1)^{s^+_{12}+l+l_z}}{2}\,
    \Theta_l^{s^-_{12}}(\mu)
    \nonumber\\
  & \quad \times
    \sum_L (-1)^L \sqrt{2L+1}
  \begin{pmatrix}
    s_1 & s_2 & L \\
    s_1 & -s_2 & -s^-_{12}
  \end{pmatrix}
  \begin{pmatrix}
    L & l_z & l \\
    s^-_{12} & 0 & -s^-_{12}
  \end{pmatrix}
    \nonumber\\
  & \hspace{9.5pc} \times
  \hat{P}^{s_1s_2;l\,l_z;L}_{X_1X_2}(k,\mu),
  \label{eq:196}\\
  \begin{Bmatrix}
    P^- \\ Q^+
  \end{Bmatrix}
  &= \pm \frac{(-1)^{s_2}}{\sqrt{2\pi}}
  \sum_{l,l_z}
    \frac{1 \pm (-1)^{s^+_{12}+l+l_z}}{2}\,
    \Theta_l^{s^+_{12}}(\mu)
  \begin{pmatrix}
    s^+_{12} & l_z & l \\
    s^+_{12} & 0 & -s^+_{12}
  \end{pmatrix}
    \nonumber\\
  & \hspace{9.5pc} \times
  \hat{P}^{s_1s_2;l\,l_z;s^+_{12}}_{X_1X_2}(k,\mu).
  \label{eq:197}
\end{align}
An important difference from the case of real space is that the
function $Q^+$ does not identically vanish in general in redshift
space, except for special cases as we consider below.

In the Universe with parity symmetry, the invariant spectrum is
invariant under the transformation of Eq.~(\ref{eq:177}), and we
have $(-1)^{s^+_{12}+l+l_z}=(-1)^{p_{X_1}+p_{X_2}}$ in
Eqs.~(\ref{eq:196}) and (\ref{eq:197}). Therefore, just in the
same way in real space, we have $P^\mathrm{EB} = P^\mathrm{BE} = 0$
when $p_{X_1}+p_{X_2}=\mathrm{even}$, and
$P^\mathrm{EE} = P^\mathrm{BB} =0$ when $p_{X_1}+p_{X_2}=\mathrm{odd}$
in a parity-conserved Universe. The inversion relations of
Eqs.~(\ref{eq:191}) and (\ref{eq:192}), and a consistency relation
of Eq.~(\ref{eq:193}) are not simply found in redshift space, partly
because of the presence of nonzero integers of $l_z$ in the
expansions.

\section{\label{sec:CorrFunc}%
  The correlation function of the projected tensor field
}

\subsection{The correlation functions in the distant-observer limit }

Within the distant-observer approximation, the two-point correlation
function of the projected field in configuration space is defined by
\begin{equation}
  \xi^{(\sigma_1\sigma_2)}_{X_1X_2}(\bm{x}-\bm{x}') =
  \left\langle
    f^{(\sigma_1)}_{X_1}(\bm{x}) f^{(\sigma_2)}_{X_2}(\bm{x}')
  \right\rangle.
  \label{eq:198}
\end{equation}
Because of the statistical homogeneity of the Cartesian coordinates in
the distant-observer approximation, the rhs is only a function of the
relative position, $\bm{x}-\bm{x}'$. This can be explicitly shown by
the Fourier transform of the above equation and Eq.~(\ref{eq:132}),
and the correlation function is a Fourier transform of the power
spectrum:
\begin{equation}
  \xi^{(\sigma_1\sigma_2)}_{X_1X_2}(\bm{x})
  = \int \frac{d^3k}{(2\pi)^3} e^{i\bm{k}\cdot\bm{x}}
  P^{(\sigma_1\sigma_2)}_{X_1X_2}(\bm{k}).
  \label{eq:199}
\end{equation}

In a similar way to the symmetric properties of the power
spectrum, Eqs.~(\ref{eq:137})--(\ref{eq:140}), the counterparts of
the correlation function are derived from
Eqs.~(\ref{eq:97})--(\ref{eq:105}). Among others, the complex
conjugate is given by
\begin{equation}
  \xi^{(\sigma_1\sigma_2)*}_{X_1X_2}(\bm{x}) =
  (-1)^{\sigma_1+\sigma_2} \xi^{(-\sigma_1,-\sigma_2)}_{X_1X_2}(\bm{x}), 
  \label{eq:200}
\end{equation}
the rotation along the line of sight is given by
\begin{equation}
  \xi^{(\sigma_1\sigma_2)}_{X_1X_2}(\bm{x})
  \xrightarrow{\mathbb{R}_\gamma}
  \xi^{\prime (\sigma_1\sigma_2)}_{X_1X_2}(\bm{x}') =
  e^{-i(\sigma_1+\sigma_2)\gamma}
  \xi^{(\sigma_1\sigma_2)}_{X_1X_2}(\bm{x}),
  \label{eq:201}
\end{equation}
the flipping rotation is given by
\begin{equation}
  \xi^{(\sigma_1\sigma_2)}_{X_1X_2}(\bm{x})
  \xrightarrow{\mathbb{F}}
  \xi^{\prime(\sigma_1\sigma_2)}_{X_1X_2}(\bm{x}) =
  e^{2i(\sigma_1+\sigma_2)\phi}
  \xi^{(-\sigma_1,-\sigma_2)}_{X_1X_2}(-\bm{x}).
  \label{eq:202}
\end{equation}
and the combined transformation of parity and flipping is given by
\begin{multline}
  \xi^{(\sigma_1\sigma_2)}_{X_1X_2}(\bm{x})
  \xrightarrow{\mathbb{PF}}
  \xi^{\prime(\sigma_1\sigma_2)}_{X_1X_2}(\bm{x}) =
  (-1)^{p_{X_1} + p_{X_2} + \sigma_1 + \sigma_2}
  \\ \times
  e^{2i(\sigma_1+\sigma_2)\phi}
  \xi^{(-\sigma_1,-\sigma_2)}_{X_1X_2}(\bm{x}).
  \label{eq:203}
\end{multline}

In the Universe with statistical isotropy, the functional form of the
power spectrum should not change under the flipping rotation, and we have
\begin{equation}
  \xi^{(\sigma_1\sigma_2)}_{X_1X_2}(-\bm{x}) =
  e^{2i(\sigma_1+\sigma_2)\phi}
  \xi^{(-\sigma_1,-\sigma_2)}_{X_1X_2}(\bm{x}),
  \label{eq:204}
\end{equation}
and the property of complex conjugate of Eq.~(\ref{eq:200}) reduces
to
\begin{equation}
  \xi^{(\sigma_1\sigma_2)*}_{X_1X_2}(\bm{x}) =
  (-1)^{\sigma_1+\sigma_2} e^{-2i(\sigma_1+\sigma_2)\phi}
  \xi^{(\sigma_1\sigma_2)}_{X_1X_2}(-\bm{x}).
  \label{eq:205}
\end{equation}

\subsection{%
  The tangential/cross correlation functions
}

The correlation function of the projected tensor field can be
decomposed into tangential ($+$) and cross ($\times$) components,
which are invariant under the two-dimensional rotation of the sky.
This decomposition is popular in the statistics of weak lensing field
of spin-2. This decomposition in configuration space is a counterpart
of the E/B decomposition in Fourier space given by
Eq.~(\ref{eq:122}). The decomposition is defined by
\begin{equation}
  f^{(\pm s)}_X(\bm{x}) =
  (\pm 1)^s e^{\pm is\phi}
  \left[
    f^{+(s)}_X(\bm{x})
    \pm i f^{\times(s)}_X(\bm{x})
  \right]
  \label{eq:206}
\end{equation}
for non-negative integers $s\geq 0$, where $\phi$ is the azimuthal
angle of $\bm{x}$. Equivalently, we have
\begin{align}
  f^{+(s)}_X(\bm{x})
  &=
    \frac{1}{2}
    \left[
    e^{-is\phi} f^{(+s)}_X(\bm{x}) +
    (-1)^s e^{is\phi} f^{(-s)}_X(\bm{x})
    \right],
  \label{eq:207}\\
  f^{\times(s)}_X(\bm{x})
  &=
    -\frac{i}{2}
    \left[
    e^{-is\phi} f^{(+s)}_X(\bm{x}) -
    (-1)^s e^{is\phi} f^{(-s)}_X(\bm{x})
    \right].
  \label{eq:208}
\end{align}
Comparing with the E/B decomposition in Fourier space,
Eq.~(\ref{eq:122}), a factor of $(\mp i)^s$ in Fourier space is
replaced by $(\pm 1)^s$ in the above, thus differs in an additional
phase factor $i^s$ from the definition in Fourier space. This addition
of the phase factor conveniently makes the correlation functions of
the decomposed field always real as shown below. According to the
property of Eq.~(\ref{eq:97}), they are equally defined by
\begin{align}
  f^{+(s)}_X(\bm{x})
  &= \mathrm{Re}\left[ e^{-is\phi}f^{(+s)}(\bm{x}) \right],
  \label{eq:209}\\
  f^{\times(s)}_X(\bm{x})
  &= \mathrm{Im}\left[ e^{-is\phi}f^{(+s)}(\bm{x}) \right].
  \label{eq:210}
\end{align}
The decomposed fields $f^{+/\times(s)}_X$ are invariant under the
two-dimensional rotation along the line of sight, Eq.~(\ref{eq:100}),
because the azimuthal angle transforms as
$\phi \rightarrow \phi - \gamma$ for the same rotation. 

In order to define the correlation functions of the decomposed field, we
should note that the averages of products such as
$\langle f^{+(s_1)}_{X_1}(\bm{x})f^{+(s_2)}_{X_1}(\bm{x}')\rangle$ are
not translationally invariant, i.e., they are not only a function of
$\bm{x}-\bm{x}'$, because the phase factors $e^{\pm is\phi}$
explicitly depend on the location of the origin in a given coordinates
system. In this paper, we define the correlation functions by choosing
the second point $\bm{x}'$ as an origin of the coordinates in defining
tangential/cross ($+/\times$) correlation functions. However, if we
choose the origin at the same location as the second point
$\bm{x}'$, the azimuthal angle of the second point is uncertain.
Therefore we infinitesimally displace the origin in the opposite
direction to the first point $\bm{x}$, so that the second point is
given by $\bm{x}'=\epsilon\bm{x}$ with $\epsilon\rightarrow +0$ and
accordingly the two points $\bm{x}$ and $\bm{x}'$ have a same
azimuthal angle. That is, we define the corresponding correlation
functions by
\begin{align}
  \xi^{++(s_1s_2)}_{X_1X_2}(\bm{x})
  &\equiv
    \lim_{\epsilon\rightarrow +0}
    \left\langle
    f^{+(s_1)}_{X_1}\left(\bm{x}\right)
    f^{+(s_2)}_{X_2}\left(\epsilon\bm{x}\right)
    \right\rangle,
  \label{eq:211}\\
  \xi^{\times\times(s_1s_2)}_{X_1X_2}(\bm{x})
  &\equiv
    \lim_{\epsilon\rightarrow +0}
    \left\langle
    f^{\times(s_1)}_{X_1}\left(\bm{x}\right)
    f^{\times(s_2)}_{X_2}\left(\epsilon\bm{x}\right)
    \right\rangle,
  \label{eq:212}\\
  \xi^{+\times(s_1s_2)}_{X_1X_2}(\bm{x})
  &\equiv
    \lim_{\epsilon\rightarrow +0}
    \left\langle
    f^{+(s_1)}_{X_1}\left(\bm{x}\right)
    f^{\times(s_2)}_{X_2}\left(\epsilon\bm{x}\right)
    \right\rangle,
  \label{eq:213}\\
  \xi^{\times+(s_1s_2)}_{X_1X_2}(\bm{x})
  &\equiv
    \lim_{\epsilon\rightarrow +0}
    \left\langle
    f^{\times(s_1)}_{X_1}\left(\bm{x}\right)
    f^{+(s_2)}_{X_2}\left(\epsilon\bm{x}\right)
    \right\rangle.
  \label{eq:214}
\end{align}
In actual data analysis, the ensemble average is replaced by the
spatial average, and in this case the azimuthal angle in the phase
factors $e^{\pm is\phi}$ vary pair by pair. With the above
definitions, the azimuthal angles of the first point $\bm{x}$ and of
second point $\epsilon\bm{x}$ are just the same. Alternatively, one
can choose the origin at the center of two points,
$(\bm{x}+\bm{x}')/2$, in which case the $+/\times$ correlation
functions are different from the above definitions by a factor of
$(-1)^{s_2}$, because the azimuthal angle of the second point differs
from the first point by an angle of 180 degrees. We use the first
definitions in the following of this paper.

Substituting Eqs.~(\ref{eq:207}) and (\ref{eq:208}) into
Eqs.~(\ref{eq:211})--(\ref{eq:213}), the relations between the
correlation functions of $+/\times$ modes and those of
Eq.~(\ref{eq:198}) are given by
\begin{align}
  \xi^{++}
  &= \frac{1}{4}
    \left[
    \left(X^{++} + X^{--}\right) + 
    \left(X^{+-} + X^{-+}\right)
    \right],
  \label{eq:215}\\
  \xi^{\times\times}
  &= -\frac{1}{4}
    \left[
    \left(X^{++} + X^{--}\right) - 
    \left(X^{+-} + X^{-+}\right)
    \right],
  \label{eq:216}\\
  \xi^{+\times}
  &= -\frac{i}{4}
    \left[
    \left(X^{++} - X^{--}\right) - 
    \left(X^{+-} - X^{-+}\right)
    \right],
  \label{eq:217}\\
  \xi^{\times+}
  &= -\frac{i}{4}
    \left[
    \left(X^{++} - X^{--}\right) + 
    \left(X^{+-} - X^{-+}\right)
    \right],
  \label{eq:218}
\end{align}
where the abbreviations
\begin{align}
  &
    X^{\pm \pm} =
    X^{\pm\pm(s_1s_2)}_{X_1X_2}(\bm{x}) \equiv
    (\pm 1)^{s_1+s_2} e^{\mp i(s_1+s_2)\phi} 
    \xi^{(\pm s_1,\pm s_2)}_{X_1X_2}(\bm{x}),
  \label{eq:219}\\
  &
    X^{\pm \mp} =
    X^{\pm\mp(s_1s_2)}_{X_1X_2}(\bm{x}) \equiv
    (\pm 1)^{s_1}(\mp 1)^{s_2} e^{\mp i(s_1-s_2)\phi} 
    \xi^{(\pm s_1,\mp s_2)}_{X_1X_2}(\bm{x}),
  \label{eq:220}\\
  &
    \xi^{++} = \xi^{++(s_1s_2)}_{X_1X_2}(\bm{x}), \quad
    \xi^{+\times} = \xi^{+\times(s_1s_2)}_{X_1X_2}(\bm{x}),
  \label{eq:221}\\
  &
    \xi^{\times+} = \xi^{\times+(s_1s_2)}_{X_1X_2}(\bm{x}), \quad
    \xi^{\times\times} = \xi^{\times\times(s_1s_2)}_{X_1X_2}(\bm{x})
  \label{eq:222}
\end{align}
are adopted just for simplicity of representation.

The symmetric properties of the correlation function,
Eqs.~(\ref{eq:200})--(\ref{eq:205}) are straightforwardly
translated to those of the functions $X^{\pm\pm}$ and $X^{\pm\mp}$.
The complex conjugates are given by
\begin{align}
  X^{\pm\pm(s_1s_2)*}_{X_1X_2}(\bm{x})
  &= X^{\mp\mp(s_1s_2)}_{X_1X_2}(\bm{x}),
  \label{eq:223}\\
  X^{\pm\mp(s_1s_2)*}_{X_1X_2}(\bm{x})
  &= X^{\mp\pm(s_1s_2)}_{X_1X_2}(\bm{x}).
  \label{eq:224}
\end{align}
Unlike the functions $Q^{\pm\pm}$ and $Q^{\pm\mp}$, they are not real
functions, while the complex conjugates change the signs of modes.
Because of this property, all of the $+/\times$ correlation functions
in Eqs.~(\ref{eq:215})--(\ref{eq:218}) are all real functions as
promised.

Under the rotation along the line of sight of Eq.~(\ref{eq:201}), the
azimuthal angle of $\bm{x}$ transforms as
$\phi \rightarrow \phi -\gamma$, and the above functions are shown to
transform as
\begin{align}
  X^{\pm\pm(s_1s_2)}_{X_1X_2}(\bm{x})
  &\xrightarrow{\mathbb{R}_\gamma}
    X^{\pm\pm(s_1s_2)}_{X_1X_2}(\bm{x}),
  \label{eq:225}\\
  X^{\pm\mp(s_1s_2)}_{X_1X_2}(\bm{x})
  &\xrightarrow{\mathbb{R}_\gamma}
   X^{\pm\mp(s_1s_2)}_{X_1X_2}(\bm{x}),
  \label{eq:226}
\end{align}
and therefore these functions do not depend on the azimuthal angle
$\phi$ of the separation vector $\bm{x}$, and depend only on the
magnitude $x$ and the polar angle $\theta$ or the direction cosine
$\mu_x = \cos\theta$. Therefore, the $+/\times$ correlation functions
of Eqs.~(\ref{eq:215})--(\ref{eq:218}) do not depend on the
azimuthal angle, and also are functions of only $x$ and $\mu_x$.

As for the combined transformation of parity and flipping rotation,
Eq.~(\ref{eq:203}), the corresponding transformations are given by
\begin{align}
  X^{\pm\pm(s_1s_2)}_{X_1X_2}(\bm{x})
  &\xrightarrow{\mathbb{PF}}
  (-1)^{p_{X_1}+p_{X_2}} X^{\mp\mp(s_1s_2)}_{X_1X_2}(\bm{x}),
  \label{eq:227}\\
  X^{\pm\mp(s_1s_2)}_{X_1X_2}(\bm{x})
  &\xrightarrow{\mathbb{PF}}
  (-1)^{p_{X_1}+p_{X_2}} X^{\mp\pm(s_1s_2)}_{X_1X_2}(\bm{x}).
  \label{eq:228}
\end{align}
Using the above transformation in Eqs.~(\ref{eq:215})--(\ref{eq:218}),
the parity transformations of the $+/\times$ correlation functions are
found to be given by
\begin{align}
  \xi^{++(s_1s_2)}_{X_1X_2}(\bm{x})
  &\xrightarrow{\mathbb{PF}}
  (-1)^{p_{X_1}+p_{X_2}} \xi^{++(s_1s_2)}_{X_1X_2}(\bm{x}),
  \label{eq:229}\\
  \xi^{\times\times(s_1s_2)}_{X_1X_2}(\bm{x})
  &\xrightarrow{\mathbb{PF}}
  (-1)^{p_{X_1}+p_{X_2}} \xi^{\times\times(s_1s_2)}_{X_1X_2}(\bm{x}),
  \label{eq:230}\\
  \xi^{+\times(s_1s_2)}_{X_1X_2}(\bm{x})
  &\xrightarrow{\mathbb{PF}}
  (-1)^{p_{X_1}+p_{X_2}+1} \xi^{+\times(s_1s_2)}_{X_1X_2}(\bm{x}),
  \label{eq:231}\\
  \xi^{\times+(s_1s_2)}_{X_1X_2}(\bm{x})
  &\xrightarrow{\mathbb{PF}}
  (-1)^{p_{X_1}+p_{X_2}+1} \xi^{\times+(s_1s_2)}_{X_1X_2}(\bm{x}),
  \label{eq:232}
\end{align}
In the Universe with parity symmetry, the $+\times$ and $\times+$
correlation functions vanish if $p_{X_1}+p_{X_2}=\mathrm{even}$, while
$++$ and $\times\times$ correlation functions vanish if
$p_{X_1}+p_{X_2}=\mathrm{odd}$.

\subsection{%
  Relations between the $\bm{+/\times}$ correlation functions and the
  E/B power spectra }

The $+/\times$ decomposition of the correlation function is related to
the E/B decomposition of the power spectrum. Before deriving the
explicit expressions for the $+/\times$ decomposition in terms of the
invariant spectrum, we first generally derive the relation between the
correlation functions and the power spectra.

The correlation functions are related to the power spectrum simply by
a Fourier transform of Eq.~(\ref{eq:199}). Applying the plane-wave
expansion formula
\begin{equation}
  e^{i\bm{k}\cdot\bm{x}}
  = 4\pi \sum_{l,m} i^l j_l(kx)
  Y_{lm}^*(\hat{\bm{k}}) Y_{lm}(\hat{\bm{x}}),
  \label{eq:233}
\end{equation}
we have
\begin{multline}
  \xi^{(\sigma_1\sigma_2)}_{X_1X_2}(\bm{x})
  = \sum_{l,m} i^l Y_{lm}(\hat{\bm{x}})
  \int d^2\hat{k}\, Y_{lm}^*(\hat{\bm{k}})
  \\ \times
  \int \frac{k^2dk}{2\pi^2} j_l(kx)
  P^{(\sigma_1\sigma_2)}_{X_1X_2}(\bm{k}).
  \label{eq:234}
\end{multline}
As obviously seen in Eqs.~(\ref{eq:215})--(\ref{eq:218}), the
$+/\times$ correlation functions are more conveniently given by
combinations $X^{++} \pm X^{--}$ and $X^{+-} \pm X^{-+}$. Combining
Eqs.~(\ref{eq:151}), (\ref{eq:152}), (\ref{eq:219}),
(\ref{eq:220}) and (\ref{eq:234}), and noting that $Q^{\pm\pm}$
and $Q^{\pm\mp}$ depend only on $k$ and $\mu$, we derive the
relations,
\begin{align}
  X^{\pm\pm(s_1s_2)}_{X_1X_2}(x,\mu_x)
  &= (-)^{s_2}
    \hat{\mathcal{F}}_{s^+_{12}}
    \left[
    Q^{\pm\pm(s_1s_2)}_{X_1X_2}(k,\mu)
    \right](x,\mu_x),  
  \label{eq:235}\\
  X^{\pm\mp(s_1s_2)}_{X_1X_2}(x,\mu_x)
  &=
    \hat{\mathcal{F}}_{s^-_{12}}
    \left[
    Q^{\pm\mp(s_1s_2)}_{X_1X_2}(k,\mu)
    \right](x,\mu_x),  
  \label{eq:236}
\end{align}
where we conveniently define a linear operator of summation-integral
type with an arbitrary integer $\sigma$,
\begin{multline}
  \hat{\mathcal{F}}_\sigma
  \left[
    Q(k,\mu)
  \right](x,\mu_x)
  \equiv
  (-i)^\sigma
  \sum_l i^l\, \Theta_l^\sigma(\mu_x)
  \int d\mu\,\Theta_l^\sigma(\mu)
  \\ \times
  \int \frac{k^2dk}{2\pi^2} j_l(kx) Q(k,\mu),
  \label{eq:237}
\end{multline}
which transforms a function
$Q(k,\mu)$ to another function $X(x,\mu_x)$. The last operator is
equivalently represented by
\begin{multline}
  \hat{\mathcal{F}}_\sigma
  \left[
    Q(k,\mu)
  \right](x,\mu_x)
  = (-i)^\sigma e^{-i\sigma\phi}
  \int \frac{d^3k}{(2\pi)^3} e^{i\bm{k}\cdot\bm{x}}
  e^{i\sigma\phi_k} Q(k,\mu),
  \label{eq:238}
\end{multline}
where $\phi$ and $\phi_k$ are azimuthal angles of $\bm{x}$ and
$\bm{k}$, respectively, although the LHS is independent of these
angles.

The above operator can be inverted by the orthogonality relation of
Eq.~(\ref{eq:183}), the completeness relation of spherical
Bessel functions,
\begin{equation}
  \int_0^\infty x^2dx\,j_l(kx)\,j_l(k'x) =
  \frac{\pi}{2k^2} \delta_\mathrm{D}(k-k'),
  \label{eq:239}
\end{equation}
and those of associated Legendre polynomials,
\begin{equation}
  \sum_l \Theta_l^m(x) \Theta_l^m(x')
  = \delta_\mathrm{D}(x-x'),
  \label{eq:240}
\end{equation}
for arbitrary integers $m$ with $|m|\leq l$. Explicitly, the
inverse operator is given by
\begin{multline}
  \hat{\mathcal{F}}_\sigma^{-1}
  \left[ X(k,\mu_x) \right](k,\mu)
  = 4\pi\,i^\sigma
  \sum_l (-i)^l\,\Theta_l^\sigma(\mu)
  \int d\mu_x\,\Theta_l^\sigma(\mu_x)
  \\ \times
  \int x^2dx\, j_l(kx) X(x,\mu_x).
  \label{eq:241}
\end{multline}
Corresponding to the representation of Eq.~(\ref{eq:238}) of the
same operator, the alternative representation of the same inverse
operator is given by
\begin{multline}
  \hat{\mathcal{F}}^{-1}_\sigma
  \left[
    X(x,\mu_x)
  \right](k,\mu)
  = i^\sigma e^{-i\sigma\phi_k}
  \int d^3x\, e^{-i\bm{k}\cdot\bm{x}}
  e^{i\sigma\phi} X(x,\mu_x).
  \label{eq:242}
\end{multline}

The inverse relations of Eqs.~(\ref{eq:235}) and (\ref{eq:236})
to represent $Q^{\pm\pm}$ and $Q^{\pm\mp}$ by $X^{\pm\pm}$ and
$X^{\pm\mp}$, respectively, are trivially given using the above
inverse operator:
\begin{align}
  Q^{\pm\pm(s_1s_2)}_{X_1X_2}(k,\mu)
  &= (-)^{s_2}
    \hat{\mathcal{F}}^{-1}_{s^+_{12}}
    \left[X^{\pm\pm(s_1s_2)}_{X_1X_2}(x,\mu_x)\right](k,\mu),  
  \label{eq:243}\\
  Q^{\pm\mp(s_1s_2)}_{X_1X_2}(k,\mu)
  &=
    \hat{\mathcal{F}}^{-1}_{s^-_{12}}
    \left[X^{\pm\mp(s_1s_2)}_{X_1X_2}(x,\mu_x)\right](k,\mu).  
  \label{eq:244}
\end{align}

The relations between the $+/\times$ correlation functions and the E/B
power spectra follow from Eqs.~(\ref{eq:147})--(\ref{eq:150}) and
(\ref{eq:215})--(\ref{eq:218}) using the above transformations.
Similarly to the combinations of power spectra $P^\pm$ and $Q^\pm$ in
Eq.~(\ref{eq:187}), it is useful to define, in abbreviated
notations,
\begin{equation}
  \xi^\pm \equiv \xi^{++} \pm \xi^{\times\times}, \qquad
  \zeta^\pm \equiv
  -i\left(\xi^{+\times} \pm \xi^{\times+}\right).
  \label{eq:245}
\end{equation}
The combinations of correlation functions $\xi^+$ and $\zeta^-$ are
related to the power spectra $P^+$ and $Q^-$ by exactly the same
transformations with Eqs.~(\ref{eq:236}) and (\ref{eq:244}),
while $\xi^-$ and $\zeta^+$ are related to $P^-$ and $Q^+$ by exactly
the same transformations with Eqs.~(\ref{eq:235}) and
(\ref{eq:243}). Thus we have, in abbreviated notations,
\begin{equation}
  \xi^{\pm}
  = (\pm 1)^{s_2}
  \hat{\mathcal{F}}_{s^{\mp}_{12}}
  \left[ P^\pm \right], \quad
  \zeta^{\pm}
  = (\mp 1)^{s_2}
  \hat{\mathcal{F}}_{s^{\pm}_{12}}
  \left[ Q^\pm \right].
  \label{eq:246}
\end{equation}
More explicitly, the relations between $+/\times$ correlation
functions and the E/B power spectra are given by
\begin{align}
  \xi^{++/\times\times} = \frac{1}{2}
  \left\{
  \hat{\mathcal{F}}_{s^-_{12}}
  \left[P^\mathrm{EE} + P^\mathrm{BB}\right]
  \pm
  (-1)^{s_2}
  \hat{\mathcal{F}}_{s^+_{12}}
  \left[P^\mathrm{EE} - P^\mathrm{BB}\right]
  \right\},
  \label{eq:247}\\
  \xi^{+\times/+\times} = \frac{1}{2}
  \left\{
  (-1)^{s_2}
  \hat{\mathcal{F}}_{s^+_{12}}
  \left[P^\mathrm{EB} + P^\mathrm{BE}\right]
  \pm
  \hat{\mathcal{F}}_{s^-_{12}}
  \left[P^\mathrm{EB} - P^\mathrm{BE}\right]
  \right\}.
  \label{eq:248}
\end{align}
The inverse relations to the above are symmetrically given by
\begin{align}
  P^\mathrm{EE/BB} = \frac{1}{2}
  \left\{
  \hat{\mathcal{F}}_{s^-_{12}}^{-1}
  \left[\xi^{++} + \xi^{\times\times}\right]
  \pm
  (-1)^{s_2}
  \hat{\mathcal{F}}_{s^+_{12}}^{-1}
  \left[\xi^{++} - \xi^{\times\times}\right]
  \right\},
  \label{eq:249}\\
  P^\mathrm{EB/BE} = \frac{1}{2}
  \left\{
  (-1)^{s_2}
  \hat{\mathcal{F}}_{s^+_{12}}^{-1}
  \left[\xi^{+\times} + \xi^{\times+}\right]
  \pm
  \hat{\mathcal{F}}_{s^-_{12}}^{-1}
  \left[\xi^{+\times} - \xi^{\times+}\right]
  \right\}.
  \label{eq:250}
\end{align}
The above equations are the explicit relations we have sought between
E/B power spectra and $+/\times$ correlation functions.

\subsection{%
  Relations to the invariant power spectrum
}

Substituting Eqs.~(\ref{eq:169}) and (\ref{eq:178}) into
Eq.~(\ref{eq:234}), the correlation functions are represented by
invariant power spectra $P^{s_1s_2;l}_{X_1X_2}(k)$ in real space
and $P^{s_1s_2;l\,l_z;L}_{X_1X_2}(k,\mu)$ in redshift space,
respectively, using the orthogonality relation of spherical harmonics.
The derivations are straightforward, and we summarize explicit
equations below.

\subsubsection{%
  Real space
}

In real space, the explicit result of representing the correlation
function in terms of the invariant spectrum is given by
\begin{multline}
  \xi^{(\sigma_1\sigma_2)}_{X_1X_2}(\bm{x})
  = (-i)^{|\sigma_1|+|\sigma_2|} \sum_l i^l\,
  Y_{l,\sigma_{12}}(\hat{\bm{x}})
  \begin{pmatrix}
    |\sigma_1| & |\sigma_2| & l \\
    \sigma_1 & \sigma_2 & -\sigma_{12}
  \end{pmatrix}
  \\ \times
  \hat{\xi}^{|\sigma_1||\sigma_2|;l}_{X_1X_2}(x),
  \label{eq:251}
\end{multline}
where $\sigma_{12}$ is given by Eq.~(\ref{eq:170}), and
\begin{equation}
  \hat{\xi}^{l_1l_2;l}_{X_1X_2}(x)
  \equiv
  \int \frac{k^2dk}{2\pi^2} j_l(kx)
  \hat{P}^{l_1l_2;l}_{X_1X_2}(k)
  \label{eq:252}
\end{equation}
is a Hankel transform of the normalized invariant spectrum defined
in Eq.~(\ref{eq:171}). We call the above function of
Eq.~(\ref{eq:252}) an invariant correlation function. The
invariant correlation function satisfies the same symmetric properties
as the invariant spectrum, Eqs.~(\ref{eq:167}) and (\ref{eq:168}).
That is, the function is real and obeys the same transformation of
parity as the invariant spectrum.

The $+/\times$ correlation functions are also represented by an
invariant spectrum. They are given by functions $X^{\pm\pm}$ and
$X^{\pm\mp}$ of Eqs.~(\ref{eq:219}) and (\ref{eq:220}). Substituting
the above Eq.~(\ref{eq:251}), explicitly we derive
\begin{align}
  X^{\pm\pm(s_1s_2)}_{X_1X_2}(\bm{x})
  &= \frac{1}{\sqrt{2\pi\,(2s^+_{12}+1)}}
  \Theta_{s^+_{12}}^{s^+_{12}}(\mu_x)
  \hat{\xi}^{s_1s_2;s^+_{12}}_{X_1X_2}(x),
  \label{eq:253}\\
  X^{\pm\mp(s_1s_2)}_{X_1X_2}(\bm{x})
  &=
  \frac{(\mp i)^{s^-_{12}}}{\sqrt{2\pi}}
  \sum_l (\pm i)^l\,
  \Theta_l^{s^-_{12}}(\mu_x)
    \nonumber\\
  & \hspace{4pc} \times
  \begin{pmatrix}
    s_1 & s_2 & l \\
    s_1 & -s_2 & -s^-_{12}
  \end{pmatrix}
  \hat{\xi}^{s_1s_2;l}_{X_1X_2}(x),
  \label{eq:254}
\end{align}
where $s^\pm_{12}$ is defined by Eq.~(\ref{eq:186}). The above
results are equally derived from Eqs.~(\ref{eq:235}) and
(\ref{eq:236}) with the operators of Eq.~(\ref{eq:237}). The
$+/\times$ correlation functions are straightforwardly given by
substituting the above equations into
Eqs.~(\ref{eq:215})--(\ref{eq:218}). In terms of the
combinations in Eq.~(\ref{eq:245}), we explicitly have
\begin{align}
  \begin{Bmatrix}
    \xi^+ \\ \zeta^-
  \end{Bmatrix}
  &= \frac{(-i)^{s^-_{12}}}{\sqrt{2\pi}} \sum_l \frac{1
    \pm (-1)^{s^-_{12}+ l}}{2}\, i^l \Theta_l^{s^-_{12}}(\mu_x)
    \nonumber\\
  & \hspace{6pc} \times
    \begin{pmatrix}
      s_1 & s_2 & l \\
      s_1 & -s_2 & -s^-_{12}
    \end{pmatrix}
  \hat{\xi}^{s_1s_2;l}_{X_1X_2}(x),
  \label{eq:255}\\
  \xi^-
  &= \frac{1}{\sqrt{2\pi\,(2s^+_{12}+1)}}
    \Theta_{s^+_{12}}^{s^+_{12}}(\mu_x)
    \hat{\xi}^{s_1s_2;s^+_{12}}_{X_1X_2}(x),
    \label{eq:256}\\
  \zeta^+
  &= 0.
  \label{eq:257}
\end{align}
The same results are equally derived from substituting
Eqs.~(\ref{eq:188})--(\ref{eq:190}) into Eq.~(\ref{eq:246}).

\subsubsection{Redshift space}

In redshift space, we first consider the simple case that the
dependence of the direction to the line of sight is completely
expanded and invariant spectrum
$\hat{P}^{l_1l_2;l\,l_z;L}_{X_1X_2}(k)$ does not have the argument of
direction cosine $\mu=\mathbf{e}_3\cdot\hat{\bm{k}}$. In this case, we
have
\begin{multline}
  \xi^{(\sigma_1\sigma_2)}_{X_1X_2}(\bm{x})
  = i^{|\sigma_1|+|\sigma_2|}
  \sum_{l}
  i^l Y_{l\sigma_{12}}(\hat{\bm{x}}) 
  \sum_{L,l_z} (-1)^L \sqrt{2L+1}
  \\ \times
  \begin{pmatrix}
    |\sigma_1| & |\sigma_2| & L \\
    \sigma_1 & \sigma_2 & -\sigma_{12}
  \end{pmatrix}
  \begin{pmatrix}
    L & l_z & l \\
    \sigma_{12} & 0 & -\sigma_{12}
  \end{pmatrix}
  \hat{\xi}^{|\sigma_1||\sigma_2|;l\,l_z;L}_{X_1X_2}(x),
  \label{eq:258}
\end{multline}
where 
\begin{equation}
  \hat{\xi}^{l_1l_2;l\,l_z;L}_{X_1X_2}(x)
  \equiv
  \int \frac{k^2dk}{2\pi^2} j_l(kx)
  \hat{P}^{l_1l_2;l\,l_z;L}_{X_1X_2}(k)
  \label{eq:259}
\end{equation}
is the Hankel transform of the normalized invariant spectrum defined
in Eq.~(\ref{eq:179}) when the argument of $\mu$ is absent. The
invariant correlation function in redshift space satisfies the same
symmetric properties as the invariant spectrum,
Eqs.~(\ref{eq:175})--(\ref{eq:177}), neglecting the argument
$\mu$ in the present case. Specifically, the index $l_z$ should be
even numbers due to Eqs.~(\ref{eq:175}).

In the general case that the invariant spectrum
$\hat{P}^{l_1l_2;l\,l_z;L}_{X_1X_2}(k,\mu)$ depends also on the
direction cosine $\mu$, the invariant correlation function is not
simply related to the invariant power spectra through the Hankel
transform of Eq.~(\ref{eq:259}). Instead, the general relation is
derived in Paper~I, and the result is given by
\begin{multline}
  \hat{\xi}^{l_1l_2;l\,l_z;L}_{X_1X_2}(x)
  = (-1)^L (2l+1)(2l_z+1)
  \sum_{l',l_z',L'} (-1)^{L'} (2L'+1)
  \\ \times
  \begin{pmatrix}
    l_z & l_z' & L' \\ 0 & 0 & 0
  \end{pmatrix}
  \begin{pmatrix}
    l & l' & L' \\ 0 & 0 & 0
  \end{pmatrix}
  \begin{Bmatrix}
    l_z & l_z' & L' \\
    l' & l & L
  \end{Bmatrix}
  \\ \times
  \int \frac{k^2dk}{2\pi^2} j_l(kx)
  \int \frac{d\mu}{2} P_{L'}(\mu)
  \hat{P}^{l_1l_2;l'l_z';L}_{X_1X_2}(k,\mu),
  \label{eq:260}
\end{multline}
where $P_{L'}(\mu)$ is the Legendre polynomial of order $L'$, and the
factor in front of the integrals is the $6j$-symbol.

The $+/\times$ correlation functions in redshift space are also
straightforward to calculate. The correlation function is given by
Eq.~(\ref{eq:258}) with Eqs.~(\ref{eq:259}) or (\ref{eq:260}). The
functions of Eqs.~(\ref{eq:219}) and (\ref{eq:220}) in this case
reduce to
\begin{align}
  X^{\pm\pm}
  &= \frac{(\mp i)^{s^+_{12}}}{\sqrt{2\pi}}
  \sum_{l} (\pm i)^l\,
  \Theta_l^{s^+_{12}}(\mu_x)
  \sum_{l_z} (\pm 1)^{l_z}
    \nonumber\\
  & \hspace{5pc} \times
  \begin{pmatrix}
    s^+_{12} & l_z & l \\
    s^+_{12} & 0 & -s^+_{12}
  \end{pmatrix}
  \hat{\xi}^{s_1s_2;l\,l_z;s^+_{12}}_{X_1X_2}(x),
  \label{eq:261}\\
  X^{\pm\mp}
  &=
  \frac{(\pm i)^{s^-_{12}}}{\sqrt{2\pi}}
  \sum_l (\pm i)^l
  \Theta_l^{s^-_{12}}(\mu_x)
  \sum_{l_z,L} (\pm 1)^{l_z}(-1)^L \sqrt{2L+1}
    \nonumber\\
  & \quad \times
  \begin{pmatrix}
    s_1 & s_2 & L \\
    s_1 & -s_2 & -s^-_{12}
  \end{pmatrix}
  \begin{pmatrix}
    L & l_z & l \\
    s^-_{12} & 0 & -s^-_{12}
  \end{pmatrix}
  \hat{\xi}^{s_1s_2;l\,l_z;L}_{X_1X_2}(x).
  \label{eq:262}
\end{align}
The above results are also derived by substituting Eqs.~(\ref{eq:194})
and (\ref{eq:195}) into Eqs.~(\ref{eq:235}) and (\ref{eq:236}),
respectively. As a consistency check, putting $l_z=0$ reproduces the
results of Eqs.~(\ref{eq:261}) and (\ref{eq:262}) in real space with a
correspondence
$\hat{\xi}^{(s_1s_2);l\,0;l}_{X_1X_2}(x) \leftrightarrow
\hat{\xi}^{(s_1s_2);l}_{X_1X_2}(x)$.

The $+/\times$ correlation functions in redshift space are given by
substituting the above equations into
Eqs.~(\ref{eq:215})--(\ref{eq:218}). In terms of the
combinations in Eq.~(\ref{eq:245}), we explicitly have
\begin{align}
  \begin{Bmatrix}
    \xi^+ \\ \zeta^-
  \end{Bmatrix}
  &=
  \frac{i^{s^-_{12}}}{\sqrt{2\pi}}
    \sum_{l,l_z}
    \frac{1 \pm (-1)^{s^-_{12}+l+l_z}}{2}
    i^l \Theta_l^{s^-_{12}}(\mu_x)
  \sum_L (-1)^L \sqrt{2L+1}
    \nonumber\\
  & \quad \times
  \begin{pmatrix}
    s_1 & s_2 & L \\
    s_1 & -s_2 & -s^-_{12}
  \end{pmatrix}
  \begin{pmatrix}
    L & l_z & l \\
    s^-_{12} & 0 & -s^-_{12}
  \end{pmatrix}
  \hat{\xi}^{s_1s_2;l\,l_z;L}_{X_1X_2}(x),
  \label{eq:263}\\
  \begin{Bmatrix}
    \xi^- \\ \zeta^+
  \end{Bmatrix}
  &= \pm
 \frac{(-i)^{s^+_{12}}}{\sqrt{2\pi}}
    \sum_{l,l_z} \frac{1 \pm (-1)^{s^+_{12}+l+l_z}}{2}
    i^l\, \Theta_l^{s^+_{12}}(\mu_x)
    \nonumber\\
  & \hspace{6pc} \times
  \begin{pmatrix}
    s^+_{12} & l_z & l \\
    s^+_{12} & 0 & -s^+_{12}
  \end{pmatrix}
  \hat{\xi}^{s_1s_2;l\,l_z;s^+_{12}}_{X_1X_2}(x).
  \label{eq:264}
\end{align}

\subsection{
  The angular correlation function
}

\subsubsection{General considerations}

In imaging surveys, redshifts of galaxies are not measured, and only
angular positions are observed. The angular correlation function of
galaxies in imaging surveys is given by projecting the three-dimensional
spatial correlation function on the sky. Within the distant-observer
approximation, the corresponding projection is given by Limber's
equation \cite{Peebles1980},
\begin{equation}
  w^{(\sigma_1\sigma_2)}_{X_1X_2}(\bm{\theta}) =
  N_0 \int dx_\parallel\,
  \xi^{(\sigma_1\sigma_2)}_{X_1X_2}\left(r_0\bm{\theta},x_\parallel\right),
  \label{eq:265}
\end{equation}
where the integrand is the correlation function in real space, $r_0$
is the fiducial distance to the sample [see Eq.~(\ref{eq:133})],
$x_\parallel = x_3$ is the coordinate along the line of sight,
$\bm{\theta}=(x_1/r_0,x_2/r_0)$ is a two-dimensional position vector
perpendicular to the line of sight, and $N_0$ is a normalization,
which is determined by a radial window function in a given region of
the survey, whose variation is neglected here just for simplicity. The
normalization $N_0$ has an inverse dimension of length.

Substituting Eq.~(\ref{eq:234}) into Eq.~(\ref{eq:265}), we derive
an expression of the angular correlation function in terms of the
power spectrum. The following formula is useful for the calculation:
\begin{equation}
  \int dx_\parallel j_l(kx) Y_{lm}(\hat{\bm{x}})
%  =  \frac{\pi}{k} (-i)^{l-m} \mathcal{P}_l^m(0) e^{im\phi}
  =  \sqrt{\frac{\pi}{2}}\,(-i)^{l-m} \Theta_l^m(0) \frac{1}{k} e^{im\phi}
  J_m(k r_0 \theta),
  \label{eq:266}
\end{equation}
where the wave vector and position vector are parametrized by
\begin{align}
  \bm{k}
  &= \left( k_\perp\cos\phi_k,k_\perp\sin\phi_k,k_\parallel\right),
  \label{eq:267}\\
  \bm{x}
  &= \left(r_0\theta\cos\phi,r_0\theta\sin\phi,x_\parallel\right),
  \label{eq:268}\\
  \bm{\theta}
  &= \left( \theta\cos\phi,\theta\sin\phi \right),
  \label{eq:269}
\end{align}
and $k_\perp = (k^2-{k_\parallel}^2)^{1/2}=k\sin\theta_k$ is the
component of the wave vector perpendicular to the line of sight. The
above formula of Eq.~(\ref{eq:266}) can be shown by applying the
plane-wave expansion formula of Eq.~(\ref{eq:233}) to an integral
\begin{equation}
  \int dx_\parallel\,e^{i\bm{k}\cdot\bm{x}}
  = 2\pi \delta_\mathrm{D}(k_\parallel)
  e^{ikr_0\theta \cos(\phi_k-\phi)},
  \label{eq:270}
\end{equation}
and using the orthogonality relation of the spherical harmonics and
the integral representation of the Bessel function,
\begin{equation}
  J_m(x) = \frac{1}{i^m} \int \frac{d\phi}{2\pi}
  e^{ix\cos\phi + im\phi}.
  \label{eq:271}
\end{equation}
The zero points of the normalized associated Legendre polynomials,
$\Theta_l^m(0)$, are nonzero only for $l\geq |m|$ and
$l-m=\mathrm{even}$, and explicitly given by
\begin{multline}
  \Theta_l^m(0)
  \\
  =
  \begin{cases}
    (-1)^{(l+m)/2}
    \sqrt{
      \frac{2l+1}{2}
      \frac{(l+m-1)!!}{(l+m)!!}
      \frac{(l-m-1)!!}{(l-m)!!}},
    & (l-m=\mathrm{even}), \\
    0,
    & (l-m=\mathrm{odd}).
  \end{cases}
  \label{eq:272}
\end{multline}

Applying the above formula of Eq.~(\ref{eq:266}) to
Eqs.~(\ref{eq:265}) and (\ref{eq:234}), we derive
\begin{multline}
  w^{(\sigma_1\sigma_2)}_{X_1X_2}(\bm{\theta})
  = \frac{N_0}{\sqrt{2\pi}}
  \sum_{l,m} i^m e^{im\phi} \Theta_l^m(0)
  \int d^2\hat{k}\, Y_{lm}^*(\hat{\bm{k}})
  \\ \times
  \int \frac{k\,dk}{2\pi} J_m(kr_0\theta)
  P^{(\sigma_1\sigma_2)}_{X_1X_2}(\bm{k}).
  \label{eq:273}
\end{multline}
The symmetric properties of the spatial correlation function,
Eqs.~(\ref{eq:200})--(\ref{eq:205}) holds for the angular
correlation function in exactly the forms with replacements
$\xi^{(\sigma_1\sigma_2)}_{X_1X_2}(\bm{x}) \rightarrow
w^{(\sigma_1\sigma_2)}_{X_1X_2}(\bm{\theta})$, where the azimuthal angle $\phi$
is common to the three-dimensional position $\bm{x}$ and
two-dimensional position $\bm{\theta}$. In particular, the rotation
along the line of sight,
\begin{equation}
  w^{(\sigma_1\sigma_2)}_{X_1X_2}(\bm{\theta})
  \xrightarrow{\mathbb{R}_\gamma}
  w^{\prime (\sigma_1\sigma_2)}_{X_1X_2}(\bm{\theta}') =
  e^{-i(\sigma_1+\sigma_2)\gamma}
  w^{(\sigma_1\sigma_2)}_{X_1X_2}(\bm{\theta}),
  \label{eq:274}
\end{equation}
indicates that only $m=\sigma_1+\sigma_2 \equiv \sigma_{12}$ should
survive in the summation of Eq.~(\ref{eq:273}), thus we have
\begin{multline}
  w^{(\sigma_1\sigma_2)}_{X_1X_2}(\bm{\theta})
  = \frac{N_0}{\sqrt{2\pi}}
  i^{\sigma_{12}} e^{i\sigma_{12}\phi}
  \sum_l \Theta_l^{\sigma_{12}}(0)
  \int d^2\hat{k}\, Y_{l,\sigma_{12}}^*(\hat{\bm{k}})
  \\ \times
  \int \frac{k\,dk}{2\pi} J_{\sigma_{12}}(kr_0\theta)
  P^{(\sigma_1\sigma_2)}_{X_1X_2}(\bm{k}).
  \label{eq:275}
\end{multline}
The summation over $l$ and the integration over polar angle $\theta_k$
of the wave vector $\bm{k}$ can be analytically evaluated by the
completeness relation of the associated Legendre polynomials,
Eq.~(\ref{eq:240}), and we find the component of wave vector parallel
to the line of sight does not contribute, as naturally expected from
Eq.~(\ref{eq:270}). However, the above expression, before
analytically reducing the summation over $l$, is still useful below.

As is clear from the analogy of
Eqs.~(\ref{eq:215})--(\ref{eq:222}) for the spatial correlation
function, the angular correlation functions of $+/\times$ modes are
given by
\begin{align}
  w^{++}
  &= \frac{1}{4}
    \left[
    \left(W^{++} + W^{--}\right) + 
    \left(W^{+-} + W^{-+}\right)
    \right],
  \label{eq:276}\\
  w^{\times\times}
  &= -\frac{1}{4}
    \left[
    \left(W^{++} + W^{--}\right) - 
    \left(W^{+-} + W^{-+}\right)
    \right],
  \label{eq:277}\\
  w^{+\times}
  &= -\frac{i}{4}
    \left[
    \left(W^{++} - W^{--}\right) - 
    \left(W^{+-} - W^{-+}\right)
    \right],
  \label{eq:278}\\
  w^{\times+}
  &= -\frac{i}{4}
    \left[
    \left(W^{++} - W^{--}\right) + 
    \left(W^{+-} - W^{-+}\right)
    \right],
  \label{eq:279}
\end{align}
where the abbreviations
\begin{align}
  &
    W^{\pm \pm} =
    W^{\pm\pm(s_1s_2)}_{X_1X_2}(\theta) \equiv
    (\pm 1)^{s_1+s_2} e^{\mp i(s_1+s_2)\phi} 
    w^{(\pm s_1,\pm s_2)}_{X_1X_2}(\bm{\theta}),
  \label{eq:280}\\
  &
    W^{\pm \mp} =
    W^{\pm\mp(s_1s_2)}_{X_1X_2}(\theta) \equiv
    (\pm 1)^{s_1}(\mp 1)^{s_2} e^{\mp i(s_1-s_2)\phi} 
    w^{(\pm s_1,\mp s_2)}_{X_1X_2}(\bm{\theta}),
  \label{eq:281}\\
  &
    w^{++} = w^{++(s_1s_2)}_{X_1X_2}(\bm{\theta}), \quad
    w^{+\times} = w^{+\times(s_1s_2)}_{X_1X_2}(\bm{\theta}),
  \label{eq:282}\\
  &
    w^{\times+} = w^{\times+(s_1s_2)}_{X_1X_2}(\bm{\theta}), \quad
    w^{\times\times} = w^{\times\times(s_1s_2)}_{X_1X_2}(\bm{\theta})
  \label{eq:283}
\end{align}
are adopted just for simplicity of representation.

The phase factors $e^{\mp i(s_1+s_2)\phi}$ and
$e^{\mp i(s_1-s_2)\phi}$ in the above expressions cancel with those in
Eq.~(\ref{eq:275}), and thus they do not depend on the azimuthal
angle $\phi$ and are functions of the absolute angular separation
$\theta=|\bm{\theta}|$. Substituting Eq.~(\ref{eq:275}) into
Eqs.~(\ref{eq:280}) and (\ref{eq:281}), and using the
completeness relation of Eq.~(\ref{eq:240}), we derive
\begin{align}
  W^{\pm\pm(s_1s_2)}_{X_1X_2}(\theta)
  &= N_0 (-1)^{s_2}
  \int \frac{k\,dk}{2\pi} J_{s^+_{12}}(kr_0\theta)
  Q^{\pm\pm(s_1s_2)}_{X_1X_2}(k,0),
  \label{eq:284}\\
  W^{\pm\mp(s_1s_2)}_{X_1X_2}(\theta)
  &= N_0
  \int \frac{k\,dk}{2\pi} J_{s^-_{12}}(kr_0\theta)
  Q^{\pm\mp(s_1s_2)}_{X_1X_2}(k,0).
  \label{eq:285}
\end{align}
where $Q^{\pm\pm(s_1s_2)}_{X_1X_2}(k,0)$ and
$Q^{\pm\mp(s_1s_2)}_{X_1X_2}(k,0)$ are the functions
$Q^{\pm\pm(s_1s_2)}_{X_1X_2}(k,\mu)$ and
$Q^{\pm\mp(s_1s_2)}_{X_1X_2}(k,\mu)$ substituted by $\mu=0$, i.e.,
the wave vector is directed perpendicular to the line of sight.
The above integrals appear as we have
\begin{equation}
  \int dx_\parallel\,
  \hat{\mathcal{F}}_\sigma\left[Q(k,\mu)\right](x,\mu_x)
  = \int \frac{k\,dk}{2\pi} J_\sigma(kr_0\theta)
  Q(k,0).
  \label{eq:286}
\end{equation}
In fact, Eqs.~(\ref{eq:284}) and (\ref{eq:285}) are also derived
by applying Eq.~(\ref{eq:286}) to Eqs.~(\ref{eq:235}) and
(\ref{eq:236}). 

\subsubsection{Relations to the invariant power spectrum}

Substituting Eq.~(\ref{eq:169}) into Eq.~(\ref{eq:275}), the
angular correlation function is straightforwardly represented in terms
of the invariant spectrum as
\begin{multline}
  w^{(\sigma_1\sigma_2)}_{X_1X_2}(\bm{\theta})
  = \frac{N_0}{\sqrt{2\pi}}
  (-i)^{|\sigma_1|+|\sigma_2|-\sigma_{12}} e^{i\sigma_{12}\phi}
  \sum_l \Theta_l^{\sigma_{12}}(0)
  \\ \times
  \begin{pmatrix}
    |\sigma_1| & |\sigma_2| & l \\
    \sigma_1 & \sigma_2 & -\sigma_{12}
  \end{pmatrix}
  \int_0^\infty \frac{k\,dk}{2\pi} J_{\sigma_{12}}(kr_0\theta)
  \hat{P}^{|\sigma_1||\sigma_2|;l}_{X_1X_2}(k),
  \label{eq:287}
\end{multline}
where the non-negative integer $l$ is summed over only for
$l-\sigma_{12}=\mathrm{even}$ and $l\geq |\sigma_{12}|$. From this equation,
Eqs.~(\ref{eq:280}) and (\ref{eq:281}) are given by
\begin{align}
  W^{\pm\pm}
  &= \frac{N_0}{\sqrt{2\pi}}
  \frac{\Theta_{s^+_{12}}^{s^+_{12}}(0)}{\sqrt{2s^+_{12}+1}}
  \int \frac{k\,dk}{2\pi} J_{s^+_{12}}(kr_0\theta)
  \hat{P}^{s_1s_2;s^+_{12}}_{X_1X_2}(k).
  \label{eq:288}\\
  W^{\pm\mp}
  &=
  \frac{N_0}{\sqrt{2\pi}}
  \sum_l
  \Theta_l^{s^-_{12}}(0)
  \begin{pmatrix}
    s_1 & s_2 & l \\
    s_1 & -s_2 & -s^-_{12}
  \end{pmatrix}
  \nonumber\\
  & \hspace{5pc} \times
  \int \frac{k\,dk}{2\pi} J_{s^-_{12}}(kr_0\theta)
  \hat{P}^{s_1s_2;l}_{X_1X_2}(k).
  \label{eq:289}
\end{align}
The $+/\times$ angular correlation functions are given by the above
equations and Eqs.~(\ref{eq:276})--(\ref{eq:279}).

In the above, one sees $W^{++}=W^{--}$ and $W^{+-}=W^{-+}$. Therefore,
the correlations $w^{+\times}$ and $w^{\times+}$ identically vanish as
seen from Eqs.~(\ref{eq:278}) and (\ref{eq:279}). We thus
have
\begin{equation}
  w^{++/\times\times} = \frac{W^{+-} \pm  W^{++}}{2}, \qquad
  w^{+\times/\times+} = 0.
  \label{eq:290}
\end{equation}
The above simple results are valid only in idealized cases that the
window function of the projection along the line of sight is
homogeneous, as noted above. In this case, the vanishing
cross-correlations of $w^{+\times/\times+} = 0$ have nothing to do
with parity conservation of the Universe, and simply due to the
projection effects of the correlation function. In the power spectra
and spatial correlation functions, the same is true for the transverse
components of wave vector and separation vector. In fact, when $\mu=0$
in Eq.~(\ref{eq:185}) of the power spectrum, the summation over $l$ is
taken only when $l+s_1+s_2=\mathrm{even}$, because of the factor
$\Theta_l^{s^-_{12}}(0)$, and therefore we have $Q^{++} = Q^{--}$ and
$Q^{+-} = Q^{+-}$, and thus
$P^\mathrm{EB}(\bm{k}) =P^\mathrm{BE}(\bm{k}) =0$ in this case when
the line-of-sight component of the wave vector vanishes, $k_3 = 0$.
The same argument also applies to the power spectrum in redshift space
and spatial correlation functions both in real space and in redshift
space. The cross-correlations of the correlation functions
$\xi^{+\times}(\bm{x})=\xi^{\times+}(\bm{x})=0$ when the line-of-sight
component of the separation vector vanishes, $x_3=0$.

One finds the integrand of Eq.~(\ref{eq:288}) is proportional to
Eq.~(\ref{eq:184}) with $\mu=0$, and also the integrand of
Eq.~(\ref{eq:289}) is proportional to Eq.~(\ref{eq:185}) with $\mu=0$.
Therefore we have identities,
\begin{multline}
  w^{++(s_1s_2)}_{X_1X_2}(\theta) \pm
  w^{\times\times(s_1s_2)}_{X_1X_2}(\theta)
  = N_0 (\pm 1)^{s_2}
  \int\frac{k\,dk}{2\pi} J_{s^\mp_{12}}(kr_0\theta)
  \\ \times
    \left[
    P^{\mathrm{EE}(s_1s_2)}_{X_1X_2}(k,0) \pm
    P^{\mathrm{BB}(s_1s_2)}_{X_1X_2}(k,0)
    \right],
  \label{eq:291}
\end{multline}
where $P^{\mathrm{EE}(s_1s_2)}_{X_1X_2}(k,0)$ and
$P^{\mathrm{BB}(s_1s_2)}_{X_1X_2}(k,0)$ are EE/BB power spectra in the
direction of the wave vector perpendicular to the line of sight,
$\mu=0$. Again, the above simple conclusion assumes that the radial
window function of the projection is homogeneous. In the case of
intrinsic alignment of galaxies with $s_1=0, s_2=2$ and $s_1=s_2=2$,
the above formula of Eq.~(\ref{eq:291}) exactly reproduces the known
result given in Refs.~\cite{Kamionkowski:1997mp,Blazek:2015lfa} with
$N_0=1$, besides the difference in sign convention\footnote{In most
  literatures of intrinsic alignment and CMB polarization, the E/B
  modes of spin-2 fields are commonly defined by $-1$ times the
  definitions in this paper. The reason for the difference is that our
  definition contains a factor $i^s$ for general spin-$s$ fields,
  i.e., Eq.~(\ref{eq:122}), and $i^s=-1$ when $s=2$. Our convention is
  more natural from the point of view of general spin fields.}.

\section{\label{sec:IPT}%
  Applications of the integrated perturbation theory}

We derived various formulas for two-point statistics of projected
tensors with arbitrary spins. So far we only use symmetric properties,
most importantly the rotational symmetry, of the statistics in
deriving the formulas, and any model of underlying dynamics has not
been assumed yet. The formulas are represented in terms of the
invariant power spectra, and they are predicted by the formalism of
the iPT of tensor fields as developed in Papers~I and II. The
predictions are made to arbitrary orders of perturbations, even though
the evaluations of higher-order corrections are complicated in
general. The lowest-order approximation of the perturbation theory is
usually called the linear perturbation theory. Nonlinear corrections
correspond to higher-order perturbation theory. In Paper~I, the
predictions of the invariant spectra in the linear perturbation theory
are given, and methods and examples of evaluating the higher-order
corrections are given in Paper~II. We apply the outcomes to the
various formulas obtained in the previous sections in this paper
below.

\subsection{%
  The linear power spectra
}

\subsubsection{General predictions for the power spectra in linear
  theory}

In the linear perturbation theory, the power spectrum of the tensor
field is proportional to the linear power spectrum of mass,
$P_\mathrm{L}(k)$, and the coefficient is analytically given without
any loop integral (Papers~I and II). The analytic results of the
invariant spectrum of the linear theory are given in Paper~I.
Substituting these results into general expressions in the previous
sections of this paper, the predictions of various two-point
statistics for projected tensors are explicitly derived. Because of
the absence of loop integrals in the linear perturbation theory, the
results are simply represented in closed forms without complicated
integrals.

The power spectrum of the three-dimensional tensor of the linear theory
with Gaussian initial conditions is given by (Paper I)
\begin{equation}
  P^{(l_1l_2)}_{X_1X_2m_1m_2}(\bm{k})
  =
  i^{l_1+l_2}
  \Pi^2(\bm{k}) \hat{\Gamma}^{(1)}_{X_1l_1m_1}(\bm{k})
  \hat{\Gamma}^{(1)}_{X_2l_2m_2}(-\bm{k})
  P_\mathrm{L}(k),
  \label{eq:292}
\end{equation}
where $\Pi(\bm{k})$ and  $\hat{\Gamma}^{(1)}_{Xlm}(\bm{k})$ are
the resummation factor and the first-order propagator, respectively,
as given below. In real space, the propagator is given by
\begin{equation}
  \hat{\Gamma}^{(1)}_{Xlm}(\bm{k}) =  
  c^{(0)}_{Xlm} + c^{(1)}_{Xlm}(\bm{k})
\label{eq:293}
\end{equation}
in real space, and 
\begin{equation}
  \hat{\Gamma}^{(1)}_{Xlm}(\bm{k}) =  
  \left(1 + f{\mu}^2\right)
  c^{(0)}_{Xlm} + c^{(1)}_{Xlm}(\bm{k})
\label{eq:294}
\end{equation}
in redshift space, where $f=d\ln D/d\ln a = \dot{D}/HD$ is the linear
growth rate, $a(t)$ is the scale factor, and $D(t)$ is the linear
growth factor. The functions $c^{(n)}_{Xlm}$ are the renormalized bias
functions of the iPT, defined by
\begin{align}
  c^{(n)}_{Xlm}(\bm{k}_1,\ldots\bm{k}_n)
  &=
    (-i)^l
  (2\pi)^{3n}
  \int \frac{d^3k}{(2\pi)^3}
  \left\langle
    \frac{\delta^n F^\mathrm{L}_{Xlm}(\bm{k})}{
      \delta\delta_\mathrm{L}(\bm{k}_1) \cdots
      \delta\delta_\mathrm{L}(\bm{k}_n)}
  \right\rangle
  \nonumber \\
  &=
    (-i)^l
  (2\pi)^{3n}
  \left\langle
    \frac{\delta^n F^\mathrm{L}_{Xlm}(\bm{q}=\bm{0})}{
      \delta\delta_\mathrm{L}(\bm{k}_1) \cdots
      \delta\delta_\mathrm{L}(\bm{k}_n)}
  \right\rangle,
  \label{eq:295}
\end{align}
where $\bm{q}$ are the Lagrangian coordinates in configuration space,
and $F^\mathrm{L}_{Xlm}(\bm{q}=\bm{0})$ in the last expression is
formally the field value at the origin $\bm{q}=\bm{0}$, although the
function does not depend on the choice of the coordinates origin after
taking the ensemble average. In the above equation,
$\delta/\delta\delta_\mathrm{L}(\bm{k})$ is the functional derivative
with respect to the linear density contrast
$\delta_\mathrm{L}(\bm{k})$, and $F^\mathrm{L}_{Xlm}(\bm{k})$ is the
Fourier transform of the three-dimensional irreducible tensor field in
distant-observer Cartesian coordinates in Lagrangian space (see
Paper~I for more details). From the definition, $c^{(0)}_{Xlm}$
corresponds to the average value of the tensor field and
$c^{(1)}_{Xlm}(\bm{k})$ is the linear response function to the initial
density fluctuations. While the linear response is a kin to the linear
bias parameter in Lagrangian space, we should note that they are not
the same. The bias parameters in perturbative treatments of the bias
correspond to coefficients of the Taylor expansion of the bias
relation. However, the renormalized bias functions do not assume the
Taylor expansion, and instead, they are based on a kind of orthogonal
expansion of nonlocal functions (see Ref.~\cite{Matsubara:1995wd} for
this point of view). As a result, the infinite number of higher-order
contributions in Taylor series is already renormalized in the
functions (see Sec.~IV~C of Paper~I). Details of the basic
construction of iPT is given in Ref.~\cite{Matsubara:2011ck}.

When the perturbations are kept exactly up to the linear order of
$P_\mathrm{L}(k)$, the resummation factor in Eq.~(\ref{eq:292}) can be
simply replaced by $\Pi(\bm{k}) = 1$. In the next order, the factor is
given by
\begin{equation}
  \Pi(k) = 
  \exp\left[
    -\frac{k^2}{12\pi^2}
    \int_0^\infty dp P_\mathrm{L}(p)
  \right]
  \label{eq:296}
\end{equation}
in real space, and
\begin{equation}
  \Pi(k,\mu) = 
  \exp\left\{
    -\frac{k^2}{12\pi^2}
    \left[ 1 + f(f+2){\mu}^2 \right]
    \int_0^\infty dp P_\mathrm{L}(p)
  \right\}
  \label{eq:297}
\end{equation}
in redshift space. When the exponential factor is expanded by linear
power spectrum $P_\mathrm{L}(k)$, the above equations obviously reduce
to $\Pi(\bm{k}) =1$ at the lowest order. While the inclusion of the
next-leading order in the resummation factor is not strictly a proper
linear perturbation theory, this factor is known to empirically
represent weakly nonlinear effects, such as the finger-of-God effect
in redshift space, smoothing effect of the BAO peak in the correlation
function, etc. We keep the exponential forms of the resummation factor
below for the sake of generality.

The renormalized bias functions of the tensor field are represented by
rotationally invariant functions as shown in Paper~I. For the above
first two functions, they are
\begin{equation}
  c^{(0)}_{Xlm}
  = \delta_{l0}\delta_{m0} c^{(0)}_X,
  \quad
  c^{(1)}_{Xlm}(\bm{k}) = \frac{(-1)^l}{\sqrt{2l+1}}
  c^{(1)}_{Xl}(k) C_{lm}(\hat{\bm{k}}),
  \label{eq:298}
\end{equation}
where $c^{(0)}_X$ and $c^{(1)}_{Xl}(k)$ are the rotationally invariant
functions. Substituting Eq.~(\ref{eq:298}) into
Eqs.~(\ref{eq:292})--(\ref{eq:294}), and using the correspondence
of Eq.~(\ref{eq:136}), we derive
\begin{multline}
  P^{(\sigma_1\sigma_2)}_{X_1X_2}(\bm{k})
  = \frac{4\pi\,(-i)^{|\sigma_1|-|\sigma_2|}A_{|\sigma_1|} A_{|\sigma_2|}}
  {\left(2|\sigma_1|+1\right)\left(2|\sigma_2|+1\right)}
  \Pi^2(k,\mu) P_\mathrm{L}(k)  
  \\ \times
  G_{X_1|\sigma_1|}(k,\mu)
  G_{X_2|\sigma_2|}(k,\mu)
  Y_{|\sigma_1|\sigma_1}(\hat{\bm{k}}) Y_{|\sigma_2|\sigma_2}(\hat{\bm{k}}),
  \label{eq:299}
\end{multline}
where
\begin{equation}
  G_{Xl}(k,\mu) \equiv
  \delta_{l0} (1 + f{\mu}^2) c^{(0)}_X
  + c^{(1)}_{Xl}(k)
  \label{eq:300}
\end{equation}
in redshift space. The corresponding result in real space is
immediately obtained by substituting $f=0$ in the above equations. The
above function, $G_{Xl}(k,\mu)$ corresponds to the first-order
invariant propagators, $\hat{\Gamma}^{(1)}_{Xl}(k)$ in real space when
$f=0$, and $\hat{\Gamma}^{(1)l\,l_z}_{Xl'}(k,\mu)$ in redshift space
when $f\ne 0$, which are introduced in Papers~I and II. That is, they
are related by $G_{Xl}(k,0) = \hat{\Gamma}^{(1)}_{Xl}(k)$ in real
space, and $G_{Xl}(k,\mu) = \hat{\Gamma}^{(1)l0}_{Xl}(k,\mu)$ in
redshift space. In linear theory, the dependence on the direction
cosine $\mu$ is present only for scalar fields, $l=0$, and irreducible
components of higher-order tensor fields do not depend on the
direction cosine $\mu$ at all. This is because the redshift-space
distortion effects do not appear in linear theory in higher-order
irreducible tensors. However, this property does not hold in the
higher-order effects beyond the linear theory (Paper~II). In fact, the
redshift-space distortion effects are still contained in the
resummation factor $\Pi(k,\mu)$.

For a scalar field, it is convenient to define the Eulerian bias
function by
\begin{equation}
  b_X(k) \equiv c^{(0)}_X + c^{(1)}_{X0}(k).
  \label{eq:301}
\end{equation}
We also define the redshift-space distortion parameter $\beta$ by
\begin{equation}
  \beta_X(k) \equiv \frac{c^{(0)}_X f}{b_X(k)}.
  \label{eq:302}
\end{equation}
With the above notations, Eq.~(\ref{eq:300}) is equivalently given by
\begin{equation}
  G_{Xl}(k,\mu) =
  \begin{cases}
    b_X(k) \left[1 + \beta_X(k) {\mu}^2\right], & (l=0), \\
    c^{(1)}_{Xl}(k), & (l\geq 1).
  \end{cases}
  \label{eq:303}
\end{equation}
When the scalar field is the number density of galaxies,
$X=\mathrm{g}$, we have $c^{(0)}_\mathrm{g} = 1$ and the above
functions $b_\mathrm{g}(k)$ and $\beta_X(k)$ correspond to the
quantities known as the Eulerian linear bias and the redshift-space
distortion parameter, respectively. In the limit of large scales,
$k \rightarrow 0$, the scale dependence of the bias is naturally
expected to disappear beyond the spatial scales of galaxy formation.

One should note that the first-order renormalized function $b_X(k)$ of
Eq.~(\ref{eq:301}) is not conceptually equivalent to the linear bias
parameter $b_1$ in the linear bias model, because the latter parameter
is a coefficient of the Taylor expansion of the bias relation
$\delta_X = b_1 \delta_\mathrm{L}$, while the invariant renormalized
bias function $b_X(k)$ of Eq.~(\ref{eq:301}) is not. For example, if
we consider a local model in which the cubic bias without linear and
quadratic terms,
$\delta^\mathrm{L}_X = b^\mathrm{L}_3{\delta_\mathrm{L}}^3/3!$ in
Lagrangian space, the renormalized bias function is still nonzero, we
derive $b_X(k) = \sigma^2 b^\mathrm{L}_3/2$ from Eq.~(\ref{eq:295}),
where $\sigma^2 = \langle {\delta_\mathrm{L}}^2\rangle$ is the
variance of the linear density field and Gaussian initial conditions
are assumed. Therefore, one should note that $b_X(k)$ is not exactly
the same as the linear bias parameter, while the resulting expression
of the linear theory agrees at the lowest order.

The product of spherical harmonics can be represented as a
superposition of a single spherical harmonics by a well-known coupling
formula, and the above expression of Eq.~(\ref{eq:299}) can be reduced
to a form of Eqs.~(\ref{eq:169}) and (\ref{eq:178}) in our previous
discussion. The invariant spectra in linear theory are given by
(Sec.~V~A~6 of Paper~I)
\begin{multline}
  P^{l_1l_2;l\,l_z;L}_{X_1X_2}(k,\mu)
  = \delta_{l_z0} \delta_{Ll}
  \frac{(-1)^{l_1}(2l+1)}{\sqrt{(2l_1+1)(2l_2+1)}}
  \begin{pmatrix}
    l_1 & l_2 & l \\
    0 & 0 & 0
  \end{pmatrix}
  \\ \times
  \Pi^2(k,\mu)
  G_{X_1l_1}(k,\mu) G_{X_2l_2}(k,\mu)
  P_\mathrm{L}(k)
  \label{eq:304}
\end{multline}
in redshift space. Here we keep the direction cosine $\mu$ unexpanded
in linear theory in redshift space. The invariant spectrum in real
space $P^{l_1l_2;l}_{X_1X_2}(k)$ is formally given from the above by a
relation
$P^{l_1l_2;l}_{X_1X_2}(k) = P^{l_1l_2;l\,0;l}_{X_1X_2}(k,\mu)|_{f=0}$.
In linear theory, the results in real space are trivially deduced from
the results in redshift space by putting $f=0$, and thus we consider
only the cases in redshift space in our discussion below.

Substituting the above invariant spectrum of Eq.~(\ref{eq:304}) into
general equations we derived in previous sections, predictions for
various power spectra and correlation functions of the iPT in the
lowest order are straightforwardly derived. For the power spectra in
linear theory, however, it is simpler to directly calculate
Eq.~(\ref{eq:299}) before applying the coupling formula of spherical
harmonics, as the spherical harmonics multiplicatively appear in the
expression. Below we explain the latter derivation. Both results agree
with each other as a matter of course.

In order to derive the predictions for linear power spectra of E/B
modes, the expressions of the power spectra, Eq.~(\ref{eq:299}) are
substituted into Eqs.~(\ref{eq:151}) and (\ref{eq:152}). The results
are given by
\begin{multline}
  Q^{\pm\pm} =
  Q^{\pm\mp} =
  \frac{2A_{s_1}A_{s_2}}{(2s_1+1)(2s_2+1)}
  \Pi^2(k,\mu) P_\mathrm{L}(k)
  \\ \times
  G_{X_1s_1}(k,\mu) G_{X_2s_2}(k,\mu)
  \Theta_{s_1}^{s_1}(\mu) \Theta_{s_2}^{s_2}(\mu).
  \label{eq:305}
\end{multline}
From the above, one finds that the functions $Q^{\pm\pm}$,
$Q^{\pm\mp}$ are all the same in the linear theory with Gaussian
initial conditions, and thus the E/B power spectra of
Eqs.~(\ref{eq:147})--(\ref{eq:150}) survive only for
$P^\mathrm{EE}=Q^{++}$, and all the other components vanish,
$P^\mathrm{BB} = P^\mathrm{EB} = P^\mathrm{BE}=0$. The associated
Legendre functions with the same integers of indices,
$\Theta_s^s(\mu)$, is explicitly given by a standard formula of
Eqs.~(\ref{eq:181}) and (\ref{eq:182}). Together with the factor
$A_s$ in Eq.~(\ref{eq:305}), we derive a simple expression,
\begin{equation}
    \sqrt{2}\,A_s\Theta_s^s(\mu)
    = \frac{(-1)^s\sqrt{2s+1}}{2^{s/2}}
    \left(
        1-{\mu}^2
    \right)^{s/2}.
    \label{eq:306}
\end{equation}
Therefore, in full notations for the E/B power spectra, we have
\begin{multline}
  P^{\mathrm{EE}(s_1s_2)}_{X_1X_2}(k,\mu) =
  \frac{(-1)^{s_{12}^+}\left( 1 - {\mu}^2 \right)^{s_{12}^+/2}}
  {2^{s_{12}^+/2}\sqrt{(2s_1+1)(2s_2+1)}}
  \\ \times
  \Pi^2(k,\mu)  P_\mathrm{L}(k)  
  G_{X_1s_1}(k,\mu) G_{X_2s_2}(k,\mu),
  \label{eq:307}
\end{multline}
and
\begin{equation}
  P^{\mathrm{BB}(s_1s_2)}_{X_1X_2}(k,\mu)
  =
  P^{\mathrm{EB}(s_1s_2)}_{X_1X_2}(k,\mu) =
  P^{\mathrm{BE}(s_1s_2)}_{X_1X_2}(k,\mu) = 0.
  \label{eq:308}
\end{equation}
The above equations are the general predictions of the iPT in the
lowest-order approximation with renormalized bias functions.

\subsubsection{Simplifications in a large-scale limit}

In the large-scale limit, $k\rightarrow 0$, the resummation factor
Eq.~(\ref{eq:297}) approaches to unity, $\Pi(k,\mu) \approx 1$. This
factor contains higher-order contributions that should be neglected
in exact linear theory where only the linear-order terms of the linear
power spectrum, $P_\mathrm{L}(k)$, should be retained. If we define
the typical scale of nonlinearity by \cite{Matsubara:2007wj}
\begin{equation}
  k_\mathrm{NL} \equiv
  \left[
    \frac{1}{6\pi^2}
    \int dp\,P_\mathrm{L}(p)
    \right]^{-1/2},
  \label{eq:309}
\end{equation}
the condition that the resummation factor is negligible is given by
$k \ll k_\mathrm{NL}$. The bias functions $c^{(1)}_{Xl}(k)$ in the
large-scale limit are naturally expected not to depend on the scales
$k$, because the scale dependencies of the functions are originated
from the dynamical scales of the galaxy formation: the properties of
galaxies are not affected by values of density field which are located
extremely far from the corresponding galaxies.

Thus we consider the following approximations below:
\begin{align}
  \Pi(k,\mu)
  &\approx 1,
  \label{eq:310}\\
  G_{Xl}(k,\mu)
  &\approx
  G_{Xl}(\mu) \equiv
  \begin{cases}
  b_X \left(1 + \beta_X {\mu}^2\right), & (l=0), \\
  c_{Xl}, & (l\geq 1),
  \end{cases}
  \label{eq:311}
\end{align}
where $b_X= \lim_{k\rightarrow 0} b_X(k)$,
$\beta_X = \lim_{k\rightarrow 0} \beta_X(k)$, and
$c_{Xl} = \lim_{k\rightarrow 0} c^{(1)}_{Xl}(k)$ are scale-independent,
limiting values in the large-scale limit. With the above
approximations, Eq.~(\ref{eq:307}) reduces to
\begin{multline}
  P^{\mathrm{EE}(s_1s_2)}_{X_1X_2}(k,\mu) =
  \frac{(-1)^{s_{12}^+}\left(1 - {\mu}^2 \right)^{s_{12}^+/2}}
  {2^{s_{12}^+/2}\sqrt{(2s_1+1)(2s_2+1)}}
  \\ \times
  G_{X_1s_1}(\mu) G_{X_2s_2}(\mu)
  P_\mathrm{L}(k),
  \label{eq:312}
\end{multline}
which is somehow a simpler expression than Eq.~(\ref{eq:307}), and
reproduces some of the previously known expressions of the linear
theory in literature. Below we only consider the autospectrum just
for simplicity, and assume $X_1=X_2\equiv X$.

In the case of $s_1=s_2=0$, Eq.~(\ref{eq:312}) reduces to
\begin{equation}
  P^{\mathrm{EE}(00)}_{X}(k,\mu) =
  {b_X}^2
  \left( 1 + \beta_X{\mu}^2\right)^2
  P_\mathrm{L}(k),
  \label{eq:313}
\end{equation}
The formula of Eq.~(\ref{eq:313}) applies to any scalar field, which
includes not only the number density field of galaxies but also fields
constructed by, e.g., sizes, colors etc.~of galaxies, and any other
astronomical objects. The appearance of Kaiser's factor,
$(1+\beta_X{\mu}^2)^2$, comes from the fact that the fields are weighted
by number density of objects by definition (see Paper ~I). As
explained above, the biasing is nonlinearly and nonlocally related to
the underlying density field of mass in general. Whatever the
relations are, the formula of Eq.~(\ref{eq:313}) in the large-scale
limit applies to objects with scalar quantities, even when they do not
have a linear term in the bias relation to the density field.

When the scalar field is just the number density of galaxies,
$X=\mathrm{g}$, Eq.~(\ref{eq:313}) reduces to
\begin{equation}
  P_\mathrm{g}(k,\mu) =
  b^2
  \left( 1 + \beta{\mu}^2\right)^2
  P_\mathrm{L}(k),
  \label{eq:314}
\end{equation}
where $b=b_\mathrm{g}$, $\beta =
\beta_\mathrm{g}$, and this equation is exactly the same as the
Kaiser's formula \cite{Kaiser:1987qv,Hamilton:1997zq} of the linear
power spectrum in redshift space. While the Kaiser's formula is
derived assuming a linear model of bias, $\delta_\mathrm{g} =
b\delta_\mathrm{L}$, our formula is still valid even when the bias
relation does not have a linear term as noted above. If we retain the
resummation factor
$\Pi^2(k,\mu)$, the form of Eq.~(\ref{eq:297}) is known to
empirically represent the nonlinear fingers-of-God effect
\cite{Eisenstein:2006nj,Matsubara:2007wj}.

Next, we consider the linear theory of the spin-2 fields, which have
been much investigated in relation to the statistics of intrinsic
alignment of galaxies, in which the spin-2 field is given by the galaxy
shapes projected onto the sky. The cross power spectrum between spin-0
and spin-2 fields with $s_1=0, s_2=2$ in the large-scale limit is
given by
\begin{equation}
  P^{\mathrm{EE}(02)}_{X}(k,\mu)
  =
  \frac{b_X c_{X2}}{2\sqrt{5}}
  \left( 1 + \beta_{X} {\mu}^2 \right)
  \left( 1 - {\mu}^2 \right)
  P_\mathrm{L}(k).
  \label{eq:315}
\end{equation}
The autopower spectrum of spin-2 fields with $s_1=s_2=2$ is given by
\begin{equation}
  P^{\mathrm{EE}(22)}_{X}(k,\mu)
  =
  \frac{{c_{X2}}^2}{20} 
  \left( 1 - {\mu}^2 \right)^2
  P_\mathrm{L}(k).
  \label{eq:316}
\end{equation}

The above equations.[(\ref{eq:315}) and (\ref{eq:316})] are comparable
to the known formulas of the linear power spectra for the intrinsic
alignment. The intrinsic alignment is usually characterized by
statistics of shape tensor $\gamma_{ij}$ of galaxies, which
corresponds to the second moment of mass or luminosity distribution of
each galaxy. In this context, the linear alignment model
\cite{Catelan:2000vm,Blazek:2011xq},
$\gamma_{ij}(\bm{x}) = b_\mathrm{K} K_{ij}(\bm{x})$, is frequently
employed, where
$K_{ij} = (\partial_i\partial_j\triangle^{-1} - \delta_{ij}/3)
\delta_\mathrm{L}$ is the linear tidal field, and the parameter
$b_\mathrm{K}$ is the so-called linear alignment parameter. In linear
theory, one does not have to distinguish perturbation variables
between Eulerian and Lagrangian space. When the traceless components
of the shape tensor, $X=\gamma$, are given by
$F ^\mathrm{L(2)}_{\gamma ij} = \gamma_{ij}$, we have
$F^\mathrm{L}_{\gamma 2m} = \sqrt{3/2}\, \gamma_{ij}
\mathsf{Y}^{(m)*}_{ij}$ from the Lagrangian-space version of
Eq.~(\ref{eq:96}). Substituting these equations into
Eq.~(\ref{eq:295}) with $n=1$, we derive
$c^{(1)}_{\gamma 2m}(\bm{k}) = - b_\mathrm{K} C_{2m}^*(\hat{\bm{k}})$,
and the invariant function of Eq.~(\ref{eq:298}) is given by
$c^{(1)}_{\gamma 2}(k) = -\sqrt{5}\,b_\mathrm{K}$. When we consider
the galaxy bias of the number density field is also linear
$\delta_\mathrm{g} = b\,\delta_\mathrm{L}$, we have $b_X(k)= b$. In
literature, the definition of the E/B modes of intrinsic alignment
differ by a negative sign compared to our definition in this paper, as
noted in the footnote at the end of Sec.~\ref{sec:CorrFunc}.
Therefore, the cross power spectrum between the galaxy density and the
E mode of intrinsic alignment in literature is given by
$P^\gamma_\mathrm{gE}(k,\mu) = -
P^{\mathrm{EE}(02)}_{\mathrm{g}\gamma}(k,\mu)$. The EE power spectrum
of intrinsic alignment is the same as that in the literature,
$P^\gamma_\mathrm{EE}(k,\mu) =
P^{\mathrm{EE}(22)}_{\gamma\gamma}(k,\mu)$, as the minus sign of E
mode cancels. Therefore, Eqs.~(\ref{eq:315}) and (\ref{eq:316}) reduce
to those derived in literature \cite{Taruya:2020tdi},
\begin{align}
  P^\gamma_\mathrm{gE}(k,\mu)
  &=
  \frac{1}{2} b b_\mathrm{K}
  \left( 1 + \beta {\mu}^2 \right)
  \left( 1 - {\mu}^2 \right)
  P_\mathrm{L}(k),
  \label{eq:317}\\
  P^\gamma_\mathrm{EE}(k,\mu)
  &=
  \frac{1}{4} {b_\mathrm{K}}^2
  \left( 1 - {\mu}^2 \right)^2
  P_\mathrm{L}(k).
  \label{eq:318}
\end{align}
We should note again that the first-order renormalized function
$c^{(1)}_{\gamma 2}(k)$ is not conceptually equivalent to the linear
alignment parameter $-\sqrt{5}\,b_\mathrm{K}$. We assume only the
linear regime of dynamical evolution, while we do not assume the
linear alignment model of galaxy shapes. Therefore, the previously
known formulas of Eqs.~(\ref{eq:317}) and (\ref{eq:318}) are
still valid on large scales when the linear alignment parameter is
replaced by the renormalized bias function, even if the relation
between the (three-dimensional) shape tensor $\gamma_{ij}$ is not
linearly proportional to the linear tidal field $K_{ij}$.

The statistics of the spin-1 tensor field have not been considered in
literature as much as those of the spin-2 field, such as the intrinsic
alignment. It is of interest to consider the correlation statistics of
the spin-1 field as well in the large-scale surveys of galaxies in the
near future. We do not have to specify what kind of spin-1 field is
considered, and what kind of mechanism the field is generated in our
discussion, as long as we describe the statistics in terms of
renormalized bias functions. The EE power spectra involving the spin-1
tensor in the large-scale limit are given by
\begin{align}
  P^{\mathrm{EE}(01)}_X(k,\mu)
  &= -\frac{b_X c_{X1}}{\sqrt{6}}
    \left( 1+\beta_X{\mu}^2 \right) \left( 1-{\mu}^2 \right)^{1/2}
    P_\mathrm{L}(k),
  \label{eq:319}\\
  P^{\mathrm{EE}(11)}_X(k,\mu)
  &= \frac{ {c_{X1}}^2 }{6}
    \left( 1-{\mu}^2 \right) P_\mathrm{L}(k),
  \label{eq:320}\\
  P^{\mathrm{EE}(12)}_X(k,\mu)
  &= -\frac{ c_{X1} c_{X2}}{2\sqrt{30}}
    \left( 1-{\mu}^2 \right)^{3/2} P_\mathrm{L}(k).
  \label{eq:321}
\end{align}

A typical example of the observable spin-1 field is the angular
momentum of galaxies, which is a pseudovector field. However, as shown
in Paper~I, the first-order renormalized bias function
$c^{(1)}_{Xl}(k)$ of the pseudotensor field vanishes in the Universe
with parity symmetry, because the parity transformation of the
function is given by $c^{(1)}_{Xl}(k) \rightarrow - c^{(1)}_{Xl}(k)$
when $X$ is a pseudotensor field. Therefore, the power spectra or
correlation functions of angular momentum in the linear order
identically vanish in the Universe with parity symmetry. In other
words, the power spectra of Eqs.~(\ref{eq:319})--(\ref{eq:321}) can be
used as possible indicators of the violation of parity symmetry in the
Universe.

Corresponding to the above discussion, there is not any conceivable
linear model of the angular momentum. For example, a linear vector
field constructed from the linear potential, $\bm{\nabla}\varphi$ with
$\varphi=\triangle^{-1}\delta_\mathrm{L}$, is not a pseudovector
field, and thus can never be a candidate of the linear model for the
angular momentum. In order to construct a pseudovector out of linear
density field, it should be at least second order or higher in
perturbations. For example, White \cite{White:1984uf} derived an
analytic evolution of the angular momentum $\bm{J}$ of protogalaxies
using the lowest-order perturbation theory in Lagrangian space
(Zel'dovich approximation), as
$J_i = - a^2 HfD \epsilon_{ijk} T_{jl}I_{lk}$, where
$T_{ij} =\partial_i\partial_j\varphi$ is the linear tidal tensor and
$I_{ij} = \int_{V} (q_i- \bar{q}_i) (q_j- \bar{q}_j)\, \rho a^3\,
d^3q$ is the inertia tensor, and $V$ is the Lagrangian volume of the
protogalaxy, $\bar{\bm{q}} = V^{-1} \int_V \bm{q} d^3q$ is its center
of mass and $\rho$ is the mass density of the volume. Tidal tensor
$T_{ij}$ is the first order in perturbations. The inertia tensor
$I_{ij}$ vanishes when the density perturbations are absent, and thus
is at least first order or higher in perturbations. Therefore, this
exemplified simple model does not contain a linear term in
perturbations as a matter of course, although the time evolution is
proportional to the linear growth factor $D$ as White originally
showed. As noted above, we need some parity-violating mechanism to
generate the angular momentum in order to have nonvanishing linear
power spectra of angular momentum.

\subsection{%
  The linear correlation functions
}

\subsubsection{General predictions of linear theory}

The linear correlation function in redshift space is given by
Eqs.~(\ref{eq:258}) and (\ref{eq:259}) in general. The invariant power
spectrum in linear theory is given by Eqs.~(\ref{eq:304}) and
(\ref{eq:179}), which is substituted into those general expressions of
the correlation function. Because of the constraints imposed by
Kronecker's deltas in Eq.~(\ref{eq:304}), the resulting expression is
much simplified in linear theory. After some calculations, the
expression reduces to
\begin{multline}
  \xi^{(\sigma_1\sigma_2)}_{X_1X_2}(\bm{x})
  = (-i)^{|\sigma_1|+|\sigma_2|}
  \sum_l i^l Y_{l\sigma_{12}}(\hat{\bm{x}})
  \\ \times
  \sum_{l'}
  \begin{pmatrix}
    |\sigma_1| & |\sigma_2| & l' \\
    \sigma_1 & \sigma_2 & -\sigma_{12}
  \end{pmatrix}
  J_{X_1X_2;l\sigma_{12}}^{|\sigma_1||\sigma_2|;l'}(x)
  \label{eq:322}
\end{multline}
where
\begin{multline}
  J_{X_1X_2;lm}^{l_1l_2;l'}(x) \equiv
  (-1)^{l_1} A_{l_1} A_{l_2}
  \sqrt{\frac{4\pi\,(2l'+1)}{(2l_1+1)(2l_2+1)}}
    \\ \times
  \begin{pmatrix}
    l_1 & l_2 & l' \\
    0 & 0 & 0
  \end{pmatrix}
  I_{X_1X_2;lm}^{l_1l_2;l'}(x)
  \label{eq:323}
\end{multline}
and
\begin{multline}
  I_{X_1X_2;lm}^{l_1l_2;l'}(x) \equiv
  \int \frac{k^2dk}{2\pi^2} j_l(kx) P_\mathrm{L}(k)
  \int d\mu\, \Theta_{l'}^m(\mu) \Theta_l^m(\mu)
  \\ \times
  \Pi^2(k,\mu) G_{X_1l_1}(k,\mu) G_{X_2l_2}(k,\mu).
  \label{eq:324}
\end{multline}
In deriving the above result, identities of $3j$- and $6j$-symbols,
\begin{align}
  \begin{pmatrix}
    l & l' & 0 \\
    0 & 0 & 0
  \end{pmatrix}
  &= \frac{(-1)^l}{\sqrt{2l+1}} \delta_{l'l},
  \label{eq:325}\\
  \begin{Bmatrix}
    l_1 & 0 & l_3 \\
    l_4 & l_5 & l_6
  \end{Bmatrix}
  &= \frac{(-1)^{l_1+l_4+l_5}}{\sqrt{(2l_1+1)(2l_4+1)}}
  \delta_{l_1l_3} \delta_{l_4l_6}
  \label{eq:326}
\end{align}
are used. From the symmetry of Eq.~(\ref{eq:182}), the function of
Eq.~(\ref{eq:323}) satisfies
\begin{equation}
  J_{X_1X_2;l,-m}^{l_1l_2;l'}(x) =
  J_{X_1X_2;lm}^{l_1l_2;l'}(x)
  \label{eq:327}
\end{equation}
and is nonzero only when $l'+l=\mathrm{even}$. 

The correlation functions of $+/\times$ modes in linear theory are
straightforwardly derived from Eq.~(\ref{eq:322}). They are given by
Eqs.~(\ref{eq:215})--(\ref{eq:222}). Substituting the above
result of Eq.~(\ref{eq:322}) into Eqs.~(\ref{eq:219}) and
(\ref{eq:220}), we have
\begin{align}
  X^{\pm\pm}
  &=
    \frac{(-i)^{s^+_{12}}}{\sqrt{2\pi (2s^+_{12}+1)}}
    \sum_l i^l \Theta_l^{s^+_{12}}(\mu_x)
    J_{X_1X_2;ls^+_{12}}^{s_1s_2;s^+_{12}}(x)
  \label{eq:328}\\
  X^{\pm\mp}
  &=
    \frac{(\mp i)^{s^-_{12}}}{\sqrt{2\pi}}
    \sum_{l,l'} i^l\, \Theta_l^{s^-_{12}}(\mu_x)
    (\pm 1)^{l'}
    \begin{pmatrix}
      s_1 & s_2 & l' \\
      s_1 & -s_2 & -s^-_{12}
    \end{pmatrix}
   J^{s_1s_2;l'}_{X_1X_2;ls^-_{12}}(x).
  \label{eq:329}
\end{align}
From Eq.~(\ref{eq:328}), we apparently have $X^{++}=X^{--}$. As we
have $s_1+s_2+l'=\mathrm{even}$ in Eq.~(\ref{eq:329}), we also have
$X^{+-}=X^{-+}$. Therefore, from
Eqs.~(\ref{eq:215})--(\ref{eq:218}), we have
\begin{equation}
  \xi^{++}=\frac{X^{+-}+X^{++}}{2}, \quad
  \xi^{\times\times}=\frac{X^{+-}-X^{++}}{2}, \quad
  \xi^{+\times}=\xi^{+\times}=0.
\label{eq:330}
\end{equation}
In terms of the combinations $\xi^\pm$ and $\zeta^\pm$ defined in
Eq.~(\ref{eq:245}), we have
\begin{equation}
  \xi^{\pm} = X^{+\mp}, \quad
  \zeta^{\pm} = 0,
  \label{eq:331}
\end{equation}
where $X^{+\mp}$ correspond to upper signs of Eqs.~(\ref{eq:329})
and (\ref{eq:328}), respectively. The same results are also derived
by applying the linear transformations of Eqs.~(\ref{eq:247}) and
(\ref{eq:248}) to the expression of E/B power spectra of
Eqs.~(\ref{eq:307}) and (\ref{eq:308}).

As a special case when $s_1=s_2=0$, Eqs.~(\ref{eq:328}) and
(\ref{eq:329}) are explicitly shown to be the same
$X^{\pm\pm} = X^{\pm\mp}$, and therefore we have
\begin{equation}
  \xi^{++(00)}_{X_1X_2}(x,\mu_x) = X^{++} =
  \sqrt{2}
  \sum_l i^l \Theta_l^0(\mu_x) I^{00;0}_{X_1X_2;l0}(x), 
  \label{eq:332}
\end{equation}
and
$\xi^{\times\times(00)}_{X_1X_2} = \xi^{+\times(00)}_{X_1X_2} =
\xi^{\times+(00)}_{X_1X_2} = 0$. As another special case when $s_1=0$
and $s_2=s\ne 0$, it can also be shown that $X^{\pm\pm} = X^{\pm\mp}$,
and therefore we have
\begin{equation}
  \xi^{++(s0)}_{X_1X_2}(x,\mu_x) = X^{++} =
  \frac{(-i)^s\sqrt{2}\,A_s}{2s+1}
  \sum_l i^l \Theta_l^s(\mu_x) I^{s0;s}_{X_1X_2;ls}(x),
  \label{eq:333}
\end{equation}
and
$\xi^{\times\times(s0)}_{X_1X_2} = \xi^{+\times(s0)}_{X_1X_2} =
\xi^{\times+(s0)}_{X_1X_2} = 0$.
The above property that only the
correlation function $\xi^{++}$ survives when $s_1=0$ or $s_2=0$ is
trivial because $f^{\times(0)}_X(\bm{x}) = 0$ as seen from
Eq.~(\ref{eq:208}). In general cases with $s_1\ne 0$ and
$s_2 \ne 0$, both $\xi^{++(s_1s_2)}_{X_1X_2}$ and
$\xi^{\times\times(s_1s_2)}_{X_1X_2}$ survive.

The above equations are sufficient to analytically derive the
expressions of correlation functions in linear theory. The dependence
on the direction cosine $\mu$ is only present for the scalar case,
$l=0$, as given in Eq.~(\ref{eq:303}) in linear theory, and only
possible dependence is just through ${\mu}^2$ in $G_{X0}(k,\mu)$. In
order to evaluate the integral of Eq.~(\ref{eq:324}) with the
resummation factor of Eq.~(\ref{eq:297}), we need to evaluate
integrals of the type
\begin{equation}
  \int_{-1}^1 d\mu\, e^{-A{\mu}^2}
   \Theta_{l'}^m(\mu) \Theta_l^m(\mu) \mu^n,
  \label{eq:334}
\end{equation}
where $n=0,2,4$. This type of integral is analytically represented by
using error function $\mathrm{erf}(x)$ for arbitrary integers
$l',l,m,n$. The cases of $n=2,4$ are deduced from the case of $n=0$ by
applying the differential operators $-\partial/\partial A$ and
$\partial^2/\partial A^2$ to the integral with $n=0$. In this way, the
integral $I^{l_1l_2;l'}_{X_1X_2;lm}(k)$ of Eq.~(\ref{eq:324}) with
the propagator Eq.~(\ref{eq:300}) in linear theory is analytically
obtained. While it is straightforward to derive the analytic
expressions for a given set of integers $l_1$ and $l_2$, the results
are somehow complicated to present, and we do not give explicit
equations here.

\subsubsection{Large-scale limit}

While the full predictions of linear theory are given by
Eqs.~(\ref{eq:328})--(\ref{eq:330}), we consider the large-scale
limit with approximations of Eqs.~(\ref{eq:310}) and
(\ref{eq:311}). In this case, the integral of Eq.~(\ref{eq:324})
simply reduces to
\begin{equation}
  I_{X_1X_2;lm}^{l_1l_2;l'}(x) \approx
  \xi_l(x)
  \int_{-1}^1 d\mu\, \Theta_l^m(\mu) \Theta_{l'}^m(\mu)
  G_{X_1l_1}(\mu) G_{X_2l_2}(\mu),
  \label{eq:335}
\end{equation}
where
\begin{equation}
  \xi_l(x) \equiv \int \frac{k^2dk}{2\pi^2} j_l(kx) P_\mathrm{L}(k).
  \label{eq:336}
\end{equation}
The integral over the direction cosine $\mu$ in Eq.~(\ref{eq:335}) can
be analytically evaluated by using a formula,
\begin{multline}
  \int_{-1}^1 d\mu\,
  \Theta_{l'}^m(\mu) \Theta_l^m(\mu) P_n(\mu)
  = (-1)^m \sqrt{(2l+1)(2l'+1)}
  \\ \times
  \begin{pmatrix}
    l & l' & n \\
    0 & 0 & 0
  \end{pmatrix}
  \begin{pmatrix}
    l & l' & n \\
    m & -m & 0
  \end{pmatrix},
  \label{eq:337}
\end{multline}
where $P_n(\mu)$ is the Legendre polynomial of order $n$. The above
formula is deduced from a known integral formula for a product of two
spherical harmonics with the same arguments. We need only the cases of
$n=0,2,4$.

For simplicity, we only consider the autopower spectrum of objects,
$X_1=X_2\equiv X$. Generalizations to cross power spectra with
different samples of objects are rather trivial. First, we consider
the case $s_1=s_2=0$. In this case, we need to evaluate an integral
$I^{00;0}_{XX;l0}(x)$ in Eq.~(\ref{eq:332}). Following the procedure
explained above using Eqs.~(\ref{eq:335})--(\ref{eq:337}), we have
\begin{multline}
  I^{00;0}_{XX;l0}(x) =
  {b_X}^2\xi_l(x)
  \left[
    \delta_{l0}
    \left(
      1 + \frac{2}{3}\,\beta_X
      + \frac{1}{5}\,{\beta_X}^2
    \right)
  \right.
  \\
  \left.
    +
    \frac{\delta_{l0}}{\sqrt{5}}
    \left(
      \frac{4}{3}\,\beta_X
      + \frac{4}{7}\,{\beta_X}^2
    \right)
    + \frac{\delta_{l4}}{3}
    \frac{8}{35}\,{\beta_X}^2
  \right].
  \label{eq:338}
\end{multline}
Substituting the above result into Eq.~(\ref{eq:332}), we have
\begin{multline}
  \xi^{++(00)}_{XX}(x,\mu_x) =
  {b_X}^2
  \left[
    \left(
      1 + \frac{2}{3}\,\beta_X
      + \frac{1}{5}\,{\beta_X}^2
    \right)
    P_0(\mu_x) \xi_0(x)
  \right.
  \\
  \left.
    -
    \left(
      \frac{4}{3}\,\beta_X
      + \frac{4}{7}\,{\beta_X}^2
    \right)
    P_2(\mu_x) \xi_2(x)
  \right.
  \\
  \left.
    +
    \frac{8}{35}\,{\beta_X}^2
    P_4(\mu_x) \xi_4(x)
  \right].
  \label{eq:339}
\end{multline}
This equation is equivalent to Hamilton's formula for the galaxy
correlation function in linear theory \cite{Hamilton:1992zz}.

Next, we consider the case $s_1=s \ne 0$ and $s_2=0$. In this case, we
need to evaluate an integral $I^{s0;s}_{X_1X_2;ls}(x)$ in
Eq.~(\ref{eq:333}). Following the procedure explained above, we
calculate
\begin{multline}
  I^{s0;s}_{XX;ls}(x)
  = b_X c_{Xs}\,\xi_l(x) \\ \times
  \left[
    \delta_{ls} \left(1 + \frac{\beta_{X_1}}{2s+3}\right) +
    \delta_{l,s+2}
    \sqrt{\frac{s+1}{2s+5}}\,\frac{2\beta_{X_1}}{2s+3}
  \right].
  \label{eq:340}
\end{multline}
Substituting the above equation into Eq.~(\ref{eq:333}), and after
some algebra, we derive
\begin{multline}
    \xi^{++(s0)}_{XX}(x,\mu_x) = \frac{b_X
      c_{Xs}}{2^{s/2}\sqrt{2s+1}\,(2s-1)!!}
    \\ \times
  \left[ \left(1 +
      \frac{\beta_X}{2s+3}\right) P_s^s(\mu_x) \xi_s(x)
  \right.
  \\
  \left.
    - \frac{2\beta_X}{(2s+1)(2s+3)} P_{s+2}^s(\mu_x)
    \xi_{s+2}(x)
  \right].
  \label{eq:341}
\end{multline}
The relevant associated Legendre polynomials are explicitly given by
$P_s^s(\mu) = (-1)^s (2s-1)!!\,(1-{\mu}^2)^{s/2}$ and
$P_{s+2}^s(\mu) = (2s+1) [(2s+3){\mu}^2/2 - 1] P_s^s(\mu)$, and the
above equation is equivalently expressed as
\begin{multline}
  \xi^{++(s0)}_{X}(x,\mu_x) = \frac{(-1)^s b_X c_{Xs}}{2^{s/2}\sqrt{2s+1}}
  \left(1-{\mu_x}^2\right)^{s/2} \\ \times \left[ \left(1 +
      \frac{\beta_X}{2s+3}\right) \xi_s(x) - \beta_X \left({\mu_x}^2
      - \frac{1}{2s+3}\right) \xi_{s+2}(x) \right].
  \label{eq:342}
\end{multline}

In the cases of $s=1$ and $s=2$, we have
\begin{multline}
  \xi^{++(10)}_{X}(x,\mu_x) = -\frac{b c_{X1}}{\sqrt{6}}
  \sqrt{1-{\mu_x}^2} \\ \times \left[ \left(1 +
      \frac{\beta}{5}\right) \xi_1(x) - \beta \left({\mu_x}^2 -
      \frac{1}{5}\right) \xi_3(x) \right],
  \label{eq:343}
\end{multline}
and
\begin{multline}
  \xi^{++(20)}_{X}(x,\mu_x)
 = - \frac{b b_\mathrm{K}}{2}
  \left(1-{\mu_x}^2\right) \\ \times \left[ \left(1 +
      \frac{\beta}{7}\right) \xi_2(x) - \beta \left({\mu_x}^2 -
      \frac{1}{7}\right) \xi_4(x) \right],
  \label{eq:344}
\end{multline}
respectively, where $\beta=\beta_X$, $b=b_{X_1}$, and
$b_\mathrm{K} = -c_{X2}/\sqrt{5}$, bearing in mind the notation used
in Eqs.~(\ref{eq:315})--(\ref{eq:318}) for the number density and
intrinsic alignment of galaxies. Equation~(\ref{eq:344}) agrees with
the results derived in literature \cite{Okumura:2019ned} in the
context of the intrinsic alignment, besides the normalization and sign
convention.\footnote{In Ref.~\cite{Okumura:2019ned}, their
  cross-correlation $\xi_{\mathrm{g}+}(\bm{r})$ is defined in a
  different way and contains a dependence on azimuthal angle $\phi$ of
  the argument $\bm{r}$, thus extra factor $\cos(2\phi)$ appears in
  their expression, compared with our expression of
  Eq.~(\ref{eq:344}).}

Finally, we consider the case $s_1\ne 0$ and $s_2\ne 0$ in the
large-scale limit. In this case, the integral of Eq.~(\ref{eq:335})
is trivially given by
\begin{equation}
  I^{s_1s_2;l'}_{XX;lm}(x)
  = \delta_{l'l} c_{Xs_1} c_{Xs_2}\xi_l(x) 
  \label{eq:345}
\end{equation}
for arbitrary integers $m$ with $|m| \leq l$. Substituting the above
into Eqs.~(\ref{eq:328}) and (\ref{eq:329}), and using the
relations of Eq.~(\ref{eq:331}), we have
\begin{multline}
  \xi^+ = \frac{\sqrt{2}\,c_{Xs_1} c_{Xs_2}\sqrt{s_1!\,s_2!}}
  {\sqrt{(2s_1+1)!!\,(2s_2+1)!!}}
  i^{s^+_{12}}  \sum_l i^l \sqrt{2l+1}
  \\ \times
  \begin{pmatrix}
    s_1 & s_2 & l \\
    0 & 0 & 0
  \end{pmatrix}
  \begin{pmatrix}
    s_1 & s_2 & l \\
    s_1 & -s_2 & -s^-_{12}
  \end{pmatrix}
  \Theta_l^{s^-_{12}}(\mu_x) \xi_l(x),
  \label{eq:346}
\end{multline}
and
\begin{equation}
  \xi^- =
  \frac{\sqrt{2}\,(-1)^{s_2} c_{Xs_1} c_{Xs_2}
    \sqrt{s^+_{12}!}}{\sqrt{(2s_1+1)(2s_2+1)\,(2s^+_{12}+1)!!}}
  \Theta_{s^+_{12}}^{s^+_{12}}(\mu_x) \xi_{s^+_{12}}(x).
  \label{eq:347}
\end{equation}
From the above, the correlation functions are given by
$\xi^{++(s_1s_2)}_{X_1X_2}=(\xi^++\xi^-)/2$,
$\xi^{\times\times(s_1s_2)}_{X_1X_2}=(\xi^+-\xi^-)/2$ and
$\xi^{+\times(s_1s_2)}_{X_1X_2}=\xi^{\times+(s_1s_2)}_{X_1X_2}=0$.

We explicitly present expressions involving the spins up to second
order. When $s_1=s_2=1$, we have
\begin{align}
  \xi^+
  &=
    \frac{{c_{X1}}^2}{9}
    \left[
    P_0(\mu_x) \xi_0(x) + P_2(\mu_x) \xi_2(x)
    \right],
    \label{eq:348}\\
  \xi^-
  &=
    - \frac{{c_{X1}}^2}{18}
    P_2^2(\mu_x) \xi_2(x).
    \label{eq:349}
\end{align}
From the above, we have
\begin{align}
  \xi^{++(11)}_{X}(x,\mu_x)
  &= \frac{{c_{X1}}^2}{18}
    \left[
    \xi_0(x) + \left(3{\mu_x}^2 - 2\right) \xi_2(x)
    \right],
  \label{eq:350}\\
  \xi^{\times\times(11)}_{X}(x,\mu_x)
  &= \frac{{c_{X1}}^2}{18}
    \left[
    \xi_0(x) + \xi_2(x)
    \right].
  \label{eq:351}
\end{align}
When $s_1=1$ and $s_2=2$, we have
\begin{align}
  \xi^+
  &=
    \frac{c_{X1}b_\mathrm{K}}{15\sqrt{6}}
    \left[
    6P_1^1(\mu_x) \xi_1(x) + P_3^1(\mu_x) \xi_3(x)
    \right],
    \label{eq:352}\\
  \xi^-
  &=
    -\frac{c_{X1}b_\mathrm{K}}{30\sqrt{6}}
    P_3^3(\mu_x) \xi_3(x).
    \label{eq:353}
\end{align}
From the above, we have
\begin{align}
  \xi^{++(12)}_{X}(x,\mu_x)
  &= -\frac{c_{X1}b_\mathrm{K}}{5\sqrt{6}}
    \sqrt{1-{\mu_x}^2}
    \left[
    \xi_1(x) + \frac{5{\mu_x}^2 - 3}{2} \xi_3(x)
    \right],
  \label{eq:354}\\
  \xi^{\times\times(12)}_{X}(x,\mu_x)
  &= -\frac{c_{X1}b_\mathrm{K}}{5\sqrt{6}}
    \sqrt{1-{\mu_x}^2}
    \left[
    \xi_1(x) + \xi_3(x)
    \right].
  \label{eq:355}
\end{align}
When $s_1=s_2=2$, we have
\begin{align}
  \xi^+
  &=
    \frac{2{b_\mathrm{K}}^2}{105}
    \left[
    7P_0(\mu_x) \xi_0(x)
  %   \right.
  %   \nonumber\\
  % & \hspace{4.5pc}
  %   \left.
    + 10P_2(\mu_x) \xi_2(x)
    + 3P_4(\mu_x) \xi_4(x)
    \right],
    \label{eq:356}\\
  \xi^-
  &=
    \frac{{b_\mathrm{K}}^2}{420}
    P_4^4(\mu_x) \xi_4(x).
    \label{eq:357}
\end{align}
From the above, we have
\begin{align}
  \xi^{++(22)}_{X}(x,\mu_x)
  &= \frac{{b_\mathrm{K}}^2}{105}
    \Biggl[
    7 \xi_0(x) + 5 \left(3{\mu_x}^2 - 1\right) \xi_2(x)
    \nonumber\\
  & \hspace{2pc}
    + \frac{3}{4}\left(35{\mu_x}^4 - 50{\mu_x}^2 + 19\right) \xi_4(x)
    \Biggr],
  \label{eq:357a}\\
  \xi^{\times\times(22)}_{X}(x,\mu_x)
  &= \frac{{b_\mathrm{K}}^2}{105}
    \Biggl[
    7 \xi_0(x) + 5 \left(3{\mu_x}^2 - 1\right) \xi_2(x)
    \nonumber\\
  & \hspace{6pc}
    + 3\left(5{\mu_x}^2 - 4\right) \xi_4(x)
    \Biggr].
  \label{eq:357b}
\end{align}
Equations~(\ref{eq:356}) and (\ref{eq:357}) agree with the results
derived in literature \cite{Okumura:2019ned} in the context of the
intrinsic alignment, besides the normalization and sign
conventions.\footnote{In Ref.~\cite{Okumura:2019ned}, their
  correlation $\xi_{-}(\bm{r})$ is differently defined and contains a
  dependence on the azimuthal angle $\phi$ of the argument $\bm{r}$,
  thus extra factor $\cos(4\phi)$ appears in their expression,
  compared with our expression of Eq.~(\ref{eq:357}).}

\subsection{%
  The angular correlation functions in linear theory
}

The angular correlation functions are generally given by
Eqs.~(\ref{eq:287})--(\ref{eq:290}). 
The invariant spectrum of the linear theory in real space is given by
(Paper~I)
\begin{multline}
  P^{l_1l_2;l}_{X_1X_2}(k) =
  \frac{(-1)^{l_1}(2l+1)}{\sqrt{(2l_1+1)(2l_2+1)}}
  \begin{pmatrix}
    l_1 & l_2 & l \\
    0 & 0 & 0
  \end{pmatrix}
  \\ \times
  G_{X_1l_1}(k) G_{X_2l_2}(k)
  \Pi^2(k)
  P_\mathrm{L}(k),
  \label{eq:358}
\end{multline}
where $G_{Xl}(k) \equiv \hat{\Gamma}^{(1)}_{Xl}(k)$ is the first-order
invariant propagator in real space. The last propagator is the same as
the one in Eq.~(\ref{eq:303}) with $f=0$:
\begin{equation}
  G_{Xl}(k) =
  \begin{cases}
  b_X(k), & (l=0), \\
  c^{(1)}_{Xl}(k), & (l\geq 1),
  \end{cases}
  \label{eq:359}
\end{equation}
where $b_X(k) = c^{(0)}_X + c^{(1)}_{X0}(k)$. Substituting
Eq.~(\ref{eq:358}) into Eqs.~(\ref{eq:171}) and
(\ref{eq:287})--(\ref{eq:290}), we obtain the expressions of the
angular correlation functions in linear theory. The expressions of
$W^{\pm\pm}$ and $W^{\pm\mp}$ are also obtained by substituting
Eq.~(\ref{eq:305}) into Eqs.~(\ref{eq:284}) and
(\ref{eq:285}). Either way, the results are given by
\begin{multline}
  W^{+\mp} =
  \frac{N_0 (-1)^{s_1}(\mp 1)^{s_2}}{2^{(s_1+s_2)/2}\sqrt{(2s_1+1)(2s_2+1)}}
  \int \frac{kdk}{2\pi} J_{s^{\mp}_{12}}(kr_0\theta)
  \\ \times
  \Pi^2(k) G_{X_1s_1}(k) G_{X_2s_2}(k) P_\mathrm{L}(k).
  \label{eq:360}
\end{multline}
Substituting the above equation into Eq.~(\ref{eq:290}), we have the
expressions for $w^{++(s_1s_2)}_{X_1X_2}(\theta)$ and
$w^{\times\times(s_1s_2)}_{X_1X_2}(\theta)$.

In the large-scale limit, where $\Pi(k) \approx 1$ and
$G_{Xl} =\lim_{k\rightarrow 0} G_{Xl}(k)$, Eqs.~(\ref{eq:360}) and
(\ref{eq:290}) show that
\begin{multline}
  w^{++/\times\times} =
  \frac{N_0}{2}
  \frac{(-1)^{s_1+s_2}G_{X_1s_1} G_{X_2s_2}}{2^{(s_1+s_2)/2}\sqrt{(2s_1+1)(2s_2+1)}}
\\ \times
  \int \frac{kdk}{2\pi}
  \left[
    J_{s^-_{12}}(kr_0\theta)
    \pm (-1)^{s_2} J_{s^+_{12}}(kr_0\theta)
  \right] P_\mathrm{L}(k).
  \label{eq:361}
\end{multline}
In the case of $s_1=s_2=2$, the integrals over $k$ in
Eq.~(\ref{eq:361}) have forms $J_0 \pm J_4$ which are familiar in
the study of two-point statistics of weak lensing field
\cite{Kaiser:1991qi,Bartelmann:1999yn}. The above equations are also
derived more easily by substituting Eqs.~(\ref{eq:307}) and
(\ref{eq:308}) into Eq.~(\ref{eq:291}).

\subsection{Outline of evaluating the nonlinear corrections}

The methods of the iPT to calculate nonlinear corrections of the
invariant spectra, $P^{l_1l_2;l}_{X_1X_2}(k)$ in real space and
$P^{l_1l_2;l\,l_z;L}_{X_1X_2}(k,\mu)$ in redshift space are detailed
in Paper~II. Substituting the results of the nonlinear corrections of
the invariant spectra into relevant equations in
Secs.~\ref{sec:FlatSky} to \ref{sec:CorrFunc}, the nonlinear
corrections of the power spectra and correlation functions are
straightforwardly calculated. A main complexity of the nonlinear
corrections which does not exist in the linear theory stems from the
coupling of Fourier modes. While higher-dimensional loop integrals
over the multiple wave vectors appear in the straightforward
expressions, the angular integrations of them are analytically
evaluated thanks to the formalism with the spherical basis, and we end
up with numerical integrations of only radial components of
wave vectors. In the case of Gaussian initial conditions, the
calculation of the $n$-loop correction of the power spectrum requires
$n+2$ integrals in general when the model of renormalized bias
functions $c^{(n)}_{Xlm}$ of Eq.~(\ref{eq:295}) is not specified in
general.

As explained in Paper~II, however, when the Lagrangian bias of tensor
field $F^\mathrm{L}_{Xlm}(\bm{q})$ is modeled by a semilocal model,
all the necessary integrations are essentially given by
one-dimensional Hankel transforms which can be evaluated very quickly
by using the FFTLog algorithm \cite{Hamilton:1999uv}. This application
of the FFTLog algorithm to the nonlinear perturbation theory of tensor
fields is naturally a tensor generalization of the similar methods for
scalar nonlinear perturbation theory developed in the literature
\cite{Schmittfull:2016jsw,Schmittfull:2016yqx,McEwen:2016fjn,Fang:2016wcf}.
Examples of the one-loop corrections of the invariant power spectrum
in real space $P^{22;l}_{XX}(k)$ of rank-2 tensor field are
numerically calculated and demonstrated in Paper~II when the tensor
field $F^\mathrm{L}_{X2m}(\bm{q})$ is a local function of linear tidal
field
$\varphi_{ij} = \partial_i\partial_j\triangle^{-1}\delta_\mathrm{L}$.
Any other tensor field with a semilocal model of bias can be similarly
evaluated using the FFTLog algorithm.

As demonstrated in Paper~II, loop corrections do not have significant
effects on large scales, as long as the bias parameters such as
$c^{(n)}_{Xlm}$ are not large enough. However, it is still useful to
calculate the amplitudes of the higher-loop corrections to the linear
predictions, as the applicable scales of the linear theory where
nonlinear effects are negligible can be estimated by evaluating the
amplitude of loop corrections. We will address this problem in the
future work with more concrete applications to the observational
cosmology, such as two-point statistics of the intrinsic alignment of
galaxies.

\section{\label{sec:Conclusions}%
  Conclusions
}

In this paper, which is a sequel to Papers~I and II, projection
effects on the sky are taken into account in the formalism of the iPT
for cosmological tensor fields. The three-dimensional tensors of
astronomical objects like galaxies are not directly observable because
we can observe the objects only from a fixed direction, i.e., the line
of sight. The observable tensors are not the original
three-dimensional tensors of objects, but are two-dimensional tensors
projected onto the sky from the original physical tensors of three
dimensions.

Following the strategy of Papers~I and II, the two-dimensional tensors
on the sky are decomposed into irreducible tensors on the spin basis.
The relation between irreducible tensors in three-dimensional space
and those in projected two-dimensional space is derived first in
spherical coordinates of a full sky. Symmetric properties including
complex conjugate, rotation, and parity in the decomposed field are
explicitly derived and summarized. The distant-observer limit of the
decomposed tensor field in the full sky is considered, and various
relations between the decomposed tensor fields in the spherical
coordinates of the full sky and those in the Cartesian coordinates of
the distant-observer limit are derived. The symmetric properties in
the Cartesian coordinates of the distant-observer limit are derived
and summarized.

The power spectra of the projected tensor fields with general spins
and their E/B decomposition are defined and considered in Cartesian
coordinates of the distant-observer limit in Fourier space. Related
formulas are derived both in real space and in redshift space. We
explicitly derive relations between the power spectra of projected
tensor fields and the three-dimensional invariant power spectra, which
are introduced in Paper~I, and evaluation methods of the invariant
spectra by iPT are developed in Papers~I and II. Thus the general
methods of the iPT are straightforwardly applied to evaluating the
power spectra of projected tensor fields on the spin basis. The
correlation functions of the projected tensor fields with general
spins and their $+/\times$ decomposition are also defined and
considered in Cartesian coordinates of the distant-observer limit in
configuration space. Linear relations between the E/B power spectra
and the $+/\times$ correlation functions for arbitrary spin fields are
generally derived, which are represented by introducing operators
$\hat{\mathcal{F}}_\sigma$ and $\hat{\mathcal{F}}^{-1}_\sigma$. We
also derive the relations between the correlation functions of
projected tensor fields and the invariant power spectra. Thus both the
power spectra and correlation functions of projected tensor fields on
the spin basis are calculated by applying the iPT up to arbitrary
orders. Corresponding results for the angular correlation functions of
projected tensor fields are also derived and presented.

Analytic expressions for the results of applying the lowest-order
results of iPT, which are counterparts of the linear perturbation
theory in literature, are derived. Known results in the literature of
two-point statistics in spin-2 tensor fields in the context of
intrinsic alignment are shown to be rederived as special cases of our
general results. The outline of calculating higher-order effects with
loop corrections in perturbation theory is given, while their
numerical evaluations are left for future work together with more
concrete situations of physical observables, such as the statistics of
angular momentum and intrinsic alignment of galaxies, etc.

As a distinguished property of the iPT, the statistical properties of
tensor values attached to the astronomical objects are related to the
underlying mass density field through the renormalized bias functions,
defined by Eq.~(\ref{eq:295}). In most of the treatments of
perturbation theory in literature (for review, see
Refs.~\cite{Bernardeau:2001qr,Desjacques:2016bnm}), the bias relations
are given by Taylor expansions, or their generalizations, of the
biased field in terms of the mass density field. In other words, the
biasing relations are treated in a perturbative manner, assuming the
fluctuations of the mass density field are small enough in the process
of biasing. In reality, however, the last assumption is not the case.
The biasing consists of fully nonlinear, nonlocal, and complicated
processes of galaxy formation and structure formation. The tensor
values attached to galaxies, such as angular momenta and shapes, are
also determined by fully nonlinear and nonlocal processes. In the iPT,
the statistics of the biased field are expanded assuming only the
smallness of the clustering on interested scales, which is represented
by the linear power spectrum $P_\mathrm{L}(k)$, but not assuming the
smallness of mass density field in the biasing process which takes
place on fully nonlinear scales. In the iPT, the latter complicated
dynamics are all encoded in the renormalized bias functions, which are
not perturbative quantities. Therefore, the formalism of Papers~I, II
and this paper is considered to overcome the shortcomings of the
perturbative treatment in other perturbation theories.

However, it is also true that determining the precise functional forms
of renormalized bias functions is not straightforward, because of the
complicated process of galaxy formation, and we need to model the
functions in numerically calculating the loop corrections for the
higher-order effects, as demonstrated in Paper~II. However, in the
lowest-order approximations, the power spectra and correlation
functions of tensor fields are all proportional to the linear power
spectrum of the same scale, and relevant first-order renormalized bias
functions are naturally expected to be constant in the large-scale
limit, ending up with a limited number of free parameters.
Beyond the lowest-order approximation, explorations of the sensible
modeling of the galaxy formation are required. 

In this paper, we focus on the two-point statistics, i.e., the power
spectra and correlation functions of projected tensor fields. We also
assume Gaussian initial conditions for simplicity. In Paper~I, we
consider higher-order statistics in the possible presence of
primordial non-Gaussianity. As shown in this paper, the statistics of
projected tensor fields are derived from the statistics of
three-dimensional tensor fields, and one can straightforwardly derive
the formulas of the higher-order statistics of projected tensor
fields, and also derive formulas with the effects of primordial
non-Gaussianity. The purpose of this paper, and other papers in the
series is to develop methods from the general point of view, not
focusing on individual targets of observations. Applications of the
present formalism to predict various quantities from observational
perspectives are left for future work. In the next Paper~IV
\cite{PaperIV}, we generalize the formalism developed in the series of
papers by relaxing the distant-observer approximation which is adopted
in Papers~I, II, and the present paper.

\begin{acknowledgments}
  I thank T.~Okumura and A.~Taruya for useful discussions. This work
  was supported by JSPS KAKENHI Grants No.~JP19K03835 and
  No.~21H03403.
\end{acknowledgments}

\appendix

\onecolumngrid

\section{\label{app:FlatSWSH}%
  Flat-sky limit of the spin-weighted spherical harmonics
}

In this Appendix, we derive an explicit expression of the flat-sky
limit of spin-weighted spherical harmonics. Similar considerations for
spin-2 polarization fields of CMB are found in the literature
\cite{Zaldarriaga:1996xe,Hu:2000ee}, and we generalize the
consideration to general spin fields.

The spin-weighted spherical harmonics $\sY{\pm s}{lm}(\theta,\phi)$
satisfy the recursion relations \cite{Newman:1966ub,Goldberg:1966uu},
\begin{equation}
  \eth_\pm\left(\sY{\pm s}{lm}\right)
  = \pm \sqrt{(l-s)(l+s+1)}\,\sY{\pm(s+1)}{lm},
  \label{eq:362}
\end{equation}
where the differential operators $\eth_+$ and $\eth_-$ correspond to
$\eth$ and $\bar{\eth}$, respectively, in common notations in
literature. The LHS of the above equation is explicitly given by
\begin{equation}
  \eth_\pm\left(\sY{\pm s}{lm}\right)
  = -
  \left(
    \frac{\partial}{\partial\theta}
    \pm \frac{i}{\sin\theta} \frac{\partial}{\partial\phi}
    - s\,\cot\theta
  \right) \sY{\pm s}{lm}.
  \label{eq:363}
\end{equation}
In the flat-sky limit, we have $\theta \ll 1$, and
$\partial/\partial\theta \pm (i/\theta)\partial/\partial\phi \approx
e^{\mp i\phi}(\partial_1 \pm i\partial_2)$, where $\partial_i$ with
$i=1,2$ are Cartesian derivatives along the first and second axes on
the unit sphere, i.e., $\theta_1 = \theta\cos\phi$,
$\theta_2 = \theta\sin\phi$ and
$\partial_i = \partial/\partial \theta_i$. Thus the above equation
reduces to
\begin{equation}
  \eth_\pm\left(\sY{\pm s}{lm}\right)
  \approx -
  \left[
    e^{\mp i\phi}\left(\partial_1 \pm i\partial_2\right)
    - \frac{s}{\theta}
  \right] \sY{\pm s}{lm}
  \label{eq:364}
\end{equation}
in the flat-sky limit. Using the last equation and Eq.~(\ref{eq:362}),
it is straightforward to show the following asymptotic formula in the
flat-sky limit by induction:
\begin{equation}
  \sY{\pm s}{lm}(\theta,\phi) \approx
  \frac{(\mp 1)^s}{(l+1/2)^s}
  e^{\mp is\phi}
  \left(\partial_1 \pm i\partial_2\right)^s
  Y_{lm}(\theta,\phi),
  \label{eq:365}
\end{equation}
where $Y_{lm}=\sY{0}{lm}$ is the usual spherical harmonics.

% \begin{figure}%[t]
% \begin{center}
% \includegraphics[width=18pc]{FigA01.eps}%{iPTdiag1.eps}
% \caption{\label{fig:iPTdiag1}
% Diagrammatic rules of the iPT: dynamics and biasing.
% }
% \end{center}
% \end{figure}

\twocolumngrid
%%%%%%%%%%%%
\renewcommand{\apj}{Astrophys.~J. }
\newcommand{\aap}{Astron.~Astrophys. }
\newcommand{\aj}{Astron.~J. }
\newcommand{\apjl}{Astrophys.~J.~Lett. }
\newcommand{\apjs}{Astrophys.~J.~Suppl.~Ser. }
\newcommand{\apss}{Astrophys.~Space Sci. }
\newcommand{\cqg}{Class.~Quant.~Grav. }
\newcommand{\jcap}{J.~Cosmol.~Astropart.~Phys. }
\newcommand{\mnras}{Mon.~Not.~R.~Astron.~Soc. }
\newcommand{\mpla}{Mod.~Phys.~Lett.~A }
\newcommand{\pasj}{Publ.~Astron.~Soc.~Japan }
\newcommand{\physrep}{Phys.~Rep. }
\newcommand{\ptp}{Progr.~Theor.~Phys. }
\newcommand{\ptep}{Prog.~Theor.~Exp.~Phys. }
\newcommand{\jetp}{JETP }
\newcommand{\jhep}{Journal of High Energy Physics}
%\newcommand{\prl}{Phys. Rev. Lett.}
%\renewcommand{\prd}{Phys.~Rev.~D}

%\bibliography{redoneloop}% Produces the bibliography via BibTeX.

\end{document}